\documentclass[%
 aip,
 jcp,
 amsmath,amssymb,
preprint,%
floatfix,
]{revtex4-1}

\usepackage{graphicx}
\usepackage{dcolumn}
\usepackage{bm}

\usepackage[utf8]{inputenc}
\usepackage[T1]{fontenc}
\usepackage{mathptmx}
\usepackage{physics}
\usepackage{subcaption}
\usepackage{csquotes}
\usepackage{float}
\usepackage{multirow}

\graphicspath{{.}{figs/}{figs/logos/}}
\newcommand{\freeenergy}{\mathcal{F}}

\newcommand{\eps}[1]{\epsilon_{\text{#1}}}
\newcommand{\sig}[1]{\sigma_{\text{#1}}}

\begin{document}
\preprint{AIP/123-QED}
\title{Wetting dynamics under periodic switching on different scales: Characterization and mechanisms} 
\author{Leon Topp}%
\affiliation{Institute for Physical Chemistry, University of Münster, Corrensstr. 28/30, 48149 Münster, Germany}%
\author{Moritz Stieneker}
 \thanks{L.\,Topp and M.\,Stieneker contributed equally to this work.}
 \affiliation{Institute for Theoretical Physics, University of Münster, Wilhelm-Klemm-Str. 9, 48149 Münster, Germany}%
 \affiliation{Center of Nonlinear Science (CeNoS), University of Münster, Corrensstr. 2, 48149 Münster, Germany}%

\author{Svetlana Gurevich}
\email{gurevics@uni-muenster.de}
 \affiliation{Institute for Theoretical Physics, University of Münster, Wilhelm-Klemm-Str. 9, 48149 Münster, Germany}
 \affiliation{Center of Nonlinear Science (CeNoS), University of Münster, Corrensstr. 2, 48149 Münster, Germany}
 
\author{Andreas Heuer}
\email{andheuer@uni-muenster.de}
\affiliation{Institute for Physical Chemistry, University of Münster, Corrensstr. 28/30, 48149 Münster, Germany}%
 \affiliation{Center of Nonlinear Science (CeNoS), University of Münster, Corrensstr. 2, 48149 Münster, Germany}
\affiliation{Center for Multiscale Theory and Computation (CMTC), University of Münster, Corrensstr. 40, 48149 Münster, Germany}

\date{\today}

\begin{abstract}

The development of substrates with a switchable wettability is on a fast pace. The limit of switching frequencies and contact angle differences between substrate states are steadily pushed further. We investigate the behavior of a droplet on a homogeneous substrate, which is switched between two wettabilities for a large range of switching frequencies. Here, we are particularly interested in the dependence of the wetting behavior on the switching frequency. We show, that results obtained on the particle level via molecular dynamics simulations and on the continuum level via the thin-film model are consistent. Predictions of simple models as the molecular theory of wetting (MKT) and analytical calculations based on the MKT also show good agreement and offer deeper insights into the underlying mechanisms.
\end{abstract}

\maketitle

\section{Introduction}
Controlling and understanding the movement of droplets is mandatory for microfluidic, e.\,g. Lab-On-A-Chip devices \cite{SqQu2005rev.mod.phys.}. Surfaces with wettability gradients \cite{Ichimura1624, ChWh1992s} and adaptive substrates \cite{karpitschka2015droplets,karpitschka2016liquid} have been the focus of investigation for quite some time. 
Recent experiments have shown that coating a substrate with a self-assembled monolayer (SAM) of photoswitchable moieties like azobenzenes or spiropyranes leads to surfaces, whose wettability can be controlled by illumination with light of a defined wavelength \cite{Ishihara1982, Rosario2002} which makes the control of the droplets motion possible by applying a light gradient \cite{Ichimura1624}. Another possible dynamical behavior one can think of are oscillatory motions. This behavior can be induced and were investigated in the context of e.g., vibrating plates or electrowetting \cite{PhysRevLett.99.144501, Sartori_2015, doi:10.1063/1.2204831}. Oscillatory motions of droplets can show interesting effects like making a droplet move up on an inclined, vibrating plate \cite{PhysRevLett.99.144501, Sartori_2015}. As such, the opportunity to change surface properties in time has a variety of applications, e.\,g. it can be used in devices, which measure liquid properties such as surface tension~\cite{Meier2000, ZOGRAFOV2014351}. It can also be applied to mix liquids inside a drop \cite{doi:10.1063/1.2204831}, which again is useful for the design of Labs-On-A-Chip\cite{SqQu2005rev.mod.phys.}.

The emergence of a variety of novel surfaces with a switchable wettability property \cite{Ishihara1982,Rosario2002, tio22005, ruehe2012, Wang1997, Sun2001}
has led to increased efforts in the theoretical realm to advance the understanding of the dynamics of liquids on such surfaces.
Theoretical investigations of the dynamical behavior of droplets on surfaces with an oscillating wettability can be performed with a variety of different models reflecting different time and spatial scales. 
 For instance, the boundary element method applied to Stokes flow~\cite{chan_mcgraw_salez_seemann_brinkmann_2017, GrSt2021softmatter,D1SM01113H}, mesoscopic models based on the lubrication approximation~\cite{HLHT2015w,GAUGLITZ19881457} and microscopic models like molecular dynamics~\cite{C3SM51508G,PhysRevLett.82.4671,doi:10.1021/la062920m} have all been successfully applied to dynamic wetting problems. The inherent complexity in such models often makes it hard to pinpoint the physical origin of the effects in question and hence, comparison with minimal models can help to understand the physical effects at work.

In this work, we present numerical solutions of a mesoscopic Thin Film (TF) model and particle-based Molecular Dynamics (MD) simulations. In particular, we characterize the wetting behavior of droplets upon periodically varying wettabilities. The degree of wettability reflects the strength of the interaction of the droplet with the switchable surface. We compare our results to the molecular kinetic theory of wetting (MKT). After its introduction by \citeauthor{BLAKE_mkt}\cite{BLAKE_mkt}  it was widely and successfully used in the context of dynamic wetting~\cite{ruijter1999, seveno2009, duvivier2013}. The MKT expresses how the deviation of a   non-equilibrium contact angle to its equilibrium value gives rise to a contact line velocity of the droplet,  promoting the approach to equilibrium. Note that in our mesoscopic and particle-based approaches the contact line velocity is inherent in the respective models. It is not, e.g., influenced by boundary conditions as in \cite{grawitter2018feedback} and does not need to be additionally imposed as in \cite{GrSt2021softmatter}.

This paper is organized as follows: In Sec.~\ref{sec:theory} we explain our simulation setup of the MD and the TF theory and introduce the mapping procedure we used to compare both methods. Additionally, we introduce the background of the MKT and present some analytical solutions when applying the MKT to switchable substrates with periodic wettability changes. For a quantitative comparison to the MKT, relevant input parameters need to be extracted from other models as presented in Sec.~\ref{sec:results_mkt}. Among others we  refer to a method introduced by \citeauthor{ruijter1999} \cite{ruijter1999}. After checking the applicability of our mapping scheme in Section~\ref{sec:md_tf_mapping}, we present the results obtained from the three methods (MD, TF, MKT) for the dynamics of a droplet on a periodically switched substrate and compare them in Section~\ref{sec:results_ps}. Finally, we conclude our results in Section~\ref{sec:conclusion}.

\section{Theoretical Background} \label{sec:theory} 
\subsection{Molecular Dynamics Simulations}\begin{figure}
\includegraphics[width=\columnwidth]{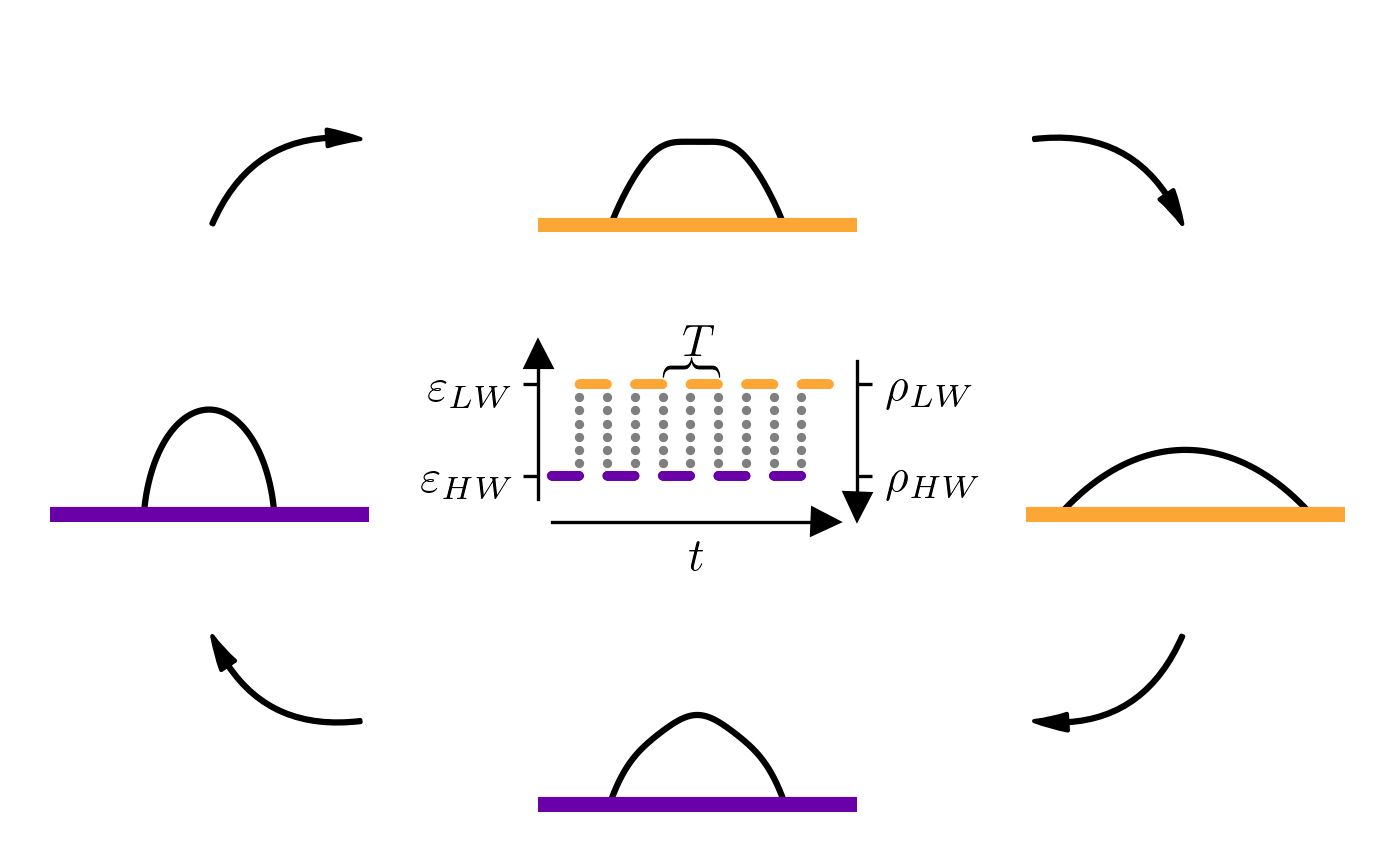}
\caption{Schematic of the periodic switching procedure. The droplet states, shown on the left and the right, respectively, correspond to the situation directly before switching to the other wettability (from violet to orange or vice versa). For very long switching periods $T$ they are the respective equilibrium states, otherwise they are non-equilibrium states. The top and bottom droplets express intermediate non-equilibrium states after the switching process.  The central panel shows, how the parameters, responsible for the substrate's wettability in MD and TF models, change with time (HW: high wettability; LW: low wettability).}
\label{fig:schematic}
\end{figure}
We use the same simulation setup as presented in \cite{StienekerToppEtAl}. 
In particular, we are performing simulations in the NVT ensemble with the framework HOOMD \cite{anderson_hoomd, phillips_hoomd_dpd}.
Two types of particles are present in the system, namely substrate particles (here denoted with "$s$"), which are frozen in two layers of a fcc(111) surface and fluid particles (denoted with "$f$").
The particles interact through the Lennard-Jones potential
\begin{equation}
  V(r_{lj}) = 4 \eps{lj} \left[ \left(\frac{\sig{lj}}{r_{\text{lj}}}\right)^{12} - \left( \frac{\sig{lj}}{r_{\text{lj}}} \right)^{6} \right],
\end{equation}
where $r_{\text{lj}}$ is the distance between two particles $\text{l}$ and $\text{j}$, $\eps{lj}$ the interaction strength between
the particles and $\sig{lj}$ the arithmetic mean of the particles' diameters $\sig{l}$ and $\sig{j}$.
The potential is truncated and shifted at a cutoff-radius of $r_{c} = 2.5 \sigma$. The interaction strength $\eps{lj}$ of two particles is calculated as the geometric mean of the self-interaction parameters of both particles $\eps{l}$ and $\eps{j}$. We set $\eps{f} = 1$ and change the wettability of the substrate by varying the parameter $\eps{s}$ which then varies the interaction $\eps{w}$ between substrate and liquid particles which by construction is given as the geometric mean $\eps{w} = \sqrt{\eps{s}\cdot \epsilon_{f} } =\sqrt{\eps{s} } $.
The schematics in Fig.~\ref{fig:schematic} visualizes the resulting behavior after periodic switching.

We set the particles diameter to $\sig{l} = \sig{s} = \sigma$ for all particles and use a time step  of $\tau = \sigma^{-1}\sqrt{\epsilon_{f}/M}/200$. The reduced temperature which is controlled by a dissipative particle dynamics (DPD) thermostat \cite{Hoogerbrugge_1992} is $\frac{k_{B}T}{\epsilon_{f}}=0.75$, where $M$ is the particle mass. Periodic boundary conditions are present in the $x$- and $y$-direction. The substrate is placed in the $xy$-plane.

The domain size in $y$-direction is chosen in such a way that Plateau-Rayleigh instabilities are suppressed and the effect of line tension is excluded by simulating a cylindrical droplet. All simulations are carried out with a total amount of $N = 4 \cdot 10^4$ fluid particles.
Every simulation setup is averaged over 50 trajectories.

To determine the contact angle of the droplet we calculate the density field of the fluid particles by averaging over the $y$-direction. From this we can determine the position of the liquid vapor interface with a tanh-fit. Then we perform a circular fit to the obtained positions of the liquid vapor interface and calculate the contact angle from this fit.

\subsection{Thin film equation theory (TF)}
On the mesoscopic scale, we model the evolution of the local height $h=h\qty(x,y,t)$ of the film or a droplet with the well-established
lubrication approximation that can be derived from the Navier-Stokes equation \cite{OrDB1997rmp}. For the simulation we employ the finite element library oomph-lib\cite{HeHa2006}.
We reduce the spatial dimension of the problem by assuming transversal symmetry in $y$-direction, i.\,e. we simulate cylindrical droplets and exclude transversal instabilities. It leads to an evolution equation in gradient dynamics form as \cite{Thie2010}
\begin{equation}
\partial_t h = \nabla \cdot\qty[M(h)\nabla \fdv{\freeenergy}{h}] \label{eq:tfe_gradient_form}
\end{equation}
with the mobility $M(h)$ and the free energy functional $\freeenergy=\freeenergy\qty[h]$. The no-slip boundary condition at the substrate leads to a mobility of $M(h)=h^3/(3\eta)$ with the dynamic viscosity $\eta$\cite{OrDB1997rmp}. The generalized pressure $P = \fdv{\freeenergy}{h}$ is given by
\begin{equation}
P(h,x,t) = -\gamma \Delta h - \Pi(h,t)\label{eq:pressure}
\end{equation}
with the surface tension $\gamma$ and the disjoining (or Derjaguin) pressure $\Pi(h,t)$. Different choices for the wetting potential and corresponding disjoining pressure are possible\cite{OrDB1997rmp}. Here, we choose
\begin{equation}
\Pi(h,t) = \qty(\frac{C}{h^6}-\frac{D}{h^3})\qty(1+\rho(t)) \label{eq:disjoining}.
\end{equation}
with the interaction strengths of long and short ranging forces $C$ and $D$ respectively. $C$ can be directly connected to the Hamaker constant $H$ by $C=H/6\pi$\citep{Engelnkemper2017}. 

Analogous to \citeauthor{HLHT2015w}\cite{HLHT2015w} we incorporated a change in wettability by modulating the disjoining pressure\cite{Thie2010}. In Eq.~\eqref{eq:disjoining} we call the parameter $\rho=\rho\qty(t)$ wettability, because its value modulates the disjoining pressure and determines the current wettability of the system.
To resolve the problem of a logarithmic energy dissipation at the contact line\cite{BEIM2009romp}, which is a consequence of the no-slip boundary condition, a precursor film with height $h_p$ is introduced \cite{Engelnkemper2017,BEIM2009romp}. Such a precursor film is also present on macroscopically \textquote{dry} parts of the substrate.
 
The oomph-lib numerical implementation is based on the non-dimensionalized form of Eq.~\eqref{eq:tfe_gradient_form}, where the quantities $h$, $x$ and $t$ are scaled in such a way, that $\gamma$, $C$, $D$ and $3\eta$ vanish from the evolution equation, yielding

\begin{equation}
\partial_t h = \nabla \cdot\qty{h^3\nabla\qty(-\Delta h -\frac{5}{3}\theta_{\text{eq}}^2\chi^2\qty(\frac{\chi^3}{h^6}-\frac{1}{h^3})\qty[1+\rho(t)] )} \label{eq:tfe}
\end{equation}
with the equilibrium contact angle $\theta_{\text{eq}}$ and the parameter $\chi =h_p/h_0$, where $h_0$ is the spatial scale. In what follows, for further analysis we subtract the precursor film height $h_p$ from the film height $h$ in Eq.\,\eqref{eq:tfe}. In all the simulations presented here, we use $\chi = 0.01$. Larger values would lead to deviations in the contact region, because the precursor film height would not be small enough compared to droplet heights anymore. Smaller values of $\chi$ would increase computation time as a higher spatial discretization would be required\cite{StienekerToppEtAl}.

In the TF model the contact region exhibits a smooth transition to the precursor film and there is no sharp contact line. Also the underlying lubrication approximation makes droplets in the TF model deviate from a strict spherical cap shape. Thus, contact angle measurements in the TF model are finicky. Even the method for the measurement can make a difference \cite{Kubochkin2021}. 
Instead of direct contact angle measurements we use the relative full width at half maximum rFWHM, which is the width at half maximum in relation to the drop height. This measure can be determined stably and does not depend on the contact region. Also, the cosine of the contact angle can be directly computed for a given rFWHM according to 
\begin{equation}
\cos(\theta) = 1 - \frac{4}{\frac{1}{rFWHM^2}+1}, \label{eq:rFWHM_to_theta}
\end{equation}
which can be derived based on the assumption of a spherical cap shape. The derivation can be found in SI.~\ref{app:rFWHM_to_theta}.

\subsection{Mapping between TF and MD}\label{sec:theory_mapping}

Following our previous work \cite{StienekerToppEtAl} the mapping occurs in two steps. First, based on equilibrium properties we have to map the interaction strengths $\eps{w}$ of the MD model and $\rho$ of the TF model such that identical equilibrium contact angles emerge.  
With the help of parameter scans in both wettability parameters we obtained an invertible mapping $\eps{w} \mapsto \rho$. To avoid confusion between these parameters we only mention the relevant $\eps{w}$ values and mean the corresponding $\rho$ value, when showing TF results. 

In the second step we determine the respective time scales when studying  the relaxation of $ \cos \theta(t)$ for a switching process from $\epsilon_1$ to $\epsilon_2$.  As the relaxation of $ \cos \theta$ does not necessarily have an exponential shape, we fit the time evolution with a stretched exponential function 
\begin{equation}
f(t) = f_{\infty} + (f_{0} - f_{\infty}) \exp\qty(-\qty(\frac{t}{\tau_0})^\beta)\label{eq:stretched_exponential}.
\end{equation}
We define the relaxation time of $f(t)$ as
\begin{equation}
\label{eq:moment}
\tau  = \frac{\tau_0}{\beta} \Gamma \left(\frac{1}{\beta}\right)
\end{equation}
where $\Gamma$ is the gamma function.

\subsection{Molecular Theory of Wetting}
The molecular kinetic theory of wetting (MKT) \cite{BLAKE_mkt} accounts for the dissipation of a moving droplet in the contact line region. This yields an equation which relates the velocity of the three phase contact line $v_{cl}$ to the time dependent cosine of the contact angle $\cos(\theta(t))$
\begin{equation}
\label{eq:mkt}
v_{cl} = \frac{\gamma}{\zeta} (\cos(\theta_{eq}) - \cos(\theta(t))).
\end{equation}
Here, $\gamma$ is the surface tension, $\zeta$ a friction coefficient and $\theta_{eq}$ the equilibrium contact angle. In general, the ratio $\gamma/\zeta$ can be obtained from measuring the contact line velocity in dependence of the contact angle. On a deeper level, as shown below, for the particle-based model the values of $\gamma$ and $\zeta$ can be separately estimated from additional MD simulations.

If we assume a circular shape of the droplet during the whole spreading or contracting process we can rewrite Eq.~\eqref{eq:mkt} as
\begin{equation}
\label{eq:mkt_circle}
\frac{d}{dt} \cos(\theta) = -k_{1} \frac{\sin(\theta)}{g(\theta)} (\cos(\theta_{eq}) - \cos(\theta))
\end{equation}
with
\begin{equation}
g(\theta) = \frac{\cos(\theta)}{\qty(2\theta - \sin(2\theta))^{\frac{1}{2}}} - \frac{\sin(\theta) (1-\cos(2\theta))}{\qty(2\theta - \sin(2\theta))^{\frac{3}{2}}}
\end{equation}
and $k_{1} = \frac{\gamma/\zeta}{\sqrt{2A}}$ where $A$ is the area of the 2D droplet (cf.~SI.~\ref{app:reform_mkt}). For the specific set of the MD simulation we have $A = 833~\sigma^{2}$ whereas in the TF model we have $A=4.04$. Consideration of both areas implies a spatial mapping of both approaches.
To good approximation $\frac{-\sin(\theta)}{g(\theta)}$ can be written as $3(1-\cos(\theta))$ in the whole range of contact angles relevant for a comparison to the TF model (cf. SI.~\ref{app:reform_mkt} for details). As a consequence, Eq.~\eqref{eq:mkt_circle} can be simplified as
\begin{equation}
\label{eq:mkt_intermediate}
\frac{d}{dt} \cos(\theta) = k_{2} (1-\cos(\theta)) (\cos(\theta_{eq}) - \cos(\theta)),
\end{equation}
where $k_{2} = 3k_{1}$. 

Now we present analytical solutions, based on Eq.~\eqref{eq:mkt_intermediate}. Details are provided in the Supplementary Information. 

{\it Single switching process}: First we study the case of the time evolution of a droplet after a single switching process. In what follows we denote $\cos(\theta(t))$ as $x(t)$. Its value before the switching process is denoted $x_0$. Here we assume that this is the equilibrium value, corresponding to the initial wettability. Furthermore, its equilibrium value, reached in the long-time limit after the switching process, is denoted $x_{eq}$. Finally, we introduce
\begin{equation}
\label{eq:normalized_x}
y(t) = \frac{x(t) - x_{eq}}{x_0 - x_{eq}}
\end{equation}
which is the normalized version of $x(t)$, with the properties $y(t=0)=1$ and $y(t=\infty)=0$. The subsequent results are expressed in terms of $y(t)$ for reasons of simplicity. For the later applications they can be easily reformulated in terms of $x(t)$.
After a short calculation (cf. SI.~\ref{app:reform_mkt}) one obtains
\begin{equation}
\label{eq:mkt_approx_solved}
y(t) = \frac{(1-x_{eq})\exp(-k_{3}t)}{1-x_{0}+\exp(-k_{3}t)(x_{0}-x_{eq})}.
\end{equation}
with $k_3 = k_2 (1 - x_{eq})$. 
When identifying the relaxation time $  \tau  $ as the integral over the normalized relaxation function $y(t)$ one gets after a straightforward calculation
\begin{equation}
 \tau   = \frac{1}{k_2}\frac{1}{x_0-x_{eq}} \ln \left (1 + \frac{x_0-x_{eq}}{1-x_0} \right) 
\label{eq:mkt_moment}
\end{equation}

For $|x_0 - x_{eq}| \ll 1$, i.e. small wettability changes, Eq.~\ref{eq:mkt_approx_solved}  can be approximated as   (cf. SI.~\ref{app:reform_mkt}) 
\begin{equation}
\label{eq:mkt_approx_solved_approximated}
y(t) \approx  \frac{1-x_{eq}}{1-x_{0}} \left[ \exp(-k_{3}t) - \frac{x_{0}-x_{eq}}{1-x_{0}} \exp(-2k_{3}t) \right].
\end{equation}

{\it Peridoic switching processes}: Here we directly start from the limit of small wettability changes. In this limit $x(t)$ is always close to $x_{eq}$ so that in Eq.~\eqref{eq:mkt_intermediate} we may approximate the prefactor $1-x(t)$ as $1 - x_{eq}$, yielding
\begin{equation}
\label{eq:mkt_small}
\frac{d}{dt} x(t) = k_{3} (x_{eq} - x(t)),
\end{equation}
with $k_{3} = k_{2} (1-x_{eq})$. Indeed, this is the same definition of $k_3$ as automatically resulting in the exact calculation in Eq.~\eqref{eq:mkt_approx_solved}.
For the sake of simplicity we again  consider the normalized version $y(t)$ of the cosine of the contact angle  so that we can write Eq.~\eqref{eq:mkt_small} as
\begin{equation}
\frac{d}{dt} y(t) = -k_{3,i}(y-a_{i}), \label{eq:mkt_small_with_x}
\end{equation}
where $k_{3,i}$ and $a_{i} \in \{0,1\}$ are the prefactor and the normalized equilibrium contact angle, respectively. Now we take into account that the wettability is periodically varied, one full cycle taking a time $T$.  Starting from a droplet equilibrated at a higher wettability (corresponding to the normalized contact angle $a=1$) we set $i=\downarrow$ and $a_{1} = 0$ for the first half of a switching period and $i=\uparrow$ and $a_{2} = 1$ for the second half of the period (cf. the schematic in Fig.~\ref{fig:schematic}). After a straightforward but slightly tedious analysis (cf.~SI.~ \ref{app:plateau_calculations} for details) the normalized cosine of the contact angle, averaged over one period, can be derived as
\begin{equation}
\label{eq:y_n}
\langle y(n) \rangle = y_{plateau} \pm \exp\qty(-K_3(n+\frac{1}{2})T/2) \cdot \hat{y}_\pm
\end{equation}
with $K_3 = k_{3,\downarrow} + k_{3,\uparrow}$. $\langle y(n)\rangle$ denotes the average of $y(t)$ between $t=nT$ and $t=(n+1)T$ and is evaluated at the time $t=(n+1/2)T$.
The $\pm$ distinguishes an initial higher wettability ($+$) from an initial lower wettability ($-$). The plateau value $y_{plateau}$, obtained in the limit of an infinite number of switching cycles $n$, is given by
\begin{equation}
\begin{split}
\label{eq:y_plateau}
    y_{plateau} = \frac{1}{2} \Bigg\lbrace  \frac{(1-\exp(-k_{3,\uparrow}T/2))(1-\exp(-k_{3,\downarrow}T/2)}{1-\exp(-K_{3}T/2)} \\
    \left(\frac{1}{k_{3,\downarrow}T/2}
  - \frac{1}{k_{3,\uparrow}T/2}\right) + 1 \Bigg\rbrace.
\end{split}
\end{equation}
This expression is independent of the starting wettability.
In the limit case for very fast switching, i.e. small $T$, Eq.~\eqref{eq:y_plateau} boils down to
\begin{equation}
  \label{eq:y_fast_switching}
 y_{plateau} = \frac{k_{3,\uparrow}}{k_{3,\downarrow}+k_{3,\uparrow}} - \frac{k_{3,\uparrow} - k_{3,\downarrow}}{96(k{3,\downarrow} + k_{3,\uparrow})} k_{3,\downarrow} k_{3,\uparrow} T^{2}.
\end{equation}
The amplitude $\hat{y}_\pm$ depends on the initial condition. If we start with the higher wettability, the amplitude $\hat{y}_+$  turns out to be
\begin{equation}
\label{eq:amplitude}
\begin{split}
\hat{y}_+ &= \exp( K_{3}T/4) \frac{\exp(-k_{3,\uparrow}T/2)-\exp(-K_{3}T/2)}{1-\exp(-K_{3}T/2)} \cdot \Bigg[\frac{1}{2k_{3,\downarrow}T/2} \\
  &\cdot  (1-\exp\qty(-k_{3,\downarrow}T/2)) + \exp\qty(-k_{3,\downarrow}T/2) \frac{1}{2 k_{3,\uparrow} T/2} \qty(1-\exp\qty(-k_{3,\uparrow}T/2)) \Bigg].
\end{split}
\end{equation}
$\hat{y}_-$ can be obtained by exchanging  $k_{3,\downarrow}$ with $k_{3,\uparrow}$. 
For the limit of very fast switching, i.e. $T \rightarrow 0$, this expression reduces to $ \hat{y}_+ =  (1 - y_{plateau})$ and $ \hat{y}_- =   y_{plateau}$, respectively.

Although this calculation has been performed for small changes of the wettability, the range of applicability can be increased. Guided by Eq.~\eqref{eq:mkt_circle} one may substitute $k_1 \sin(\theta)/g(\theta)$ by $ k_1 \sin(\langle \theta \rangle )/g(\langle \theta \rangle)$.
$k_{3,i}$ in Eq.~\eqref{eq:y_n} and all corresponding formulas can be substituted by this value. Convenient is the choice of the contact angle $\theta$ as calculated via  $\cos \langle \theta \rangle  =(1/2)(\cos \theta_{eq}(\epsilon_{HW} ) + \cos \theta_{eq}(\epsilon_{LW}))$ which is the average angle for large switching times $T$. The resulting value, is then denoted $\tilde{k}_{3,i}$ which would substitute the corresponding value of $ {k_{3,i}}$. Note that for $T \rightarrow 0$ the value of $y_{plateau}$ is insensitive to this choice because it only depends on the ratio of the $k_1$-values.

\section{Results}

\subsection{Relation between contact line velocity and contact angle}

\label{sec:results_mkt} 
The friction coefficient $\zeta$, appearing in the MKT, can be directly computed from quantities calculated from MD simulations as shown by \citeauthor{mkt_zeta_calc}\cite{mkt_zeta_calc} (further on denoted as '$\zeta_{R}$' with $R$ for de Ruijter). 
It can be expressed as
\begin{equation}
\zeta_{R} = n k_{b} T / K_{0} \lambda, \label{eq:zeta_from_parameters}
\end{equation}
where $n$ is the number of absorption sites per unit area on a solid, $K_0$ is the equilibrium frequency for particle displacements parallel to the solid and $\lambda$ is the characteristic length of the displacement.
We adopt the scheme of that reference to determine $K_{0}$ as the inverse of the time where half of the particles have moved between the first and second liquid layer in $z$-direction.  We set $\lambda = 1~\sigma$ since this is the distance between the first two liquid layers over the substrate and $n$ is taken to be the density of the first liquid layer. The measured values of $K_{0}$, $n$ and the resulting value of $\zeta_{R}$ via Eq.~\eqref{eq:zeta_from_parameters} can be found in SI \ref{app:mkt_values}.

The obtained value of $\gamma = (0.477 \pm 0.005)~\epsilon/\sigma^{2}$ is comparable to values from the literature for similar systems \cite{Toxvaerd2007}. It was calculated by simulating a slab of liquid, which was placed in the $xy$-plane between its own vapor phase, and integrating the difference between the normal $p_{n}$ and tangential $p_{t}$ part of the pressure tensor over the box length in $z$-direction \cite{calc_surf_tension}
\begin{equation}
  \gamma = \int \mathrm{d}z~p_{n}(z) - p_{t}(z).
\end{equation}

\begin{figure*}
\begin{tabular}{ll}
(a)&(b)\\
\includegraphics[width=0.45\textwidth]{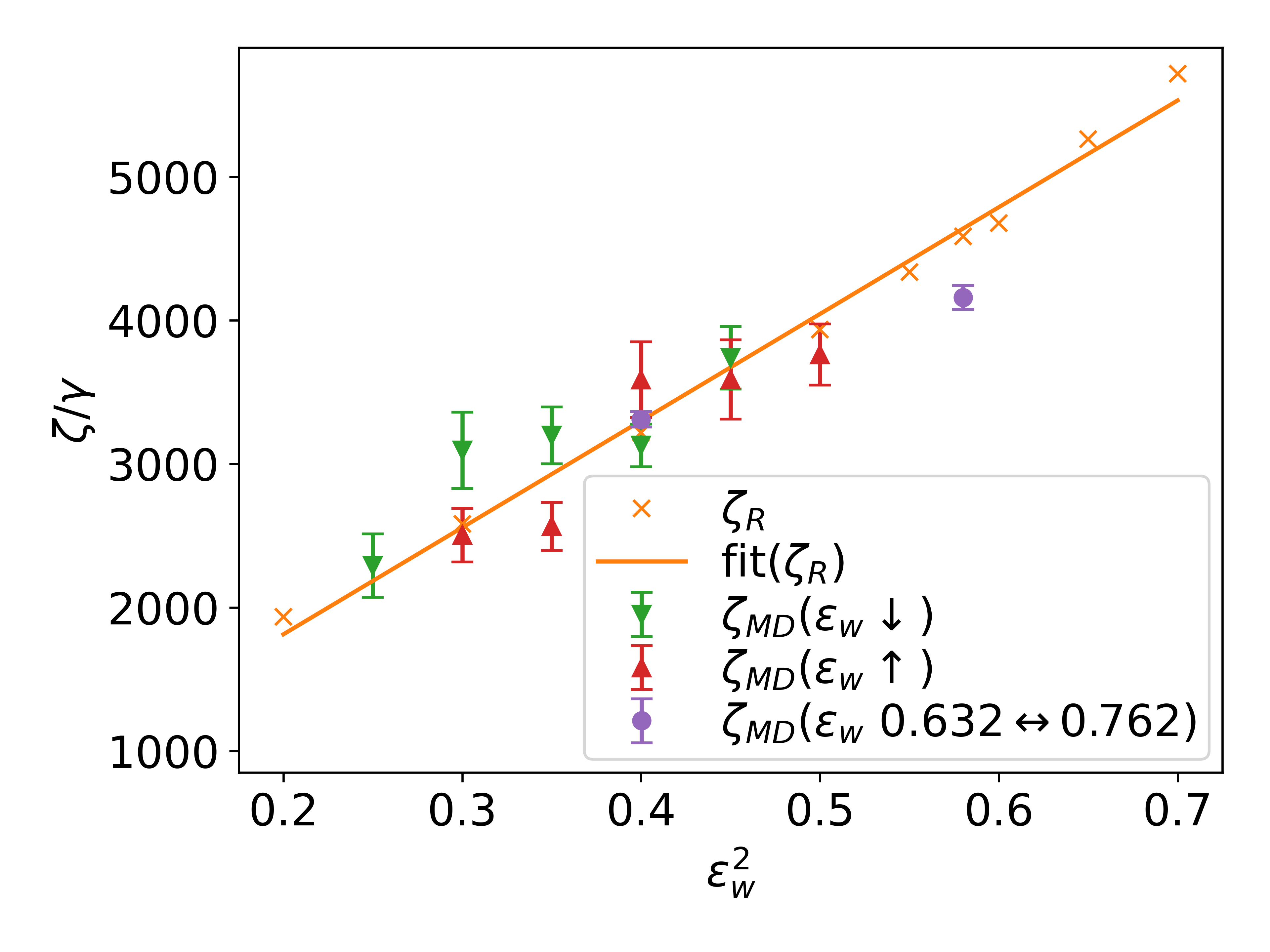} &
\includegraphics[width = 0.45\textwidth]{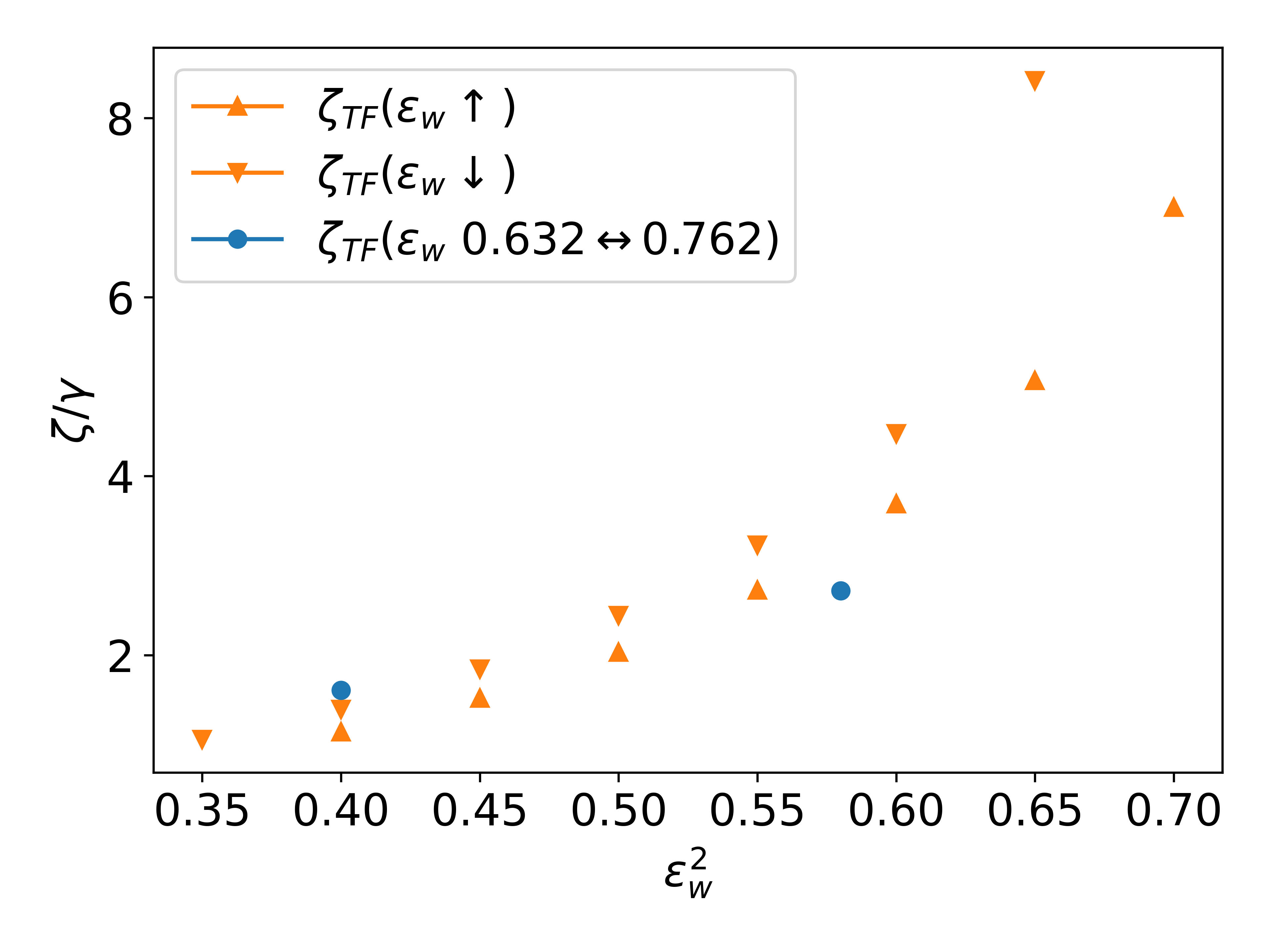} 
\end{tabular}
\caption{(a) Values of $\zeta_{R}$ and $\zeta_{MD}$ plotted against $\epsilon_{w}^2$. The solid line is a linear fit through the values of $\zeta_{R}$. $\blacktriangle$ and $\blacktriangledown$ marks values of $\zeta_{MD}$ calculated from simulations where the wettability of the surface is changed by $\Delta \epsilon_{w} = 0.05$ from a higher wettability $\blacktriangledown$ and from a lower wettability $\blacktriangle$, respectively. $\bullet$ marks values obtained from fits of Eq.~\eqref{eq:mkt} for switching between $\epsilon_{w} = 0.632$ and $\epsilon_{w} = 0.762$. The error bars show one standard deviation errors. (b) $\zeta_{TF}$ plotted against $\epsilon_{w}^2$ for TF simulations. }
\label{fig:zeta_vs_eps}
\end{figure*}

 Finally, the resulting ratios $\zeta/\gamma$ are shown in Fig.~\ref{fig:zeta_vs_eps} in dependence of the squared wetting energy $\epsilon_w^2$ which to a good approximation displays a linear behavior.

\begin{figure*}
\begin{tabular}{ll}
(a) & (b) \\
\includegraphics[width = 0.45\textwidth]{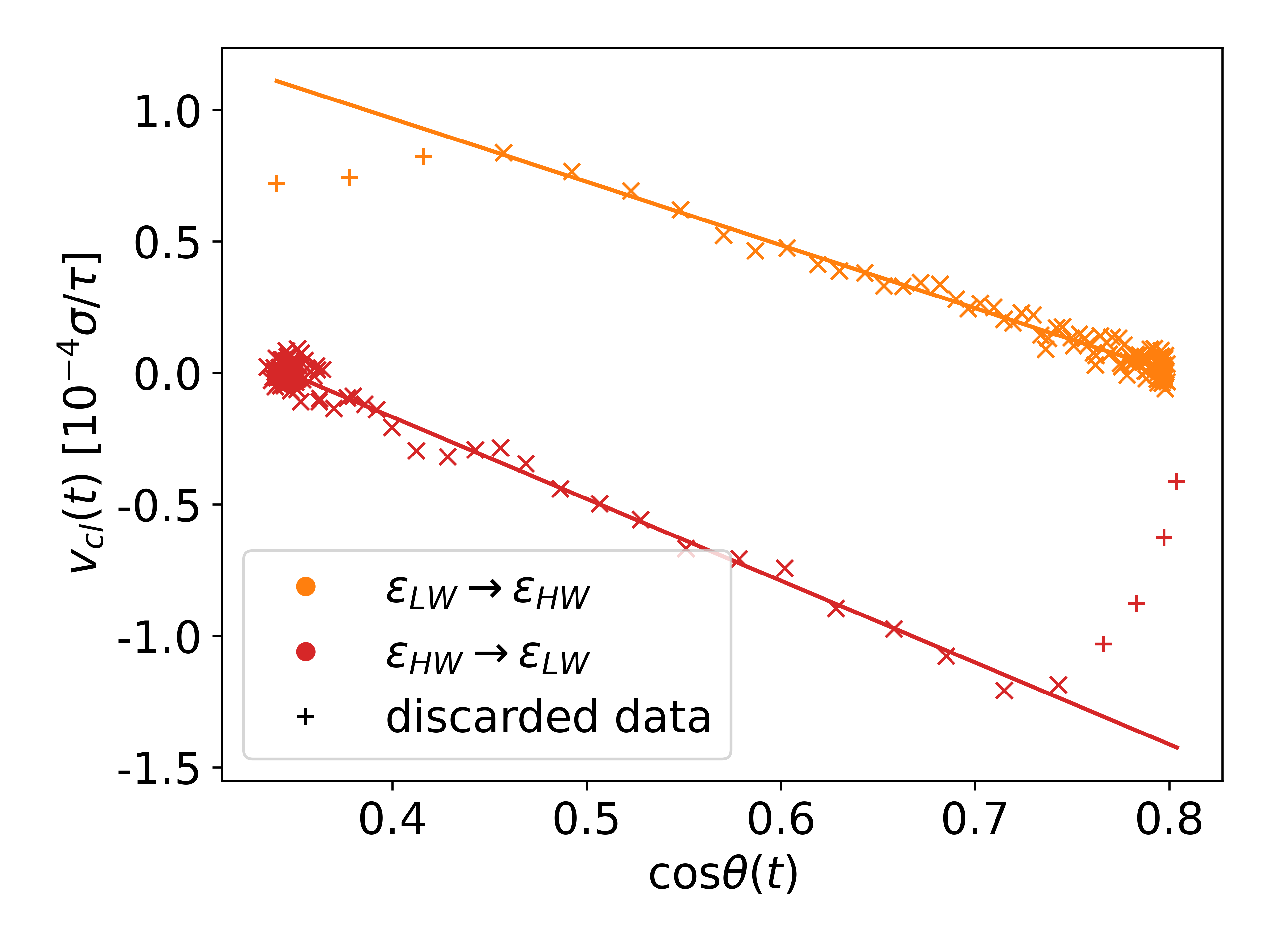} &
\includegraphics[width = 0.45\textwidth]{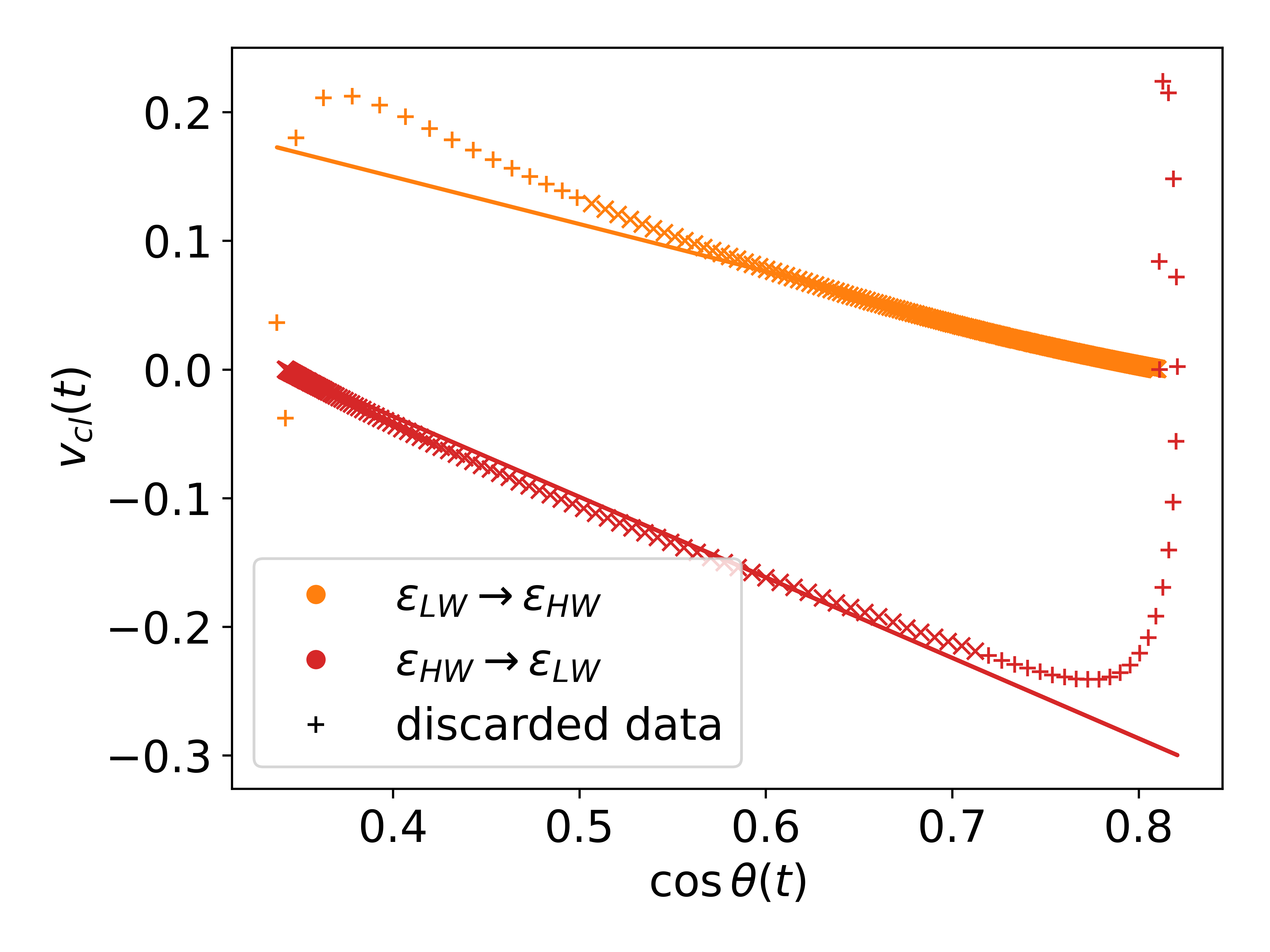} \\
(c) & (d)\\
\includegraphics[width = 0.45\textwidth]{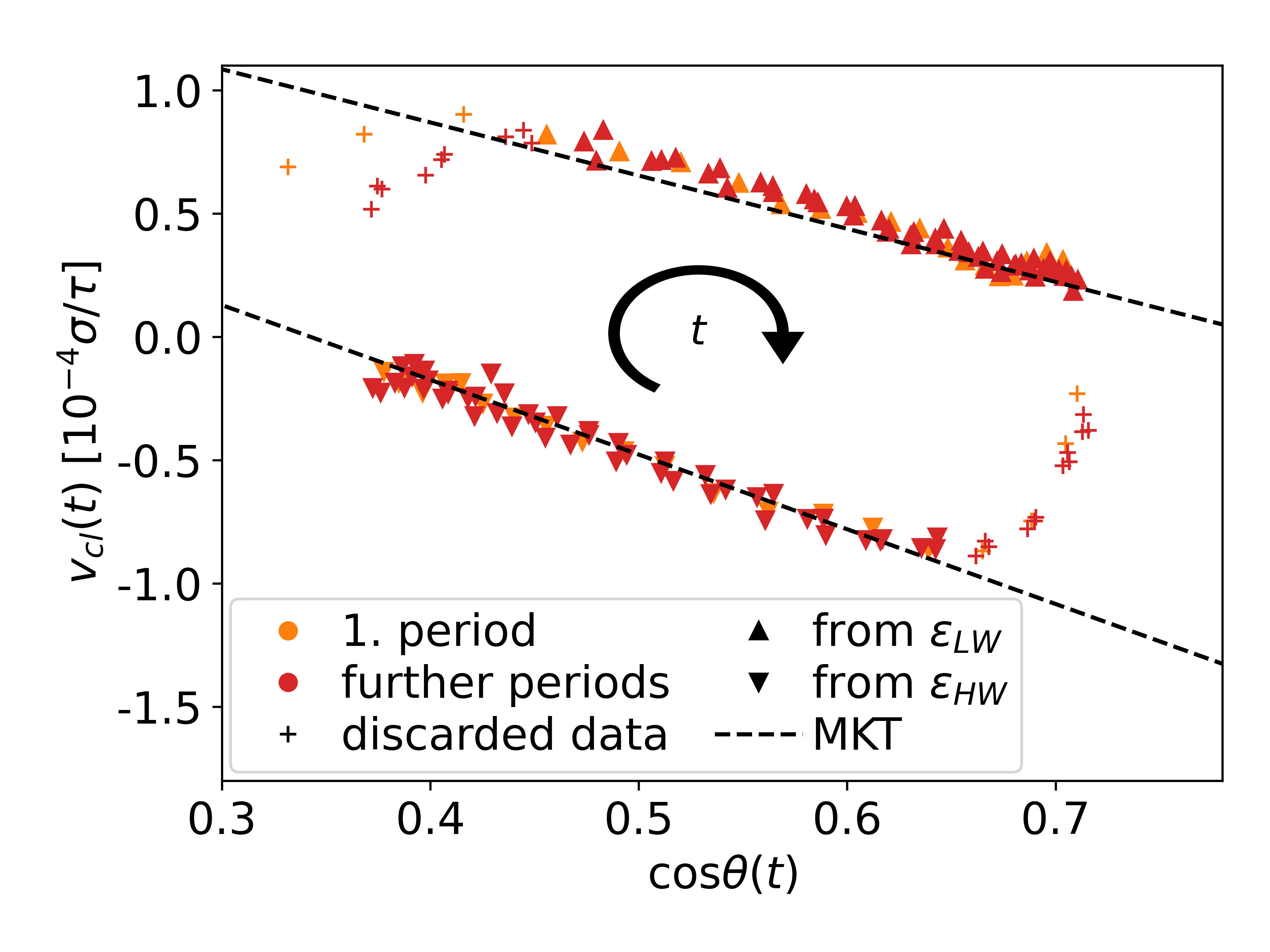} &
\includegraphics[width = 0.45\textwidth]{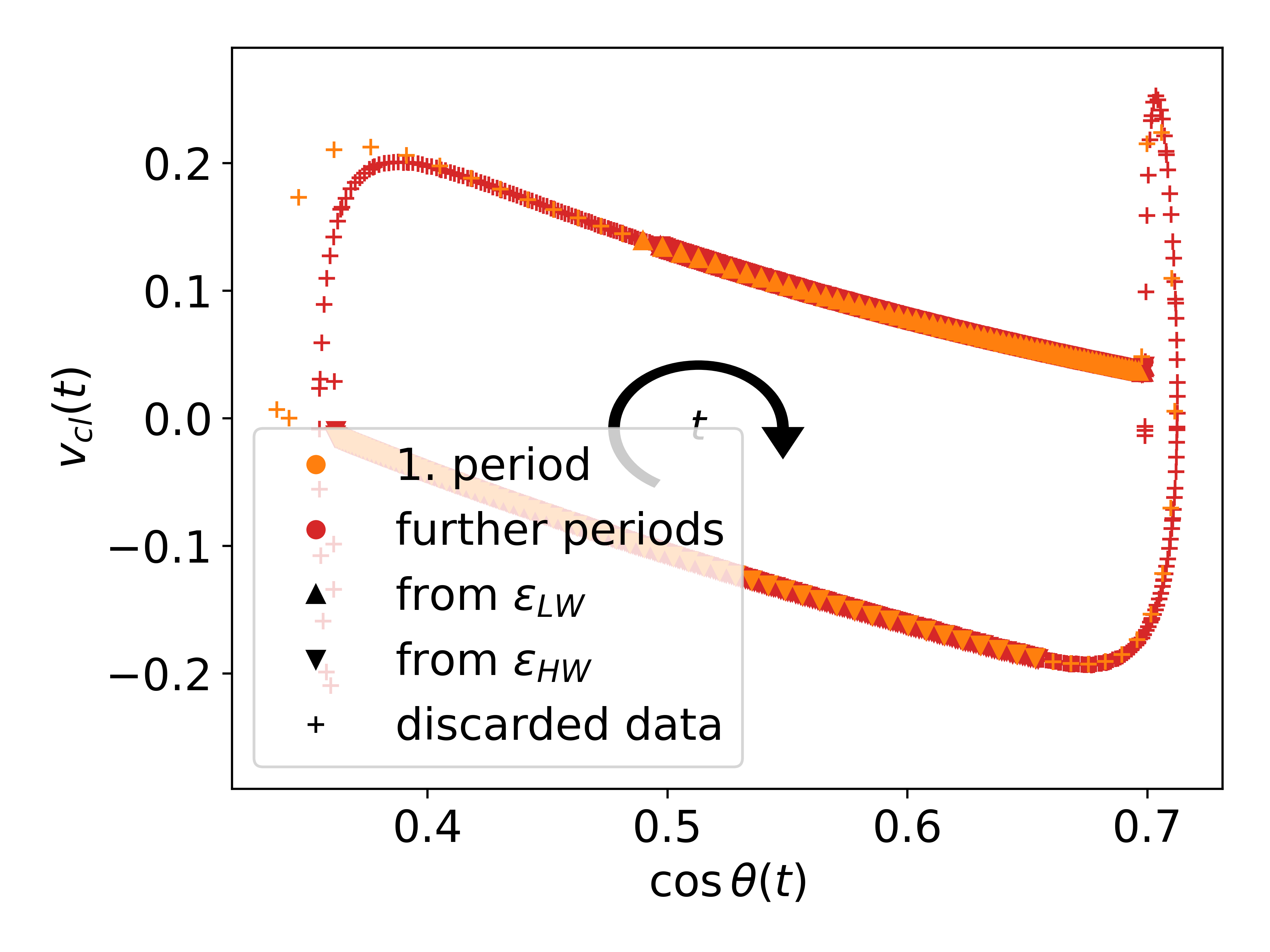}\\
\end{tabular}
\caption{Velocity of the contact line $v_{cl}$ plotted against $\cos \theta(t)$ for (a) the relaxation of a droplet on a surface with a wettability change from $\epsilon_{HW} = 0.762$ to $\epsilon_{LW} = 0.632$ and the reverse process. A line is fitted to the data to compute $\zeta_{MD}$ from its slope according to the MKT theory. The first few data points (plus sign) were discarded for these fits. Data points are spaced equidistantly with $\Delta t = 10^{4}$ MD steps, (b) TF simulations corresponding to the MD results in (a) with $\Delta t = 0.1$.  (c) $v_{cl}$ plotted against $\cos \theta(t)$ for a droplet on a surface with a periodically switched wettability from $\epsilon_{HW} = 0.762$ to $\epsilon_{LW} = 0.632$ with the initial droplet equilibrated on a surface with a wettability of $\epsilon_{LW}$. The switching period was $T = 2  \cdot 10^{6}$ MD steps. The data points are again spaced equidistantly with $\Delta t = 10^{4}$ MD steps. The dashed lines are plots of Eq.~\eqref{eq:mkt} with values of $\zeta_{R}$ for wettabilities of $\epsilon_{HW}$ and $\epsilon_{LW}$, respectively. (d) The TF equivalent of (c) with $T=15.8$ and $\Delta t = 0.1$. Note that there is no noise in the TF model.}
\label{fig:mkt_fit_switch}
\end{figure*}

Another option to determine $\zeta$ is to calculate the slope of a linear fit of $v_{cl}$ versus $\cos\theta(t)$ (denoted with '$\zeta_{MD}$' or $\zeta_{TF}/\gamma$).
This slope corresponds to the prefactor $\frac{\gamma}{\zeta}$ of equation~\eqref{eq:mkt} so that $\zeta_{MD}$ can be determined by dividing the interface tension by the slope of the fit. Therefore, we equilibrated a droplet on a substrate with a lower solid-liquid  interaction of $\epsilon_{LW}$ and switched it instantaneously to a higher solid-liquid interaction $\epsilon_{HW}$ or vice versa and calculated the contact angle during the relaxation process of the droplet. As can be seen in Fig.~\ref{fig:mkt_fit_switch}~a) the first few data points do not show a linear behavior of the contact line velocity in contrast to the behavior at later time steps. This is particularly pronounced when switching from $\epsilon_{HW}$ to $\epsilon_{LW}$. As a consequence the first few data points for both switching directions cannot be used for the linear fit to determine $\zeta_{MD}$, as indicated by the solid line in the figure.
The droplet starts adapting to changes in the substrate's wettability in the contact region and this influences the contact angle obtained from the overall droplet properties. Partly, this effect may be a result of calculating the contact angle from a circle fit in a situation where major deviations from a circle are present. Note that for practical reasons a fit dependent on the whole droplet profile is common in the analysis of experimental data \cite{ROTENBERG1983169, KWOK1999167}. 

However, here we argue that these deviations are not just an fitting issue.  The data points in Fig.~\ref{fig:mkt_fit_switch}~a) have been written out in regular time intervals. Thus, if the true contact angle would follow the MKT prediction from the very beginning one would estimate for the transition from high to low wettability that $\cos(\theta(t=0)) \sim 0.85$ which is significantly larger than the actual equilibrium angle of $\sim 0.80$, describing the droplet just before switching. Thus, directly after the switching event the droplet reacts much slower than predicted by MKT. This effect is much weaker when switching from low to high wettability. This has a very natural explanation: in particular for the transition to the low wetting state the whole droplet has to rearrange so that the contact line can significantly move towards the center of the droplet. Since the MKT assumes that the dynamics of the contact line is driven by local forces, this effect cannot be captured. Remarkably, the data points approach the MKT-type linear behavior for contact angles very close to the initial equilibrium angle. To conclude, the deviations are not due to fitting issues but just express the presence of a kind of {\it dead time} during which the droplet hardly changes its shape.

Naturally, $\zeta_{MD}$ can be determined for multiple pairs of $\epsilon_{LW}$ and $\epsilon_{HW}$ (cf. SI~\ref{app:plots_vcl_vs_cos}). The resulting values are shown in Fig.~\ref{fig:zeta_vs_eps}~a) where $\zeta_{R}/\gamma$ and $\zeta_{MD}/\gamma$ are plotted against $\epsilon_{w}^{2}$. It shows that $\zeta$ increases with an increasing wettability since the particles are attracted more by the surface and thus the friction increases. Also it emphasizes the applicability of the scheme to calculate $\zeta_{R}$ since the values of $\zeta_{R}$ and $\zeta_{MD}$ agree very well. Furthermore, we would like to stress that within the statistical uncertainties $\zeta_{MD}$ does not depend on the switching direction and just reflects the interaction of the droplet with the substrate close to the contact line. This is in accordance with the physical picture of the MKT approach.

As shown in Fig.~\ref{fig:mkt_fit_switch}~b) the same phenomena are observed when analyzing the velocity of the contact line for the TF data. Due to the absence of noise effects the effects are seen even more clearly. This holds, first, for the highly linear dependence of the velocity of the contact line on $\cos(\theta(t))$ (including some minor systematic deviations) but in particular also for the immobility of the droplet directly after switching. 
In analogy to the MD data we find again a mostly linear relation between $\zeta_{TF}$ and $\epsilon_{w}^2$. Only for high values of the squared wettability energy $\epsilon_{w}^2$ the values of $\zeta_{TF}/\gamma$ start to deviate from the linear behavior and also start to depend on the switching direction. 

To check whether the MKT is also applicable to the case of periodic switching 
we simulate periodic switching by first equilibrating a droplet on a substrate with a constant interaction strength of $\epsilon_{LW}$ or $\epsilon_{HW}$, respectively. Then, at time $t=0$ we start to periodically change the wettability of the substrate between $\epsilon_{LW}$ and $\epsilon_{HW}$ each $T/2$ time steps.

Figure~\ref{fig:mkt_fit_switch}~c) shows the result for such a periodically switched substrate. Linear fits of the data points after each change in wettability show comparable slopes to the single switch scenario in Fig.~\ref{fig:mkt_fit_switch}~a). This holds for later switching cycles as well. Also the dashed lines, which show Eq.~\eqref{eq:mkt} with values of $\zeta_{R}$ for the corresponding wettabilities, confirm again that the values of $\zeta_{R}$ and $\zeta_{MD}$ agree well. The same phenomena are seen for the TF data in Figure~\ref{fig:mkt_fit_switch}~d).

\begin{figure}
    \centering
    \begin{tabular}{ll}
         (a)&(b)  \\
         \multirow{3}{*}[5cm]{\includegraphics[width=0.48\textwidth]{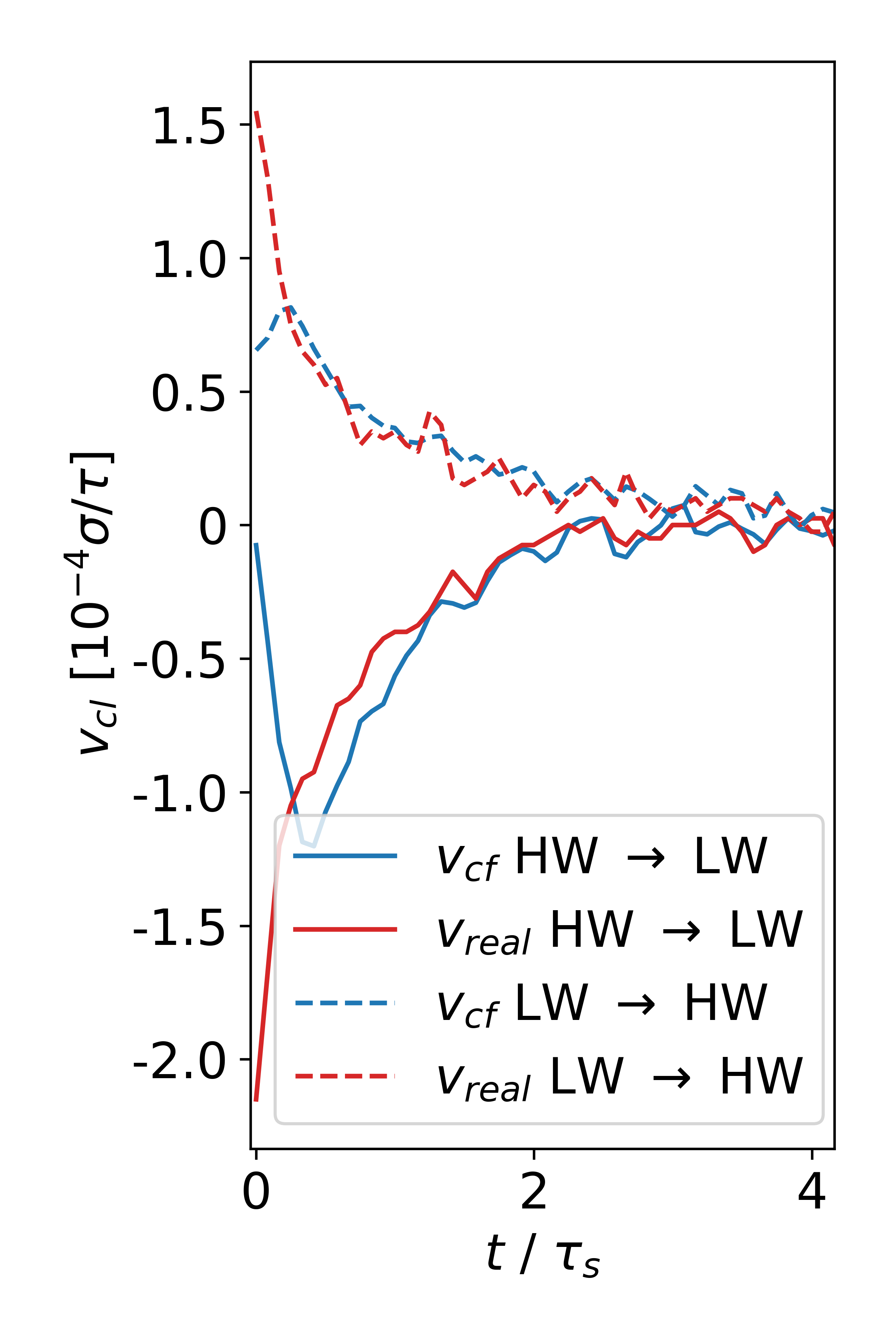}}
         & \includegraphics[width=0.43\textwidth]{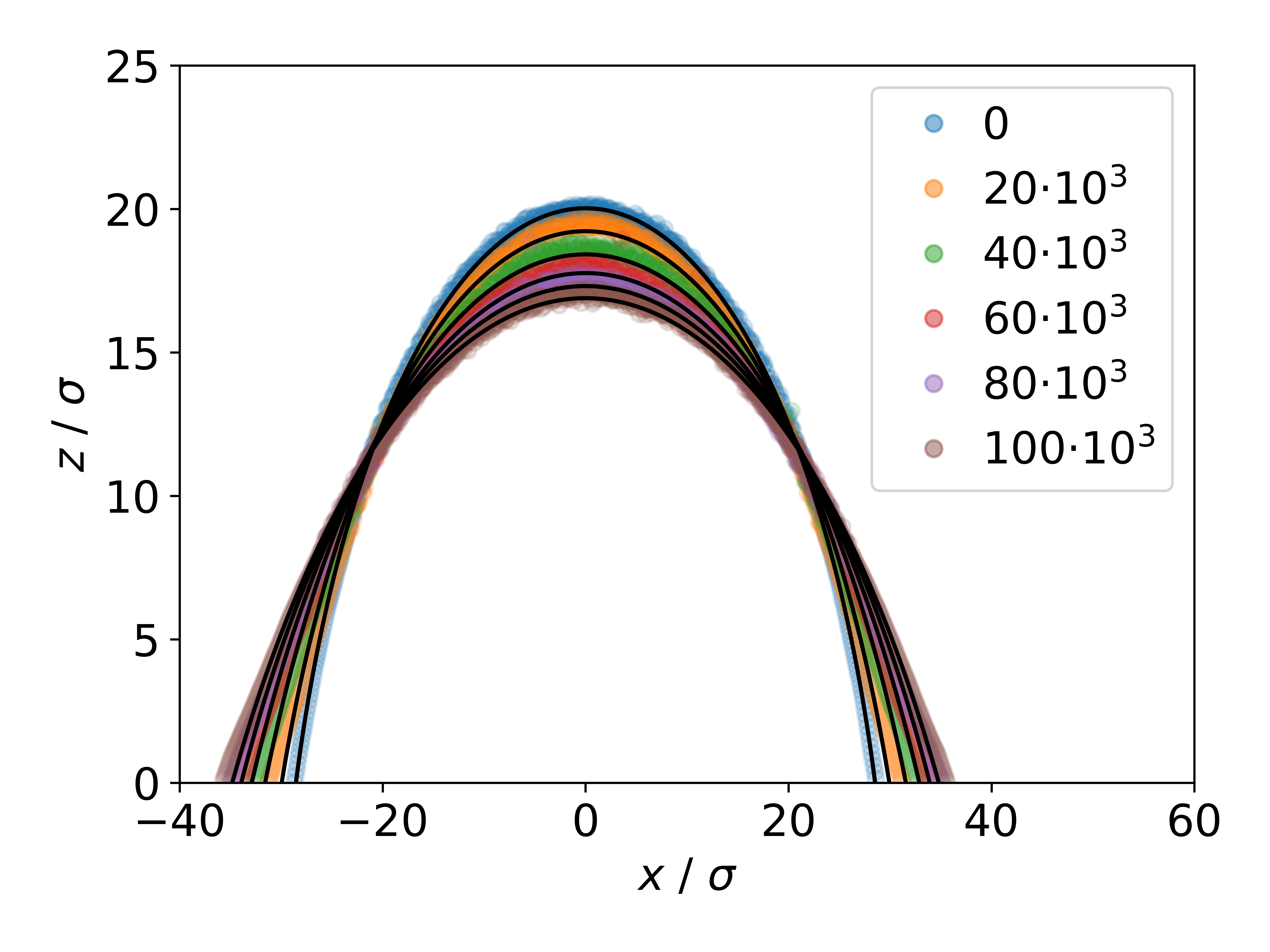}\\
         & (c)\\
         & \includegraphics[width=0.43\textwidth]{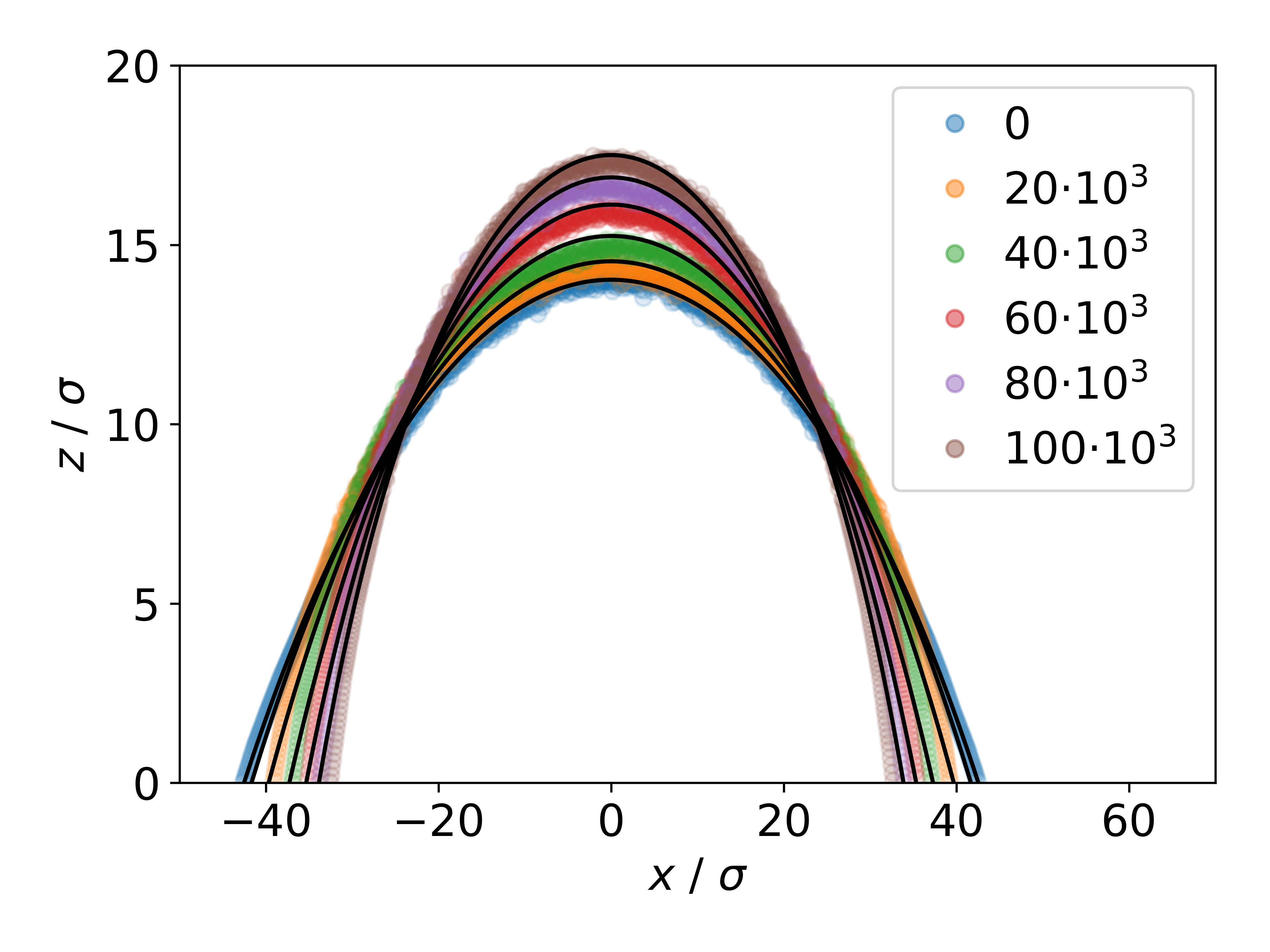}
    \end{tabular}
    \caption{(a) Microscopic contact line velocity compared to the effective contact line velocity extracted from circular fits for different switching directions. (b) Droplet profiles compared to circular fits.}
    \label{fig:real_contactline_velocity}
\end{figure}

So far, the contact line has been determined from a circle fit to the droplet. In this sense it can be regarded as an effective contact line. However, in atomistic simulations it is possible also to determine the microscopic contact line. In particular, if the droplet shape is highly non-circular some deviations between the microscopic and effective contact line may be expected. To illustrate the differences, we plotted in Fig.~\ref{fig:real_contactline_velocity} the velocities as a function of simulation time. Indeed, this initial non-monotonous velocity behavior of the effective contact line is not reproduced by the microscopic contact line. Here a monotonous behavior is observed. To compensate for this initial difference, for intermediate times the microscopic contact line has to be slower than the effective contact line. Closer to the new equilibrium position, where the shape is nearly circular, the velocities basically agree. As also seen in Fig.~\ref{fig:real_contactline_velocity} this can be qualitatively seen when plotting the droplet shape and the resulting circle fit for different initial times. As already expected from Fig.~\ref{fig:zeta_vs_eps} the deviations between both contact lines are higher for the transition from high to low wettability

\subsection{Relaxation behavior after a single switching event}\label{sec:md_tf_mapping}

Naturally, knowledge about the velocity of the contact line should contain the relevant information to predict the relaxation behavior of $\cos(\theta(t))$. This is first explicitly explored for a single switching event for different pairs of wettabilities. For this purpose we determine $\cos(\theta(t))$ after a single switching event and fit the resulting relaxation curve by a stretched exponential. The resulting relaxation times, using Eq.~\eqref{eq:moment}, are listed in Tab.~\ref{tab:moments} both for the TF and the MD data. As a comparison we use the input from the MKT analysis, namely the values of $\zeta_{MD}/\gamma$ and $\zeta_{TF}/\gamma$ to predict the expected switching times according to Eq.~\eqref{eq:mkt_moment}. They are also listed in Tab.~\ref{tab:moments} as $\tau_\zeta$. Indeed, one can find a reasonable agreement between both approaches. The remaining deviations may result from the introduction of the $(1-\cos(\theta))$-factor to enable the analytical calculation as well as from the dead time effects, discussed above. Naturally, the times slightly depends on the switching direction.

\begin{table}
  \caption{\label{tab:moments}Relaxation times $\tau_{rel}$ obtained from a fit of a stretched exponential to the values of $\cos(\theta)$ versus $t$ for switching from $\epsilon_{1}$ to $\epsilon_{2}$ for MD and TF simulations and relaxation times $\tau_{zeta}$ according to Eq.~\eqref{eq:mkt_moment} with values of $\zeta_{MD}/\gamma$ and $\zeta_{TF}/\gamma$ as input parameters for MD and TF simulations, respectively.}
\begin{tabular*}{0.98\columnwidth}{@{\extracolsep{\fill}}cccccc}
\hline
\mbox{$\epsilon_{1}$} & \mbox{$\epsilon_{2}$} & \mbox{$\tau_{rel}$ (MD)} & \mbox{$\tau_{\zeta}$ (MD)} & \mbox{$\tau_{rel}$ (TF)} & \mbox{$\tau_{\zeta}$ (TF)} \\
      \hline
      0.632 & 0.671 & $6.84 \cdot 10^{4}$ & $7.55 \cdot 10^4$ & 2.37 & 2.19\\ 
      0.671 & 0.632 & $6.84 \cdot 10^{4}$ & $6.92 \cdot 10^4$ & 2.23 & 2.02\\ 
      0.632 & 0.762 & $12.02 \cdot 10^{4}$ & $11.82 \cdot 10^4$ & 5.34 & 5.87 \\ 
      0.762 & 0.632 & $11.33 \cdot 10^{4}$ & $10.52 \cdot 10^4$ & 4.74 & 3.54  \\ 
      \hline
\end{tabular*}
\end{table}


\begin{figure}
\centering
\includegraphics[width = 0.98\columnwidth]{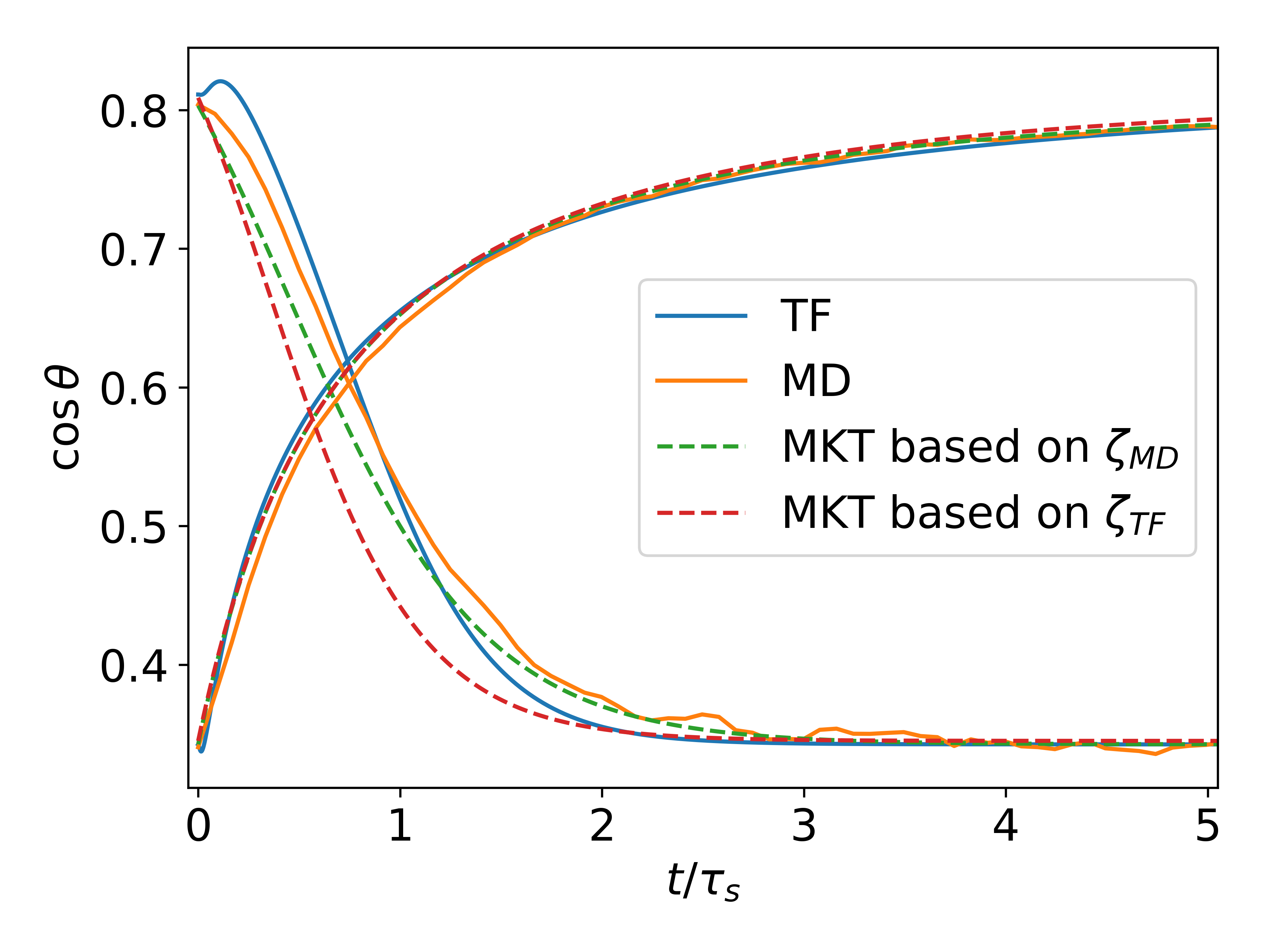}
\caption{$\cos(\theta)$ obtained from TF, MD and MKT simulation plotted against $t$ for a single switch. The corresponding values of $\epsilon_{w}$ are 0.632 and 0.762.}\label{fig:mapping_single_switch} 
\end{figure}

Next,  we specifically analyse our standard example $\epsilon_{LW}=0.632 $ and $\epsilon_{HW}=0.762$ and show the relaxation curves for the transitions from  $\epsilon_{LW}$ to $\epsilon_{HW}$ and vice versa in Fig.~\ref{fig:mapping_single_switch}. In order to compare both approaches (TF and MD) we relate the time scales to the respective relaxation  time $\tau_{rel}$ when switching from $\epsilon_{LW}$ to $\epsilon_{HW}$. This time is denoted $\tau_s$. In this dimensionless representation, corresponding to a kind of mapping,  the MD and the TF data agree very well. Indeed, for all subsequent analysis we will always express the times in terms of $t/\tau_s$.

Note that in Fig.~\ref{fig:mapping_single_switch} not only the time scales but also the degree of non-exponentially is comparable between the MD and TF approach. Indeed, as discussed in SI.\ref{app:beta} and qualitatively seen in Fig.~\ref{fig:mapping_single_switch}, when switching to the state of lower wettability one observes a stretched exponential whereas in the opposite case a compressed exponential is observed. This can be fully rationalized by the properties of the analytical solution in Eq.~\eqref{eq:mkt_approx_solved_approximated} (see SI.\ref{app:beta}).

Furthermore, the MKT predictions, obtained from integration of Eq.~\ref{eq:mkt_circle}, are included. 
The MKT resembles the MD and TF values for switching remarkably well except for an offset. This offset of the MKT  mirrors that initially $v_{cl}$ does not depend linearly on $\cos \theta$ since the short time behavior of the contact angle from a circle fit is not included in the MKT. However, if the MKT values were shifted by $\sim 0.2~\tau_{s}$ which is approximately the time of the nonlinear behavior at the beginning of each switching event, the agreement with MKT would work very well. This shows that MKT can reproduce MD and TF results but with much less computational effort, once the dead time effects are taken into account. 

\subsection{Periodic Switching}\label{sec:results_ps}
\begin{figure}
\begin{tabular}{ll}
(a) & (b)\vspace{-0.21cm} \\
\vspace{-0.9cm} \includegraphics[width = 0.45\textwidth]{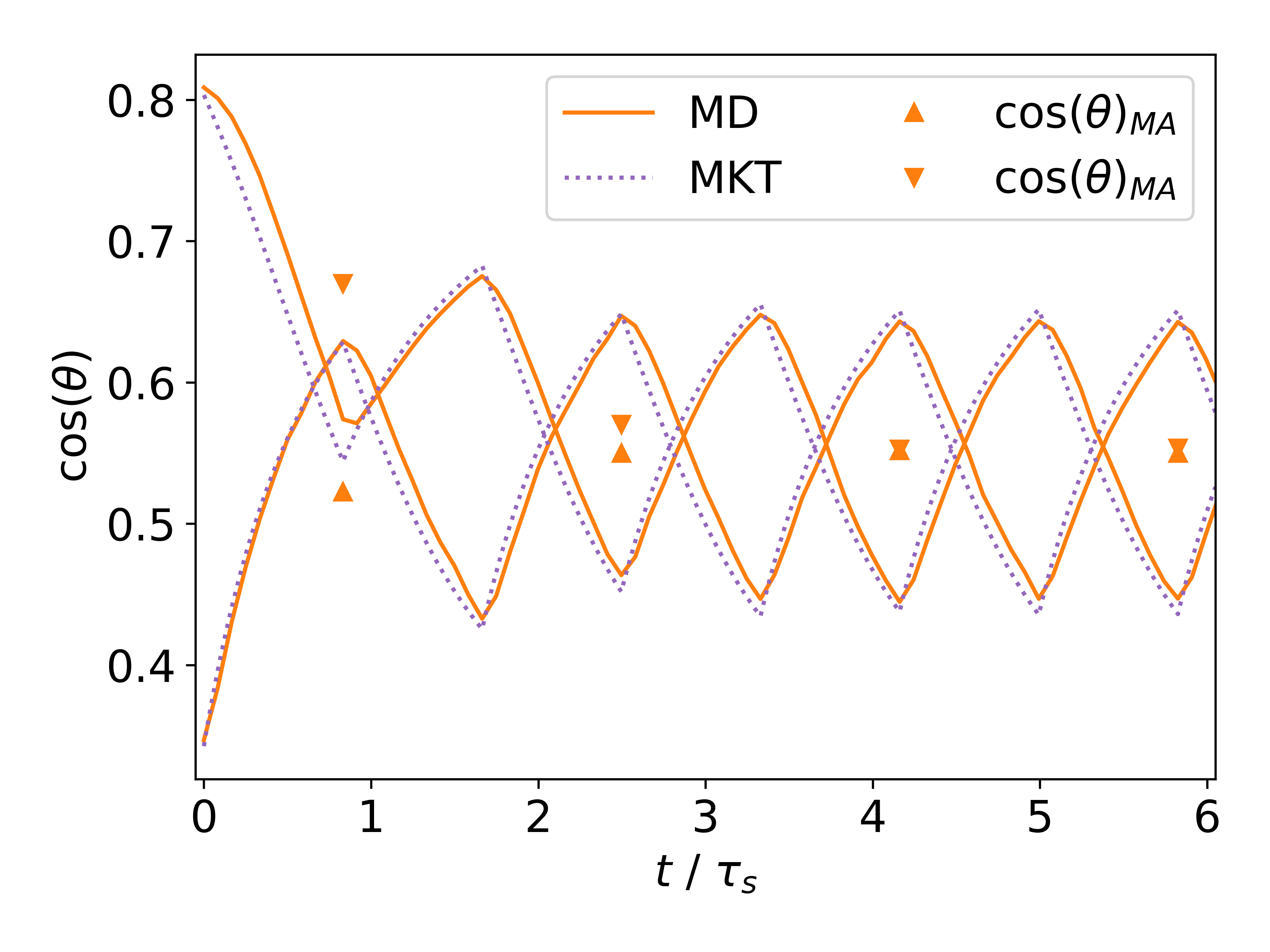} & \includegraphics[width = 0.45\textwidth]{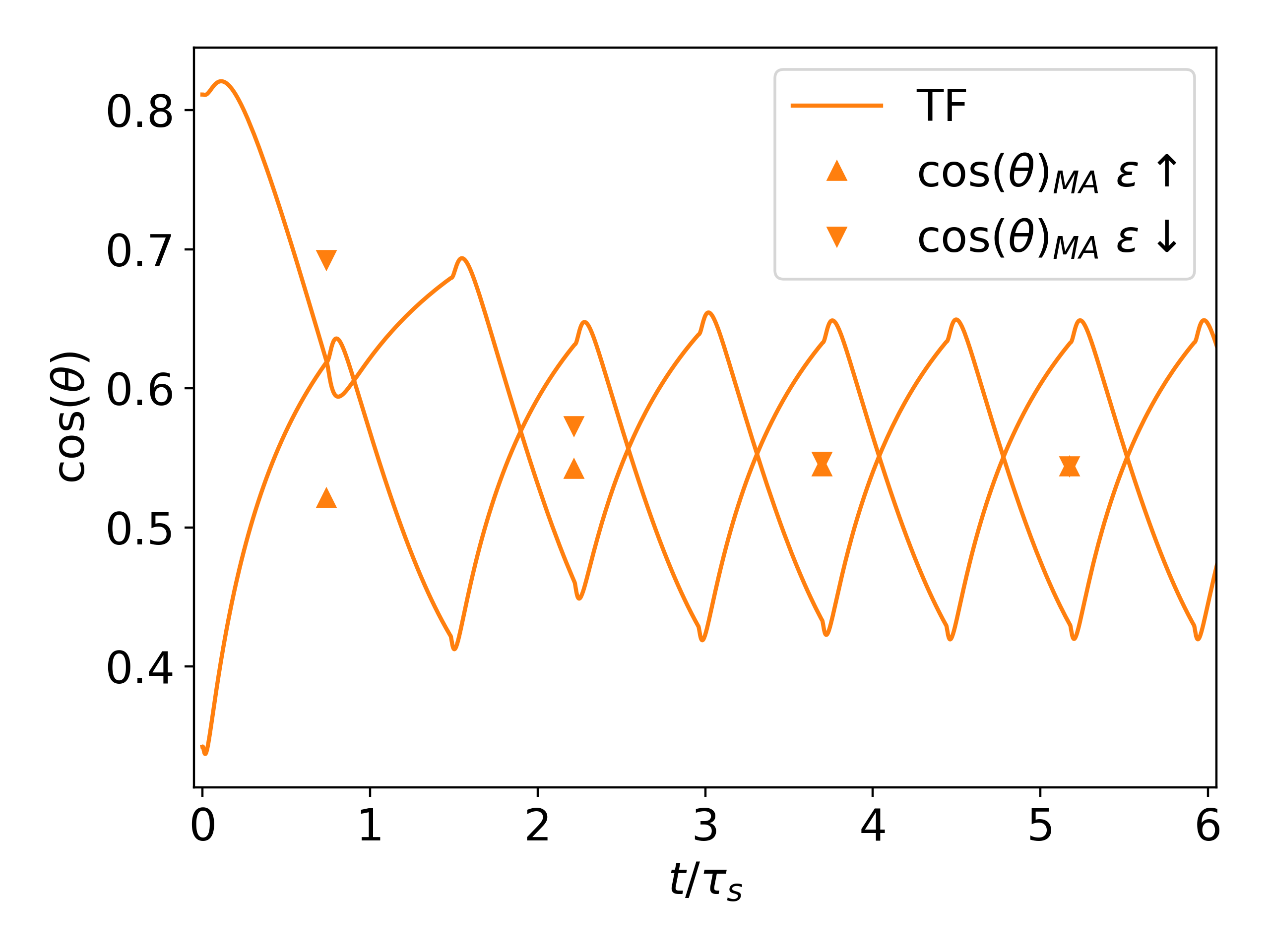} \\
(c) & (d)\vspace{-0.21cm} \\
\vspace{-0.9cm}\includegraphics[width = 0.45\textwidth]{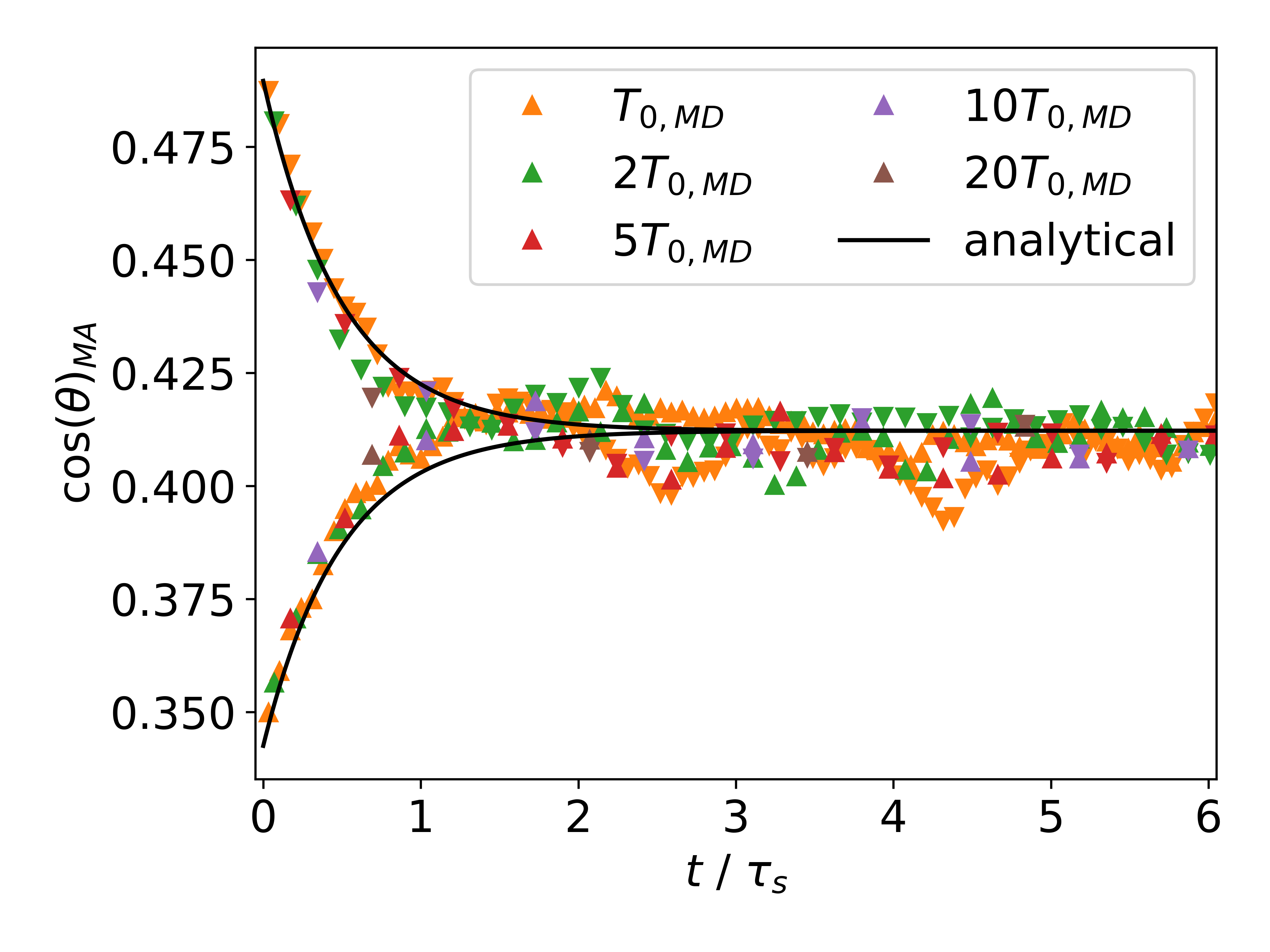} &
\includegraphics[width = 0.45\textwidth]{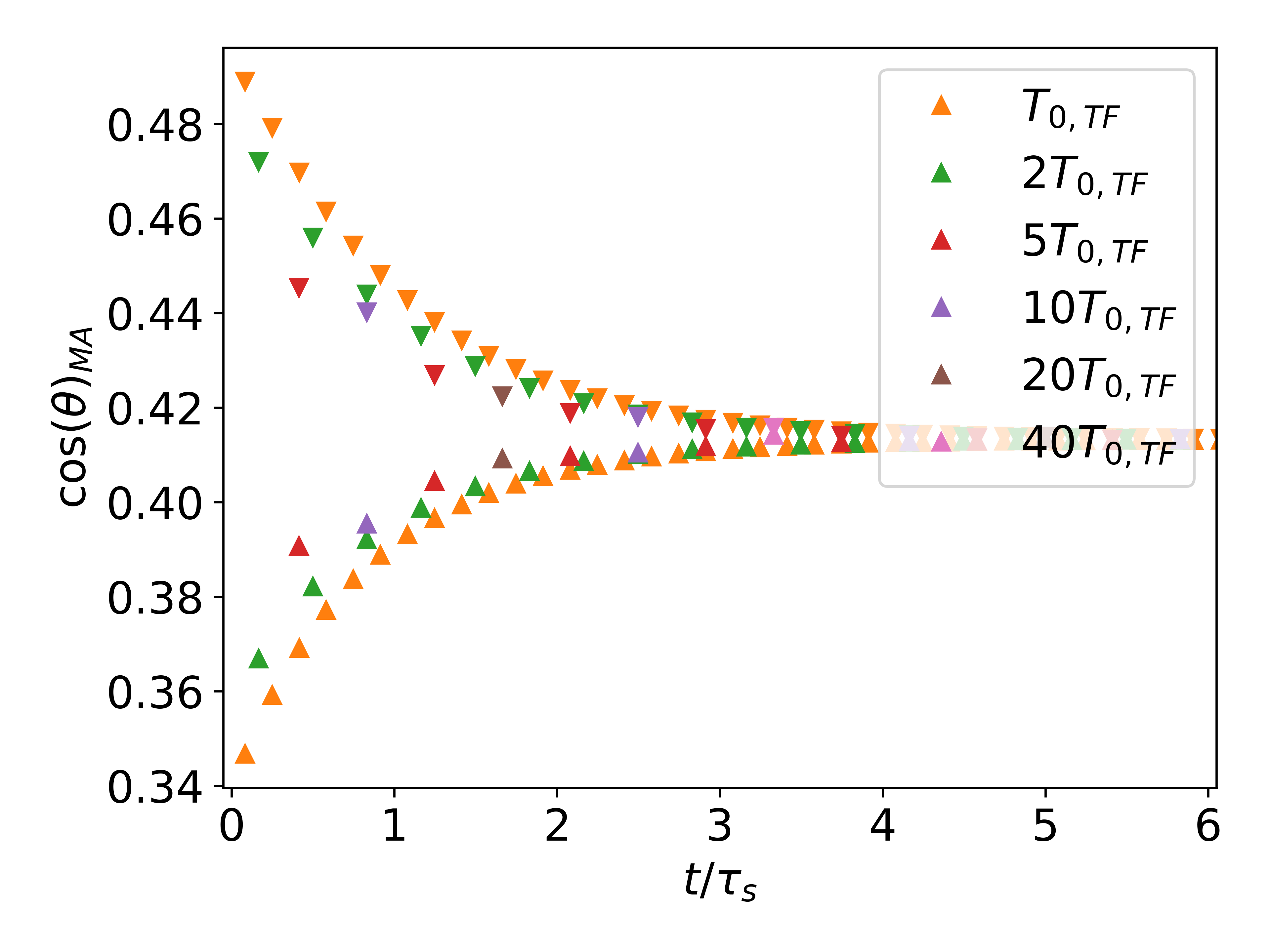}\\
(e) & (f)\vspace{-0.21cm} \\
\vspace{-0.9cm}\includegraphics[width = 0.45\textwidth]{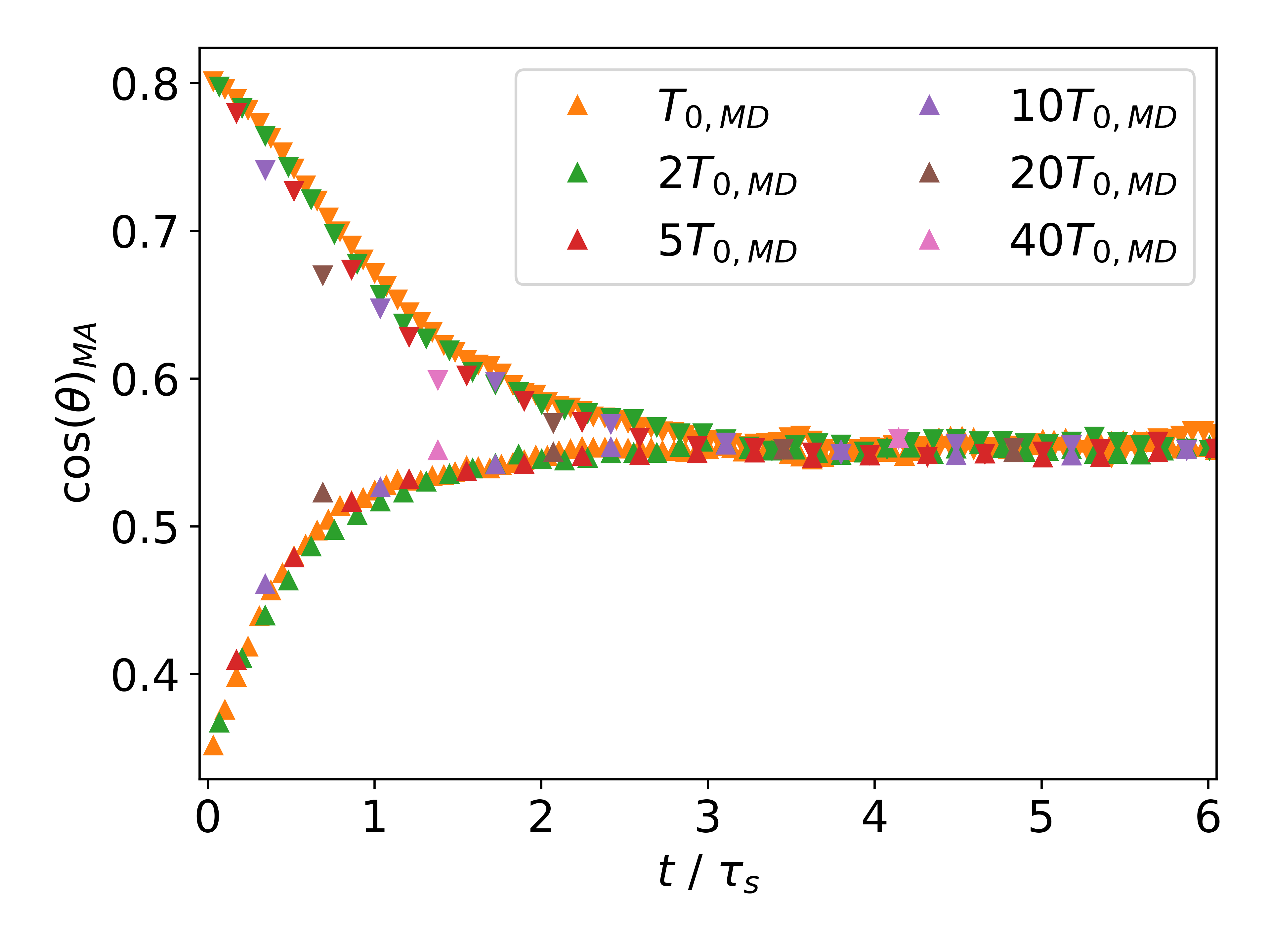} &
\includegraphics[width = 0.45\textwidth]{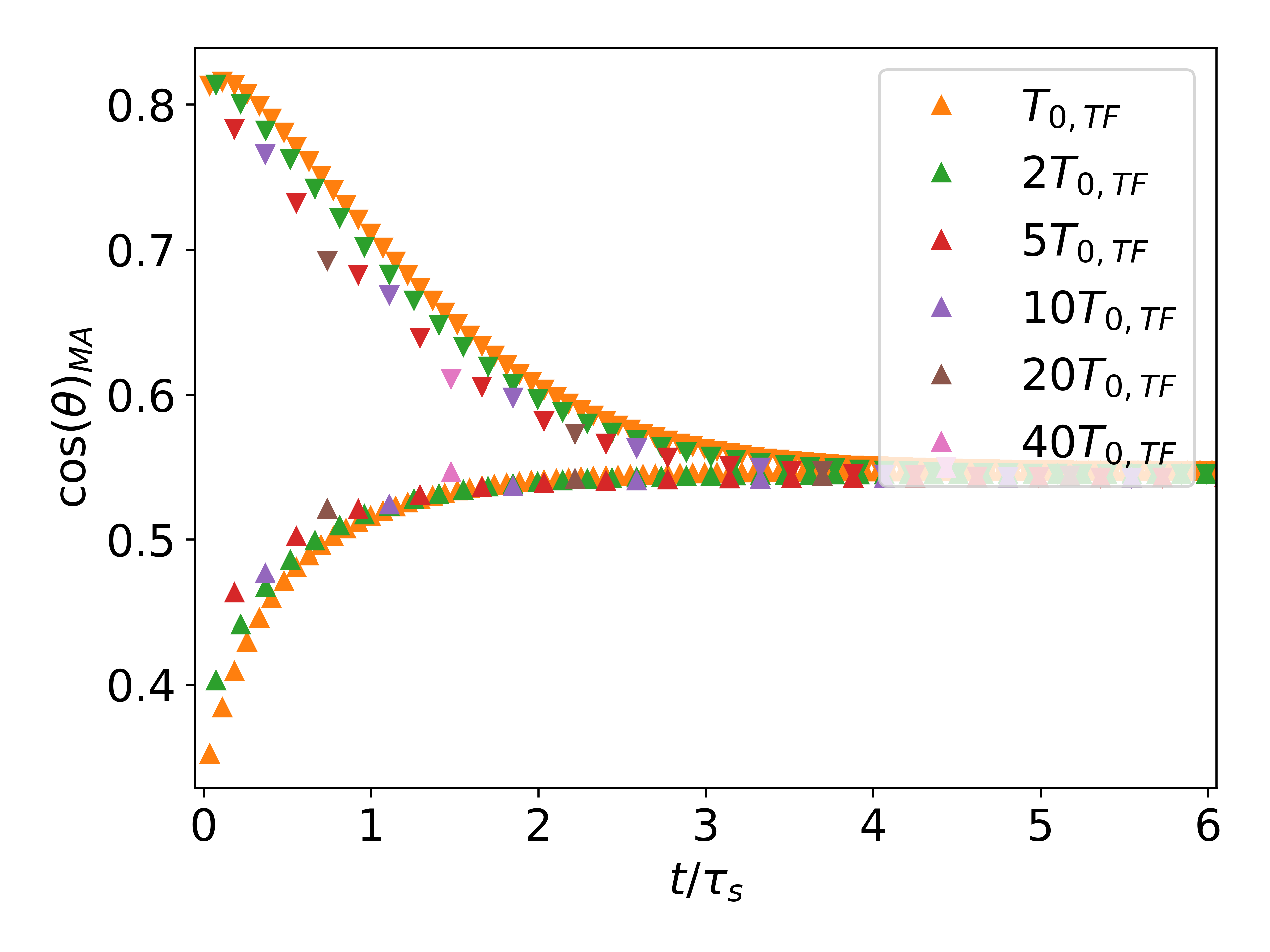}\\
\end{tabular}
\caption{(a) The solid line shows the cosine of the contact angle $\cos(\theta)$ plotted against $t$ for a switching period of $T = 1.524\tau_s$. The triangles mark the average cosine of the contact angle over one period $\cos(\theta)_{MA}$ and the dashed line shows the results obtained from the MKT. (b) Corresponding results in the TF model for a switching period of $T = 1.478 \tau_{s}$. (c) and (e) $\cos(\theta)_{MA}$ for different periods from 1 to 40 $T_{0,MD}$, where $T_{0,MD} = 7.62 \cdot 10^{-2} \tau_{s}$ while periodically switching the surface between $\epsilon_{LW}$ and $\epsilon_{HW}$. (d) and (f) TF simulations with corresponding wettability parameters, where $T_{0,TF} = 7.39\cdot 10^{-2} \tau_s$. In (c) and (d) the chosen wettabilities are $\epsilon_{LW} = 0.632$, $\epsilon_{HW} = 0.671$ whereas in (e) and (f) the corresponding values read $\epsilon_{LW} = 0.632$, $\epsilon_{HW} = 0.762$. In (c) the MKT prediction for the limit of fast switching has been added.}
\label{fig:periodic_switching_average}
\end{figure}

A periodic switching procedure as described above yields an oscillating state around a plateau value of $\cos(\theta)_{plateau}$ after an initial relaxation. This is shown for MD simulations  in Fig.~\ref{fig:periodic_switching_average}~(a) and the TF simulations in Fig.~\ref{fig:periodic_switching_average}~(b) where the wettability of the surface is switched with a period of $T = 1.524\tau_s$. 

For the MD values we have added the MKT results. Again, we see a very good agreement except for an offset during the relaxation towards the high wettability state. One consequence of these deviations is that the prediction of the plateau value of MKT is slightly too low.

From these data we determine the average values of $\cos(\theta)$ during the individual switching periods  as shown in Fig.~\ref{fig:periodic_switching_average}~(a) and (b). They are denoted $\cos(\theta)_{MA}$. These simulations have been repeated for different switching periods as well as for two different pairs of wetting energies, reflecting the case of small changes of the contact angle in Fig.~\ref{fig:periodic_switching_average}~(c)  and (d) as well as the case of large changes of the contact angle in (e) and (f). One can already see in (c) that the MKT-prediction for very fast switching agrees very well with the actual MD data.

\begin{figure}
\begin{tabular}{ll}
(a) & (b)\\
\includegraphics[width= 0.45\textwidth]{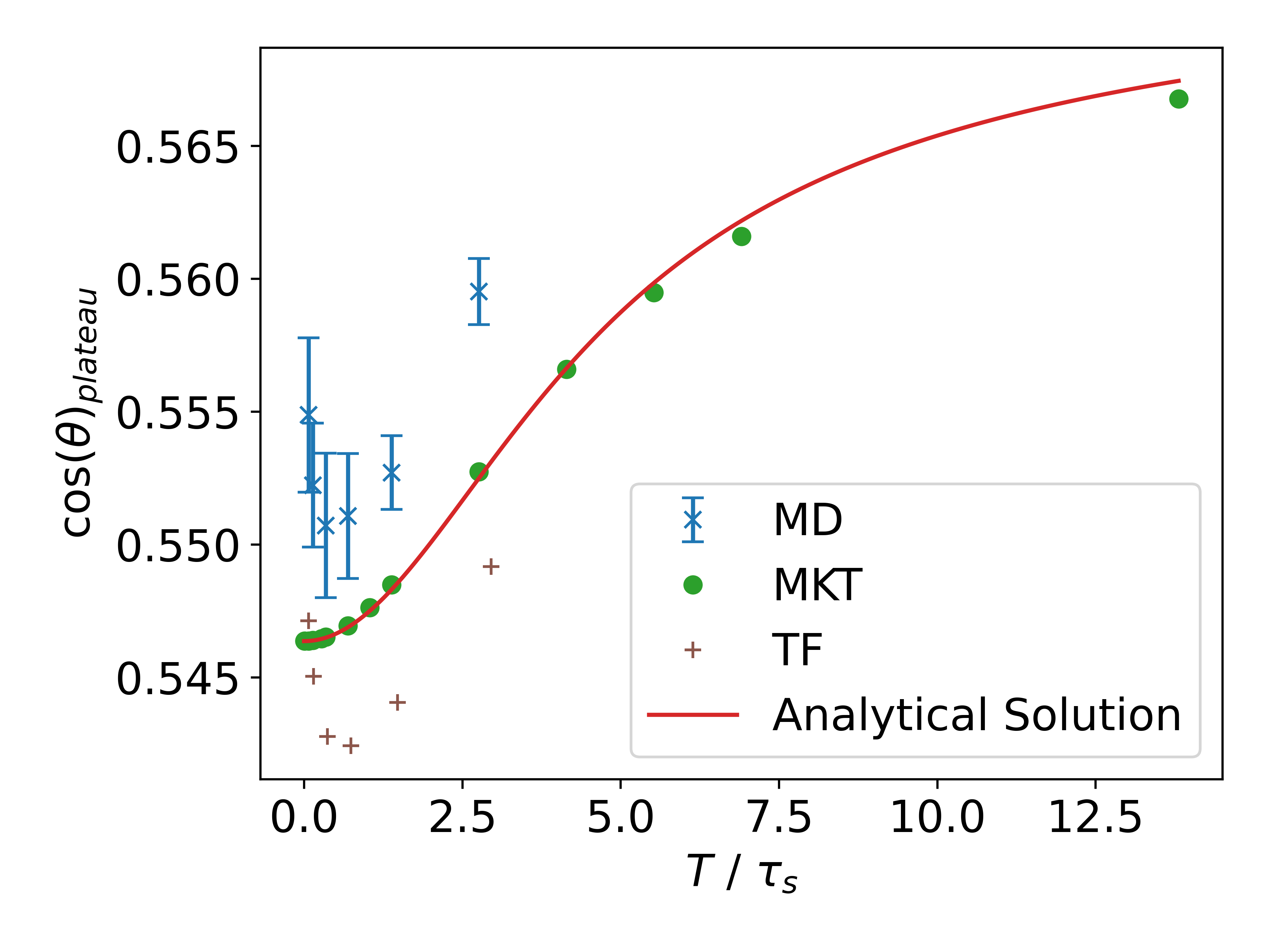} &
\includegraphics[width=0.45\textwidth]{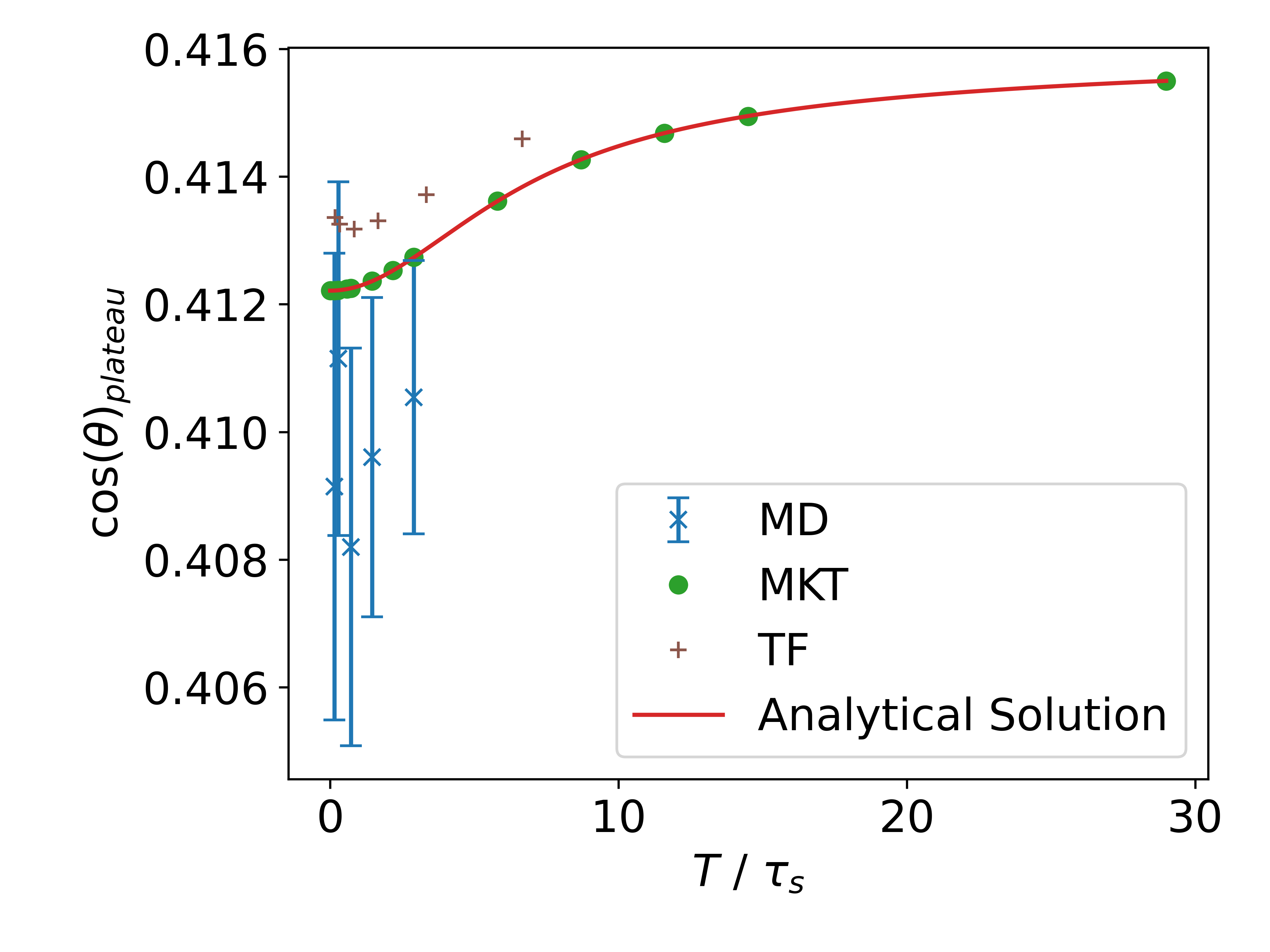}\\
\end{tabular}
\caption{(a) $\cos(\theta)_{plateau}$ obtained from MD, MKT, TF and from Eq.~\eqref{eq:y_plateau} plotted against $T$ for switching between (a) $\epsilon_{LW} = 0.632$ and $\epsilon_{HW} = 0.762$ and (b) $\epsilon_{LW} = 0.632$ and $\epsilon_{HW} = 0.671$.} 
\label{fig:plateau_vs_T}
\end{figure}

Now we analyze the different properties of $\cos(\theta(t))_{MA}$, seen in Figs.~\ref{fig:periodic_switching_average}~(c)-(f), namely the plateau value, the relaxation time, and the amplitude as a function of the switching time $T$.

 {\it Plateau values:} To estimate the plateau values we averaged the values of $\cos(\theta)_{MA}$ during the last cycles after the relaxation is complete and show the resulting values  $\cos(\theta)_{plateau}$ against $T$ in Fig.~\ref{fig:plateau_vs_T}.  There, small changes in $\cos(\theta)_{plateau}$ in dependence of $T$ can be seen for MD and MKT as well as for TF simulations (note the scaling of the $y$-axis, $\cos(\theta)$ for droplets equilibrated at the corresponding high and low wettabilities are (a) $\sim 0.34$ and $\sim 0.80$ and (b) $\sim 0.34$ and $\sim 0.49$). Indeed, since the relaxation times (see Tab.~\ref{tab:moments}) are nearly identical when switching to the lower or the higher wettability value) the smallness of the dependence on the switching period is compatible with Eq.~\ref{eq:y_fast_switching}. Naturally, in the limit of large $T$ the plateau value just corresponds to the average of the two equilibrium values of $\cos(\theta)$.
 
 For the subsequent discussion we restrict ourselves to (a) because there the statistical uncertainties of the MD data are sufficiently small. It turns out that for $T/\tau_s$ the increase of the plateau value for the MD and TF data is nicely reflected by the MKT prediction. Indeed, in agreement with the analytical prediction, the MKT prediction initially displays a quadratic dependence on the length $T$ of the period. For the good agreement between the numerically obtained MKT data and the analytical MKT solution it was important to use $\tilde{k}_{3,i}$ instead of $ {k}_{3,i}$ in order to compensate for the approximation of only small changes of the contact angle (see above). However, two deviations are present. First, for the MD and TF data the plateau value is in general higher and, second, for $T/\tau_s \le 0.6$ the plateau value starts to increase again when approaching the limit of very fast switching. 
 Here we argue that this is a consequence of the dead time effect.   To check this hypothesis, we included this effect and set in our numerical solution of the MKT equation $\gamma/\zeta = 0$ for the first few time steps after each switching event. The result can be found in SI.~\ref{sec:app_mkt_st}. Indeed, a minimum can be seen. This effect can be also understood from the analytical expression. In the limit of fast switching we have $ y_{plateau} = \frac{k_{3,\uparrow}}{k_{3,\downarrow}+k_{3,\uparrow}}$. Due to the dead time effect, which is particularly pronounced for the transition from high to low wettability, the value $k_{3,\downarrow}$ would need to be substituted by an effective relaxation rate which is smaller because of the initial presence of the dead time effect. This effect naturally becomes more prominent for very short switching times. As a consequence $ y_{plateau}$ increases when $T$ becomes very small.
 
\begin{figure*}
\begin{tabular}{ll}
(a) & (b) \\
\includegraphics[width = 0.45\textwidth]{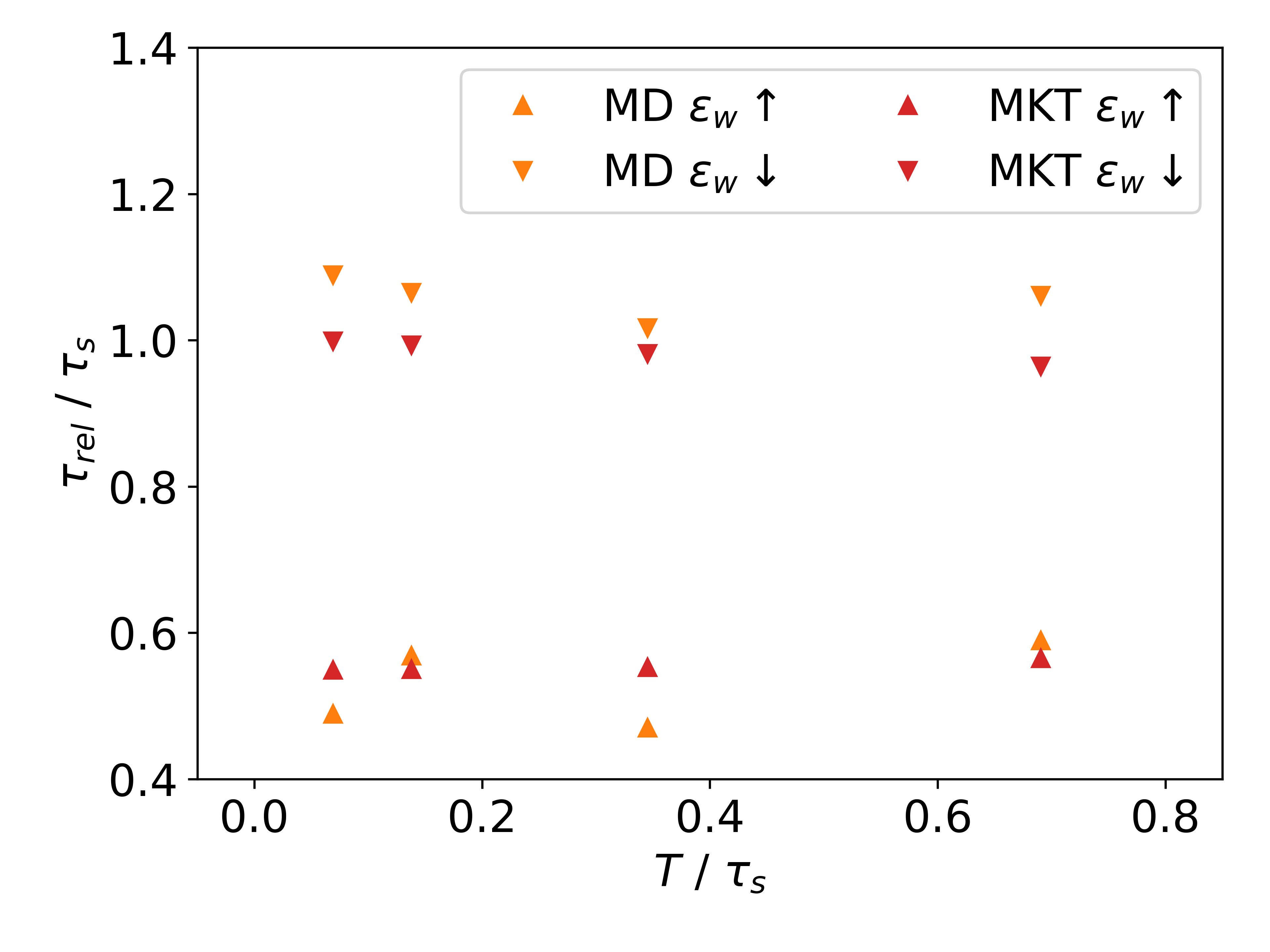} &
\includegraphics[width = 0.45\textwidth]{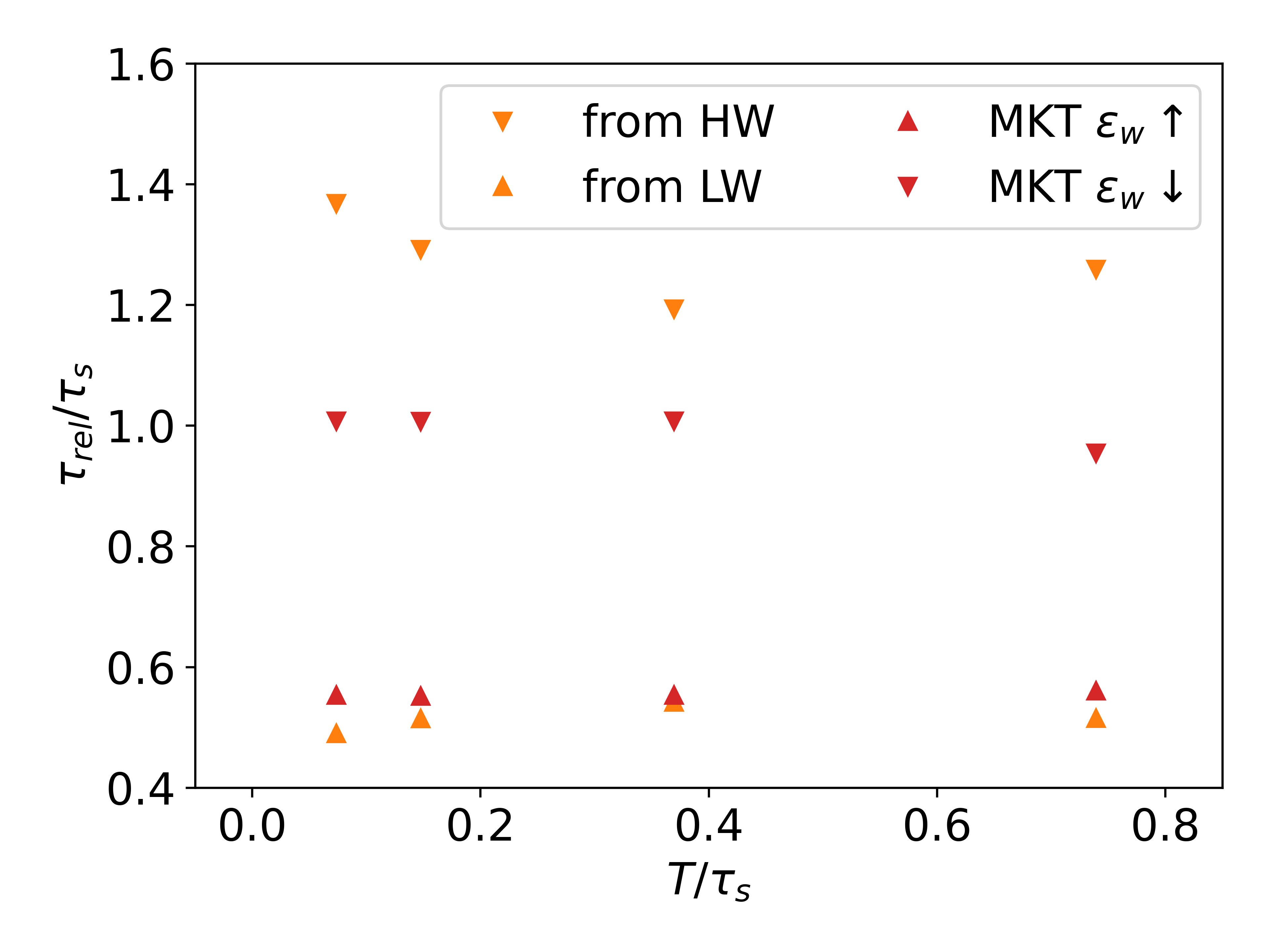}\\
\end{tabular}
\caption{(a) and (b) Relaxation times $\tau_{rel}$ resulting from stretched exponential fits to $\cos(\theta)_{MA}$ versus $t$ from MD and TF data respectively for switching between $\epsilon=0.632$ and $\epsilon=0.762$. }
\label{fig:moments}
\end{figure*}
 {\it Relaxation times:} To obtain relaxation times for the MD and TF simulations as well as for the numerical solution of the MKT approach (Eq.~\eqref{eq:mkt_circle}) we again fitted stretched exponentials to the data and calculated the respective time scales with Eq.~\eqref{eq:moment}. Importantly, during a very broad range of different switching periods there is hardly any variation of the relaxation times, neither for MKT nor for the MD and TF data, respectively. Furthermore, one observes that the relaxation times for the initial transitions to higher and lower wettabilities, respectively, become much closer to each other if the two wettabilities are closer to each other. Indeed, this is fully compatible with our analytical solution Eq.~\ref{eq:y_n} in the limit of small wettability changes where the relaxation time does not depend on the time scale of the switching period and is identical for both initial switching  directions. 

  \begin{figure*}
\begin{tabular}{ll}
(a) & (b) \\
\includegraphics[width = 0.45\textwidth]{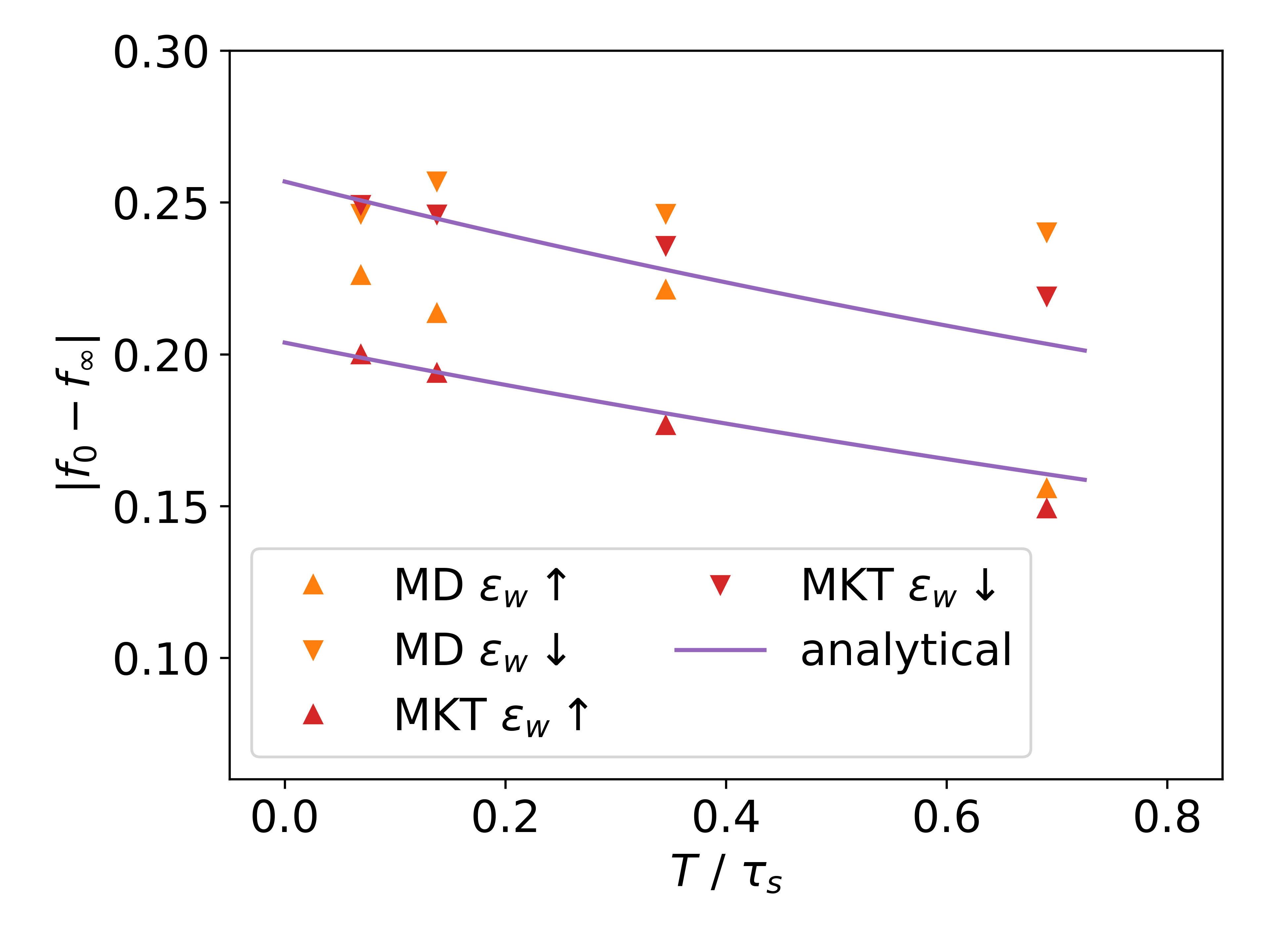} &
\includegraphics[width = 0.45\textwidth]{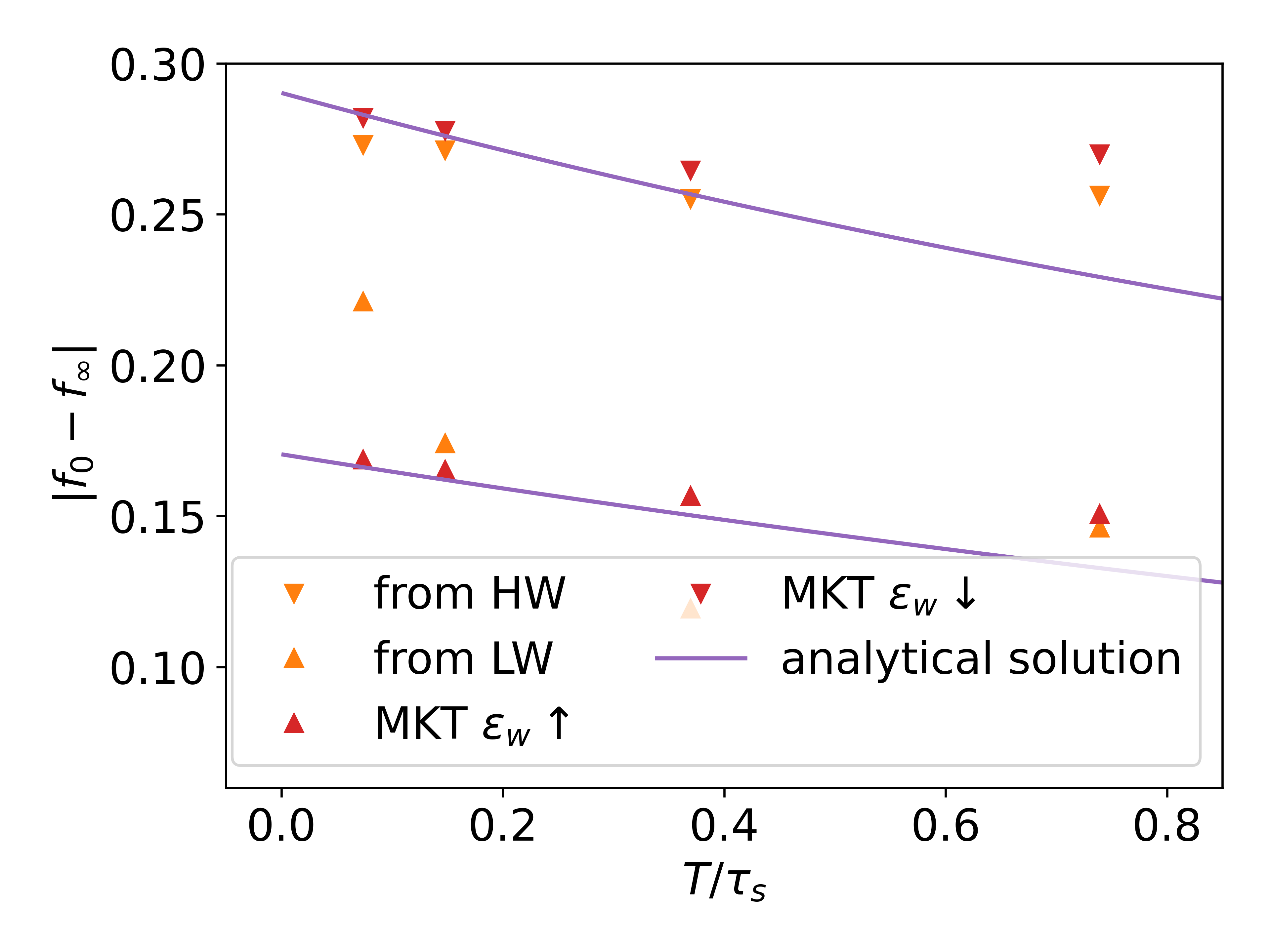} \\
\end{tabular}
\caption{(a) and (b) amplitudes resulting from stretched exponential fits to MD and TF data respectively for switching between $\epsilon=0.632$ and $\epsilon=0.762$. }
\label{fig:amplitude}
\end{figure*}

 {\it Amplitudes:} Finally, we study the dependence of the amplitude on the switching period. The analytical solution given in Eq.~\eqref{eq:amplitude} overlaps very well with the results from the MD and MKT simulations even for higher changes in wettability. In Fig.~\ref{fig:amplitude} it can be seen that the amplitude decreases with $T$. In (d) one would expect the circles agree well with the analytical solution. The lower circle belongs to a simulation, where the effective wettability is higher than the initial wettability. In such a case $\cos(\theta)$ exhibits a small dip before it starts to increase. This cannot be grasped by a stretched exponential function. This results in  this small deviation from the analytical solution. For higher contact angle differences as shown in (a) and (b) the underlying approximations of the analytical solution become notable, but there is still a qualitative agreement.

\section{Conclusion} \label{sec:conclusion}
In this paper we characterized the wetting dynamics of a droplet on a periodically switched surface by microscopic MD and  mesoscopic TF simulations. Through the mapping of the respective energy scales the equilibrium behavior can be mapped. Via additional rescaling of the respective spatial and temporal scales, also the non-equilibrium behavior of both approaches can be compared in dimensionless units. Indeed, both approaches display a very similar relaxation behavior when analysing the response to a single switch or to a periodically switched substrate. 

Important additional insight could be gained by interpreting the results within the molecular kinetic theory of wetting (MKT). This was realized on different levels. 

(1) We explicitly checked the assumption of the MKT, namely the presence of a driving force for the contact line dynamics which is proportional to the difference of the cosine of the contact angle to its equilibrium value. This was fulfilled for a large range of contact angles and allowed for the definition of a friction term, translating the driving force into the dynamics of the contact line. Only for times directly after a switching event, the contact angle hardly changed due to additional reorganization of the droplet shape. To a good approximation this can be described as a kind of dead time. 

(2) We formulated the time evolution of the contact angle based on the MKT approach, taking into account the preservation of the droplet volume. The input for the single model parameter of that approach (apart from the equilibrium properties) was taken from (1). For this comparison it turned out to be very helpful to average the relaxation behavior over single switching periods and to analyse the time evolution of these averaged values. After careful comparison of different features of this time evolution (long time stationary behavior, i.e. plateau value, relaxation time, and amplitude of relaxation) the results from MKT turned out to be very close to those obtained from MD and TF simulations. 
Beyond the conceptual insight gained from this comparison the MKT approach may substitute MD simulations for some applications as it is orders of magnitude more efficient with respect to simulation time.
Finally, this detailed comparison also allowed us to identify the impact of the initial dead time of the droplet relaxation. Beyond the slight modification of the plateau value of the contact angle the initial behavior gives rise to a non-monotonous dependence of the plateau value of the contact angle when approaching very short switching periods.

(3) We managed to obtain analytical solutions of the MKT equations. The approximations were particularly uncritical when the change in contact angle between both states were not too large. Indeed, some of the numerically seen features such as the independence of the relaxation time on the switching period could be reproduced analytically.

(4) Using the scheme from \citeauthor{mkt_zeta_calc}\cite{mkt_zeta_calc} we could additionally extract the required MKT parameters directly from analysis of MD simulations close to the contact line. This underlines that the microscopic mechanisms, proposed in the MKT close to the contact line, are indeed relevant to describe the wetting dynamics of the droplet.

The substrates in this work have been switched homogeneously. The control of pattern formation with the help of pre-structured substrates has been the focus of research in the past\cite{HLHT2015w, TBHT2017tjocp}. The combination of switchable, pre-structured substrates promises to offer even more detailed influence on the pattern formation may be an interesting topic for future work.


\section*{Conflicts of interest}
There are no conflicts to declare.
\begin{acknowledgments}
We wish to acknowledge the financial support of the DFG Schwerpunktprogramm SPP 2171.
\end{acknowledgments}

\renewcommand{\appendixname}{Supporting Information}
\newpage
\appendix
\section{Calculation of the contact angle from rFWHM}\label{app:rFWHM_to_theta}
For a spherical cap shaped droplet one has
\begin{equation}
  r^{2} = \left(\frac{\sigma}{2}\right)^{2} + \left( r -\frac{h}{2}  \right)^{2}
\end{equation}
 where $h$ is the height, $r$ the radius of the droplet, and $\sigma$ the width at half height. Therefore, the radius $r$ can be expressed as
\begin{equation}
  \label{eq:radius}
  r = \frac{1}{4h}(\sigma^{2} + h^{2}).
\end{equation}
Further, basic trigonometry yields
\begin{equation}
  \label{eq:theta}
  \cos(\theta) = 1- \frac{h}{r}
\end{equation}
with the contact angle $\theta$.
By plugging Eq.~\eqref{eq:radius} into Eq.~\eqref{eq:theta} we get
\begin{equation}
  \cos(\theta) =  1- \frac{h}{\frac{1}{4h} \left(\sigma^{2} + h^{2} \right)}
\end{equation}
and finally we obtain
\begin{equation}
  \cos(\theta) =  1- \frac{4}{ \frac{1}{rFWHM^{2}}+1}.
\end{equation}
with the definition of the relative full width at half maximum rFWHM =$h/\sigma$.

\section{Reformulation of MKT equation}\label{app:reform_mkt}

\begin{figure}[H]
\includegraphics[width=0.85\columnwidth]{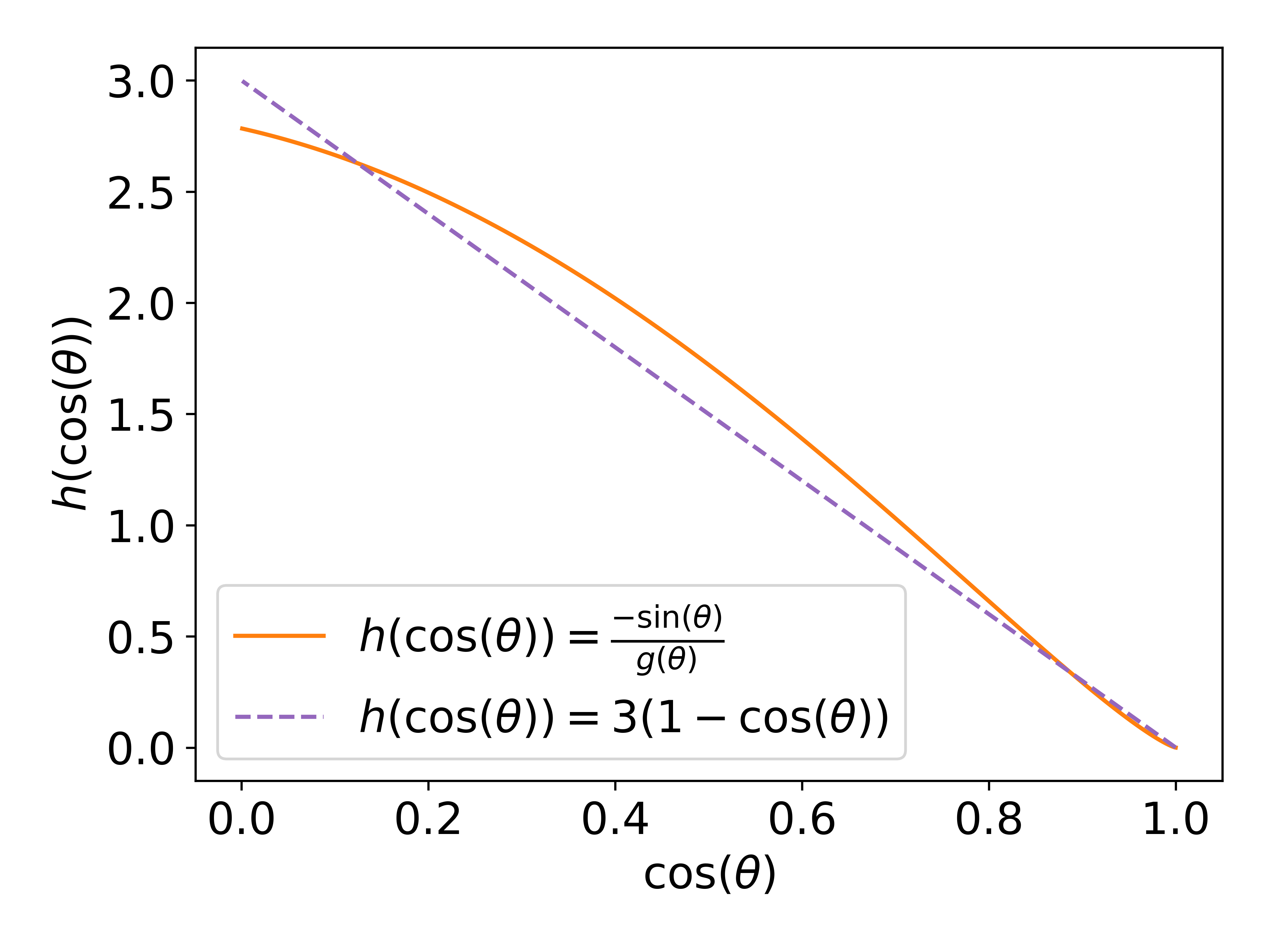}
\caption{$\frac{-\sin(\theta)}{g(\theta)}$ in comparison with $3(1-\cos(\theta))$ plotted versus $\cos(\theta)$.}
\label{fig:app_approx}
\end{figure}

\begin{figure}[H]
\begin{tabular}{ll}
(a) &(b)\\
\includegraphics[width = 0.48\columnwidth]{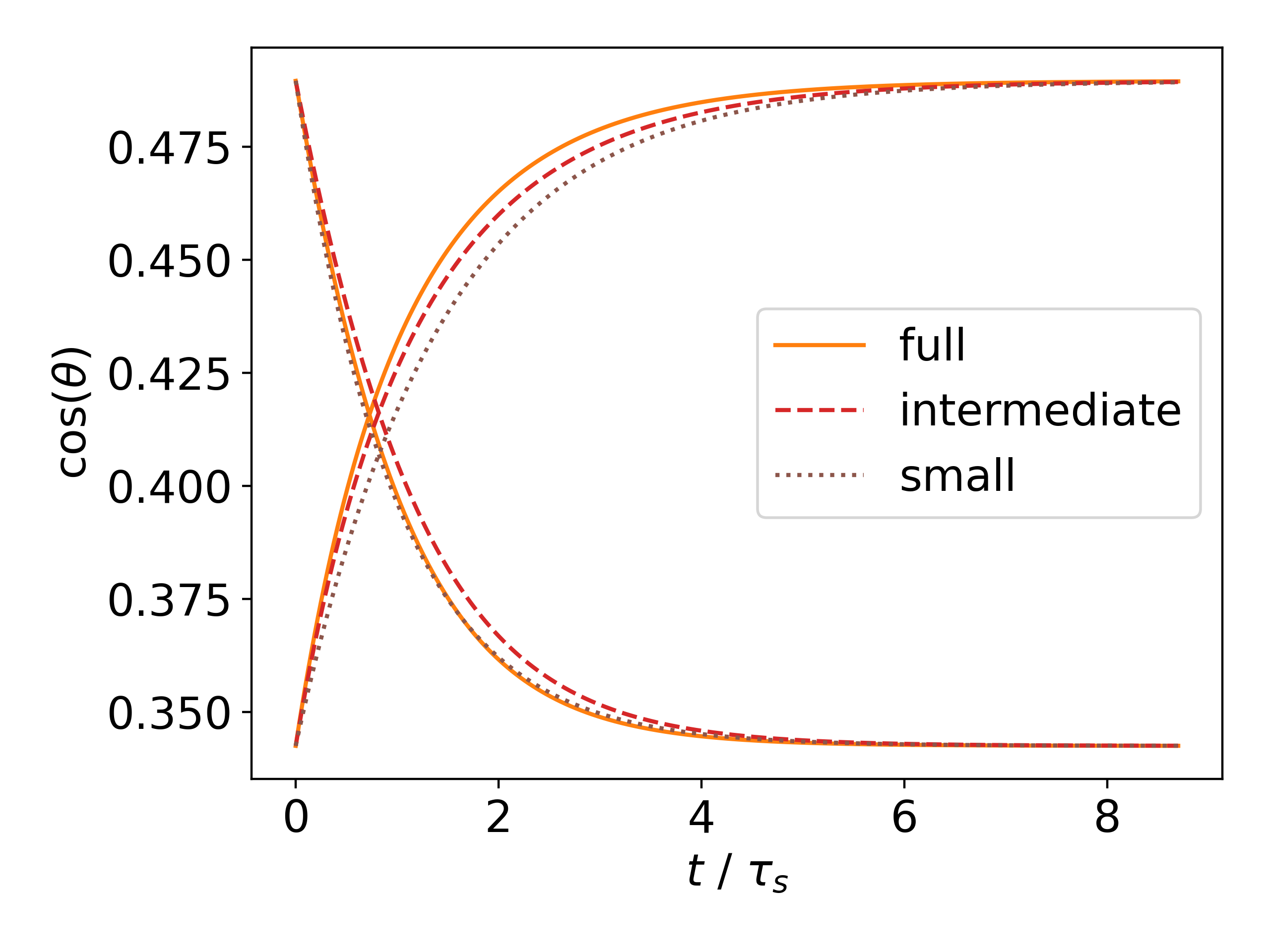} &
\includegraphics[width = 0.48\columnwidth]{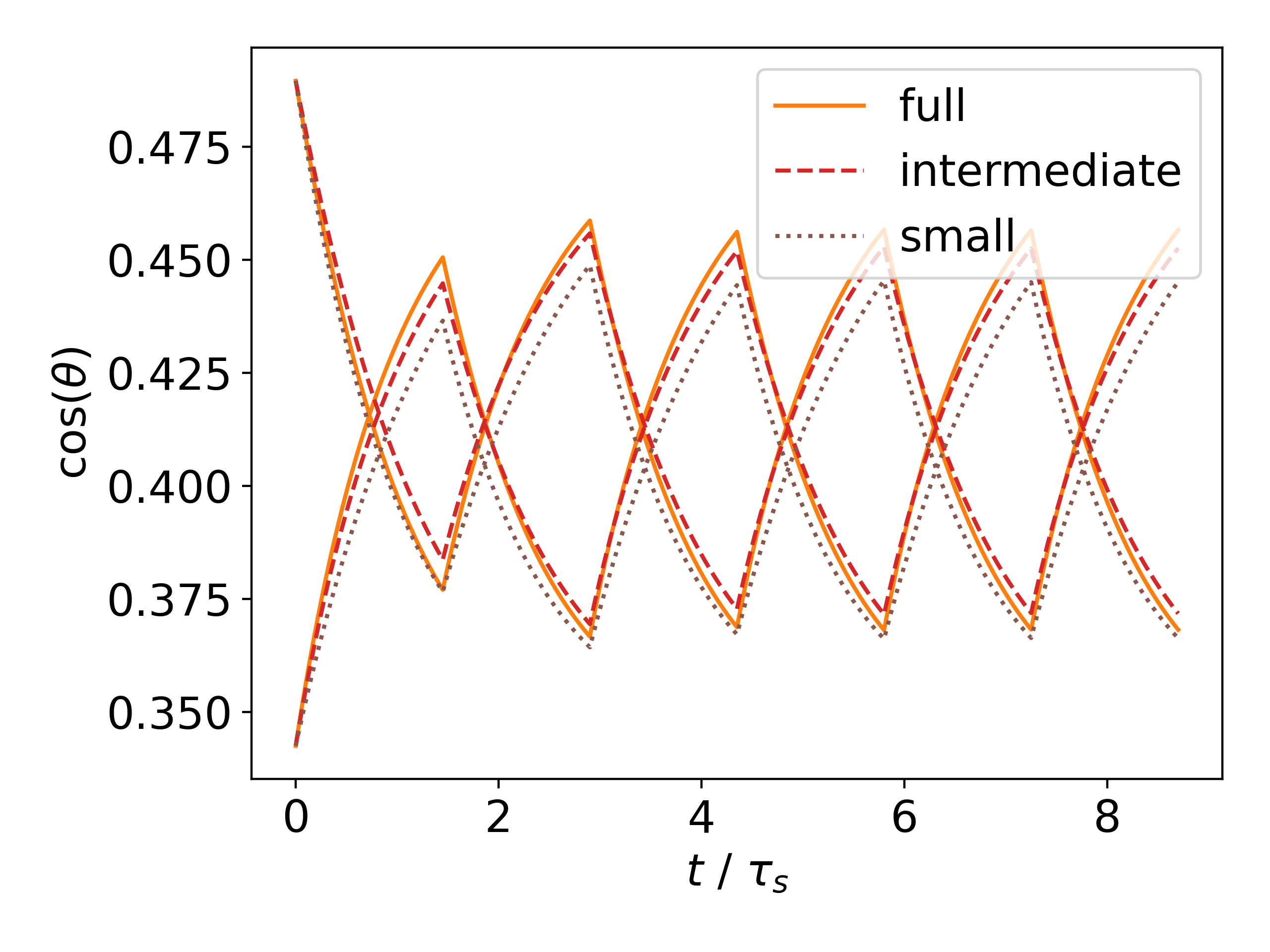}\\
\end{tabular}
\caption{$\cos(\theta)$ plotted versus $t$ for the different approximations of the MKT for (a) a single switch and (b) periodic switching between $\epsilon_{w} = 0.632$ and $\epsilon_{w} = 0.671$. (full: Eq.\ref{eq:app_mkt_circle}, intermediate: Eq.\ref{eq:mkt_intermediate_app}; small: Eq.\ref{eq:mkt_small_app})}
\label{fig:app_mkt_approx_040-045}
\end{figure}

\begin{figure}[H]
\begin{tabular}{ll}
(a) & (b) \\
\includegraphics[width = 0.48\columnwidth]{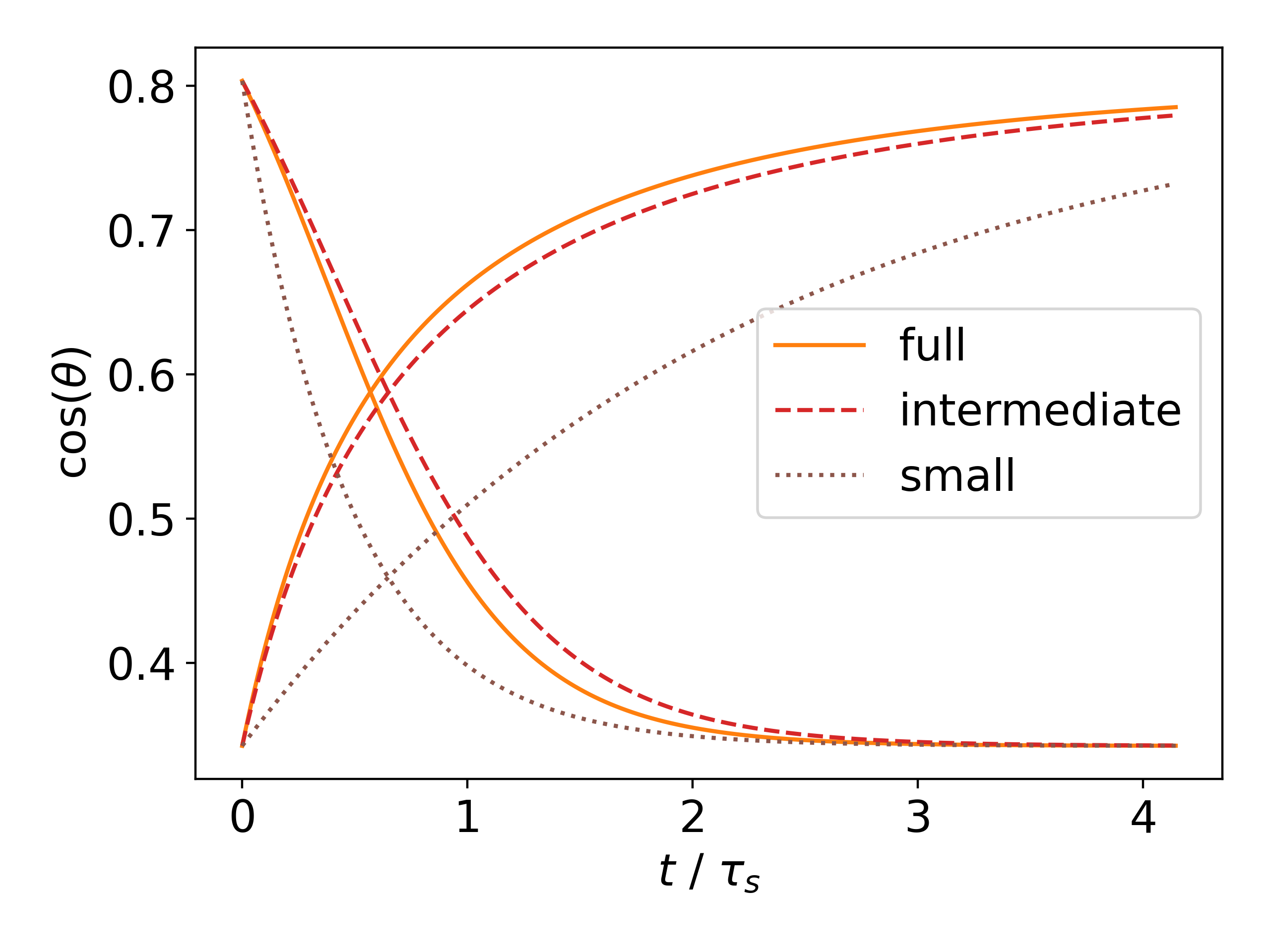} &
\includegraphics[width = 0.48\columnwidth]{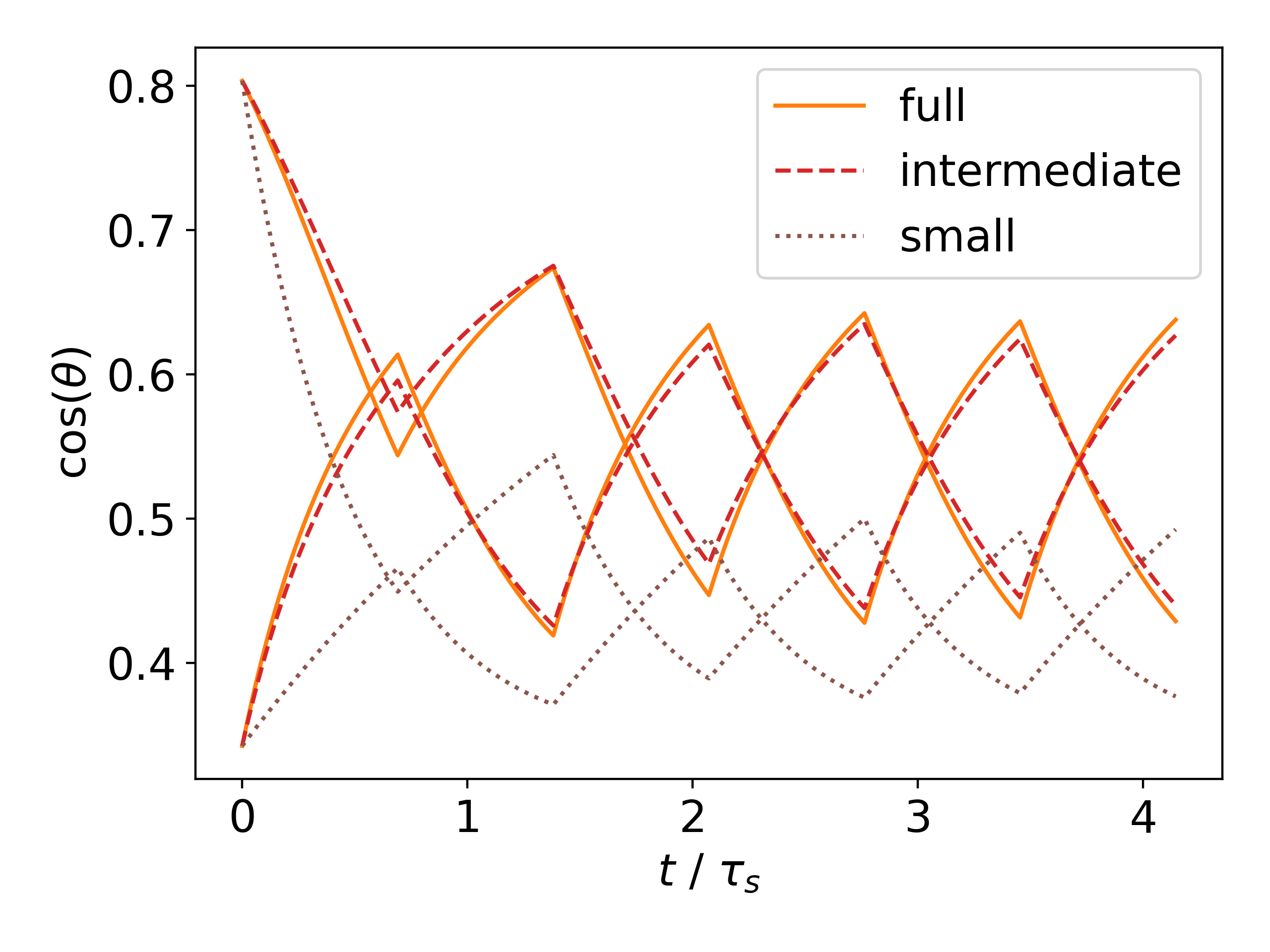}\\
\end{tabular}
\caption{$\cos(\theta)$ plotted versus $t$ for the different approximations of the MKT for (a) a single switch and (b) periodic switching between $\epsilon_{w} = 0.632$ and $\epsilon_{w} = 0.762$. (full: Eq.\ref{eq:app_mkt_circle}, intermediate: Eq.\ref{eq:mkt_intermediate_app}; small: Eq.\ref{eq:mkt_small_app})}
\label{fig:app_mkt_approx_040-058}
\end{figure}

The relation between the half chord length $r$ and the half of the central angle $\theta$ of a circular segment is given by
\begin{equation}
  r = \sqrt{2A} \frac{\sin(\theta)}{\sqrt(2\theta - \sin(2\theta))} = \sqrt{2A}f(\theta)
\end{equation}
where $A$ is the area of the circular segment.
Here, we $r$ is the radius of a droplet on a surface and $\theta$ is the contact angle.
Then we can write
\begin{equation}
  v_{cl} = \frac{dr}{dt} = \sqrt{2A} \frac{df}{d\theta}\frac{d\theta}{dt} = -\sqrt{2A} \frac{df}{d\theta} \frac{1}{sin(\theta)} \frac{d\cos(\theta)}{dt}
\end{equation}
where
\begin{equation}
  g(\theta) = \frac{df}{dt} = \frac{\cos(\theta)}{\sqrt{2\theta - \sin(2\theta)}} - \frac{\sin(\theta) (1-\cos(2\theta))}{\sqrt{2\theta - \sin(2\theta)}^{3}}.
\end{equation}
The MKT relation reads
\begin{equation}
  \label{eq:app_mkt_eq}
  v = k_{0} (\cos(\theta_{eq}) - \cos(\theta))
\end{equation}
with $k_{0} = \frac{\gamma}{\zeta}$ and can thus be rewritten as 
\begin{equation}
  \label{eq:app_mkt_circle}
 \frac{d}{dt} \cos(\theta) = -k_{1} \frac{\sin(\theta)}{g(\theta)} (\cos(\theta_{eq}) - \cos(\theta))
\end{equation}
with $k_{1} = \frac{k_{0}}{\sqrt{2A}}$.
As can be seen in Fig.~\ref{fig:app_approx} $\frac{-\sin(\theta)}{g(\theta)}$ can be approximated as $3(1-\cos(\theta))$. Thus, Eq.~\eqref{eq:app_mkt_circle} can be rewritten for intermediate changes in contact angles as
\begin{equation}
  \label{eq:mkt_intermediate_app}
  \frac{d}{dt} \cos(\theta) = k_{2} (1-\cos(\theta)) (\cos(\theta_{eq}) - \cos(\theta))
\end{equation}
and for small changes as
\begin{equation}
  \label{eq:mkt_small_app}
  \frac{d}{dt} \cos(\theta) = k_{3} (\cos(\theta_{eq}) - \cos(\theta))
\end{equation}
where $k_{3} = k_{2} (1-\cos(\theta_{eq}))$ since $\cos(\theta)$ can be substituted by the constant $\cos(\theta_{eq})$.
In Fig.~\ref{fig:app_mkt_approx_040-045} and \ref{fig:app_mkt_approx_040-058} the two approximations and the solution from Eq.~\eqref{eq:app_mkt_circle} are plotted for a single switching event and periodic switching.

From Eq.~\eqref{eq:mkt_intermediate_app} after separation of variables one can write
\begin{equation}
  \frac{dx}{(1-x)(x-x_{eq})} = -k_{2} dt
\end{equation}
Here and in the following, we will write $x$ and $x_{eq}$ for $\cos(\theta)$ and $\cos(\theta_{eq})$, respectively, for reasons of better readability.
Note that
\begin{equation}
  \frac{1}{(1-x)(x-x_{eq})} = \frac{1}{1-x_{eq}} \left[ \frac{1}{1-x} + \frac{1}{x-x_{eq}} \right]
\end{equation}
Therefore, one obtains for the solution of the differential equation
\begin{equation}
  -\ln(1-x) + \ln(x-x_{eq}) = -k_{2}(1-x_{eq})t+C
\end{equation}
with the integration constant $C = -\ln(1-x_{0}) + \ln(x_{0} - x_{eq})$. Furthermore, we use the abbreviation $k_{3} = k_{2}(1-x_{eq})$.
Then we can rewrite
\begin{equation}
  \frac{x-x_{eq}}{1-x} = \frac{x_{0}-x_{eq}}{1-x_{0}} \exp(-k_{3}t)
\end{equation}
Finally, one needs to solve this equation for $x(t)$. After a short calculation one obtains
\begin{equation}
  x(t) = \frac{x_{eq} + \epsilon(t)}{1+\varepsilon(t)}
\end{equation}
with
\begin{equation}
  \varepsilon(t) = \frac{x_{0}-x_{eq}}{1-x_{0}} \exp(-k_{3}t)
    \end{equation}
    
For the normalized relaxation function $y(t) = (x(t) - x_{eq})/(x_0-x_{eq})$ this yields
\begin{equation}
y(t) = \frac{(1-x_{eq})\exp(-k_{3}t)}{1-x_{0}+\exp(-k_{3}t)(x_{0}-x_{eq})}.
\end{equation}
For small but finite differences of $x_{0} - x_{eq}$ (indicated by a small parameter $\Delta \epsilon_{w}$) in order to learn more about the impact of increasing differences between the initial and the final state. For this purpose we take into account terms until $\epsilon^{2}$. This yields
\begin{equation}
\begin{split}
  x(t) = [x_{eq} +  \varepsilon(t)[1-\varepsilon(t)+\varepsilon(t)^{2}+ ...] \\
  \approx x_{eq} + (1- x_{eq})(\varepsilon(t)- \varepsilon(t)^{2})+...
\end{split}
\end{equation}

Then we may write
\begin{equation}
  \label{eq:mkt_approx_solved_appendix}
  y(t) = \frac{1-x_{eq}}{1-x_{0}} \left[ \exp(-k_{3}t) - \frac{x_{0}-x_{eq}}{1-x_{0}} \exp(-2k_{3}t) \right]
\end{equation}

\section{Analytical calculation of wetting properties upon periodically switching}\label{app:plateau_calculations}

We start from Eq.~\eqref{eq:mkt_small_with_x} with a contact angle independent $k_{i}$:
\begin{equation}
  \frac{d}{dt} y(t) = -k_{3,i}(y-a_{i})
\end{equation}
Its general solution reads
\begin{equation}
  y(t) = a_{i}(1-\exp\qty(-k_{3,i}t)) + y(0)\exp\qty(-k_{3,i}t)
\end{equation}
As before, $y$ is a normalized version of the cosine of the contact angle.
Here, we first consider that before the first switching event the system is in the state of higher wettability. Then, in the first part of the switching experiment ($t\in[0,T/2)$) we have $a_{\downarrow} =0$, in the second half $a_{\uparrow}= 1$ ($t \in [T/2,T)$). The period is denoted as $T$.  Thus, for the first half switching period one obtains
\begin{equation}
  y(t) = y(0)\exp(-k_{3,\downarrow}t)
\end{equation}
and for the second half
\begin{equation}
  y(t) = (1-\exp(-k_{3,\uparrow}t)) + y(0) \exp(-k_{3,\downarrow}T/2) \exp(-k_{3,\uparrow}T/2)
\end{equation}
The average over the first time interval is thus given by 
\begin{equation}
  \langle y_{1}\rangle  = y(0) \frac{1}{k_{3,\downarrow}T/2}\qty(1-\exp\qty(-k_{3,\downarrow}T/2))
\end{equation}
and that over the second time interval
\begin{equation}
\begin{split}
  \langle y_{2} \rangle =& 1 - \frac{1-\exp(-k_{3,\uparrow}T/2)}{k_{3,\uparrow}T/2} + y(0) \exp(-k_{3,\downarrow}T/2)\frac{1}{k_{3,\uparrow}T/2} (1- \exp\qty(-k_{3,\uparrow}T/2)) \\ 
  =& 1- \frac{(1-\exp\qty(-k_{3,\uparrow}T/2))}{k_{3,\uparrow}T/2} (1-y(0)\exp(-k_{3,\downarrow}T/2))
\end{split}
\end{equation}
The average over both time intervals finally reads
\begin{equation} \label{eq:app_average_first_period}
\begin{split}
  \langle y(0) \rangle =& \frac{1}{2} - \frac{1-\exp(-k_{3,\uparrow} T/2)}{2k_{3,\uparrow}T/2} + y(0) \Bigg[ \frac{1}{2k_{3,\downarrow}T/2} (1-\exp\qty(-k_{3,\downarrow}T/2)) \\
   &+ \exp\qty(-k_{3,\downarrow}T/2) \frac{1}{2 k_{3,\uparrow} T/2} (1-\exp\qty(-k_{3,\uparrow}T/2)) \Bigg]
\end{split}
\end{equation}
 Naturally, Eq.~\eqref{eq:app_average_first_period} also holds to express the average $\langle y(n) \rangle $ during the time $t = 2nT/2$ and $t = 2(n+1)T/2$ in dependence of $y(t=n\cdot T)$. Thus, we first need to find an explicit expression for $y(n)$. With the general solution given above, one can directly write (using the abbreviation $K_{3} = k_{3,\downarrow} + k_{3,\uparrow}$)
\begin{equation}
  y(t=T) = (1- \exp(-k_{3,\uparrow}T/2)) + y(0) \exp(-K T/2) \equiv C + D y(0)
\end{equation}
In general one has
\begin{equation}
  y(t=n\cdot T)) = C+ D y(t=(n-1)\cdot T)
\end{equation}
This recursive relation has a straightforward solution which reads (setting $y(0) = 1$)
\begin{equation}
\begin{split}
  y(t=n\cdot T) &= C (1+ D + D^{2} + ... + D^{n-1}) + D^{n} = C \frac{1-D^{n}}{1-D}+ D^{n} \\
  &= \frac{1-\exp(-k_{3,\uparrow}T/2)}{1-\exp(-K_{3}T/2)} (1- \exp\qty(-K_{3}nT/2)) + \exp\qty(-K_{3}nT/2)
\end{split}
\end{equation}
Thus, we finally have
\begin{equation}
\begin{split}
  \langle y(n) \rangle &= \frac{1}{2} - \frac{1-\exp(-k_{3,\uparrow} T/2)}{2k_{3,\uparrow}T/2} \\
  &+ \left[ \frac{1-\exp(-k_{3,\uparrow}T/2)}{1-\exp(-K_{3}T/2)} (1- \exp\qty(-K_{3}nT/2)) + \exp\qty(-K_{3}nT/2) \right] \\
  &\cdot \left[ \frac{1}{2k_{3,\downarrow}T/2} (1-\exp\qty(-k_{3,\downarrow}T/2)) + \exp\qty(-k_{3,\downarrow}T/2) \frac{1}{2 k_{3,\uparrow} T/2} (1-\exp\qty(-k_{3,\uparrow}T/2)) \right]
 \end{split}
\end{equation}

In the long-time limit one finds the plateau value
\begin{equation}
\begin{split}
  \lim_{n \rightarrow \infty} y(t=n\cdot T) &= \frac{1}{2} \Bigg\lbrace 1 + \frac{1-\exp(-k_{3,\uparrow}T/2)}{1-\exp(-K_{3}T/2)} \frac{1}{k_{3,\downarrow}T/2} (1- \exp(-k_{3,\downarrow}T/2)) \\
 & - \frac{1-\exp(-k_{3,\uparrow}T/2)}{k_{3,\uparrow}T/2} \left( 1 - \frac{1-\exp(-k_{3,\uparrow}T/2)}{1-\exp(-K_{3}T/2)} \exp(-k_{3,\downarrow}T/2) \right)\Bigg\rbrace \\
  &= \frac{1}{2} \Bigg\lbrace 1 + \frac{1-\exp(-k_{3,\uparrow}T/2)}{1-\exp(-K_{3}T/2)} \frac{1}{k_{3,\downarrow}T/2} (1- \exp\qty(-k_{3,\downarrow}T/2)) \\
  &- \frac{1-\exp(-k_{3,\uparrow}T/2)}{k_{3,\uparrow}T/2} \frac{1-\exp(-k_{3,\downarrow}T/2)}{1-\exp(-K_{3}T/2)} \Bigg\rbrace \\
  &= \frac{1}{2} \Bigg\lbrace 1 + \frac{(1-\exp(-k_{3,\uparrow}T/2))(1-\exp(-k_{3,\downarrow}T/2)}{1-\exp(-K_{3}T/2)} \Bigg(\frac{1}{k_{3,\downarrow}T/2} \\
 & - \frac{1}{k_{3,\uparrow}T/2}\Bigg) \Bigg\rbrace \equiv y_{plateau}
\end{split}
\end{equation}
For very fast switching this boils down to
\begin{equation}
  y_{plateau} = \frac{k_{3,\uparrow}}{K_3} - \frac{k_{3,\uparrow} - k_{3,\downarrow}}{24K_3} k_{3,\downarrow} k_{3,\uparrow} (T/2)^{2}
\end{equation}
The general equation for $y(n)$ can be rewritten with $y_{plateau}$
\begin{equation}
\begin{split}
  \langle y(n) \rangle &= y_{plateau} + \exp(-K_{3}nT/2) \left( 1- \frac{1-\exp(-k_{3,\uparrow}T/2)}{1-\exp(-K_{3}T/2)} \right) \\
  &\cdot \left[ \frac{1}{2k_{3,\downarrow}T/2} (1-\exp\qty(-k_{3,\downarrow}T/2)) + \exp\qty(-k_{3,\downarrow}T/2) \frac{1}{2 k_{3,\uparrow} T/2} (1-\exp\qty(-k_{3,\uparrow}T/2)) \right] \\
&\equiv y_{plateau} + \exp\qty(-K(n+\frac{1}{2})T/2) \cdot \hat{y}_+
\end{split}
\end{equation}
with
\begin{equation}
\begin{split}
 \hat{y}_+ &= y \exp\qty(\frac{K_{3}T}{4}) \frac{\exp(-k_{3,\uparrow}T/2)-\exp(-K_{3}T/2)}{1-\exp(-K_{3}T/2)} \\
  &\cdot \left[ \frac{1}{k_{3,\downarrow}T} (1-\exp\qty(-k_{3,\downarrow}T/2)) + \exp\qty(-k_{3,\downarrow}T/2) \frac{1}{ k_{3,\uparrow} T} (1-\exp\qty(-k_{3,\uparrow}T/2)) \right] \\ 
\end{split}
\end{equation}

When starting from the state with lower wettability one ends up with 
\begin{equation}
   \langle y(n) \rangle = y_{plateau} - \exp\qty(-K(n+\frac{1}{2})T/2) \cdot \hat{y}_-
 \end{equation}
$\hat{y}_-$ is identical to $\hat{y}_+$ after exchange of $k_{3,\downarrow}$ with $k_{3,\uparrow}$.

For $k_{3,\downarrow} = k_{3,\uparrow} = k_{3}$ these expressions simplify to 
\begin{equation}
\begin{split}
  y(n) &= \frac{1}{2} + \exp\qty(-k_{3}(2n+1)T/2) \frac{1-\exp(-k_{3}T/2)}{1-\exp(-2k_{3}T/2)} \frac{1}{2k_{3}T/2} \\
  &+(1 - \exp(-k_{3}T/2))(1+\exp(-k_{3}T/2))  \\
  &=\frac{1}{2} + \exp(-k_{3}(2n+1)T/2) \frac{1}{2k_{3}T/2}(1-\exp(-k_{3}T/2))
\end{split}
\end{equation}

\section{$\zeta$} \label{app:mkt_values}
Table Tab.~\ref{tab:zeta_values} shows values of $\zeta_{R}$ extracted from the MD model with two different methods. $\zeta$ can be obtained via $K_0$, which is the inverse time needed for half the particles to move from the first to the second layer, and $n$, the density of the first liquid layer\cite{mkt_zeta_calc}.
 $\zeta_{MD}$ denotes the $\zeta$ value extracted from an analysis of the contact line velocity in dependence of the cosine of the contact angle. A visualization is shown in Figure \ref{fig:zeta_vs_eps}. 
 
\begin{table}[h]
  \caption{\label{tab:zeta_values}Values of $K_{0}$, $n$, and $\zeta_{R}$ directly calculated from MD simulations as well as values of $\zeta_{MD}$ extracted from the analysis of the contact line velocity dependence on the cosine of the contact angle.}
\begin{tabular*}{0.98\columnwidth}{@{\extracolsep{\fill}}ccccc}
\mbox{$\epsilon$} & \mbox{$K_{0} / \tau^{-1}$} & \mbox{$n /\sigma^{-3}$}& \mbox{$\zeta_{R}$} \\
      \hline
      0.447 & $4.62 \cdot 10^{-4}$ & 0.57 & $0.93 \cdot 10^{3}$  \\ 
      0.548 & $3.81 \cdot 10^{-4}$ & 0.62 & $1.23 \cdot 10^{3}$  \\ 
      0.632 & $3.24 \cdot 10^{-4}$ & 0.66 & $1.53 \cdot 10^{3}$  \\ 
      0.707 & $2.76 \cdot 10^{-4}$ & 0.69 & $1.88 \cdot 10^{3}$  \\ 
      0.742 & $2.55 \cdot 10^{-4}$ & 0.70 & $2.07 \cdot 10^{3}$  \\ 
      0.762 & $2.44 \cdot 10^{-4}$ & 0.71 & $2.19 \cdot 10^{3}$  \\ 
      0.809 & $2.40 \cdot 10^{-4}$ & 0.71 & $2.23 \cdot 10^{3}$  \\ 
      0.775 & $2.17 \cdot 10^{-4}$ & 0.73 & $2.51 \cdot 10^{3}$  \\ 
      0.837 & $2.03 \cdot 10^{-4}$ & 0.74 & $2.73 \cdot 10^{3}$  \\ 
   \end{tabular*}
\end{table}
 
In the TF model the contact angle velocity can be analyzed analogously. The resulting values can be found in Tab.~\ref{tab:zeta_values_tf}. 
 
\begin{table}[h]
\centering
  \caption{\label{tab:zeta_values_tf}Values of $\zeta_{TF}$ within the TF model extracted from the analysis of the contact line velocity dependence on the cosine of the contact angle}
\begin{tabular*}{0.98\columnwidth}{@{\extracolsep{\fill}}ccc}
\mbox{Initial $\epsilon$} & \mbox{Final $\epsilon$} & \mbox{$\zeta_{TF}$}\\
      \hline
0.762 & 0.632 & 1.60 \\ 
0.671 & 0.632 & 1.39 \\ 
0.632 & 0.671 & 1.52 \\ 
0.632 & 0.762 & 2.72 \\ 
\end{tabular*}
\end{table}

\section{Additional plots of the contact line velocity $v_{cl}$ versus $\cos(\theta)$}
\begin{figure}[H]
\centering
\includegraphics[width=0.7\columnwidth]{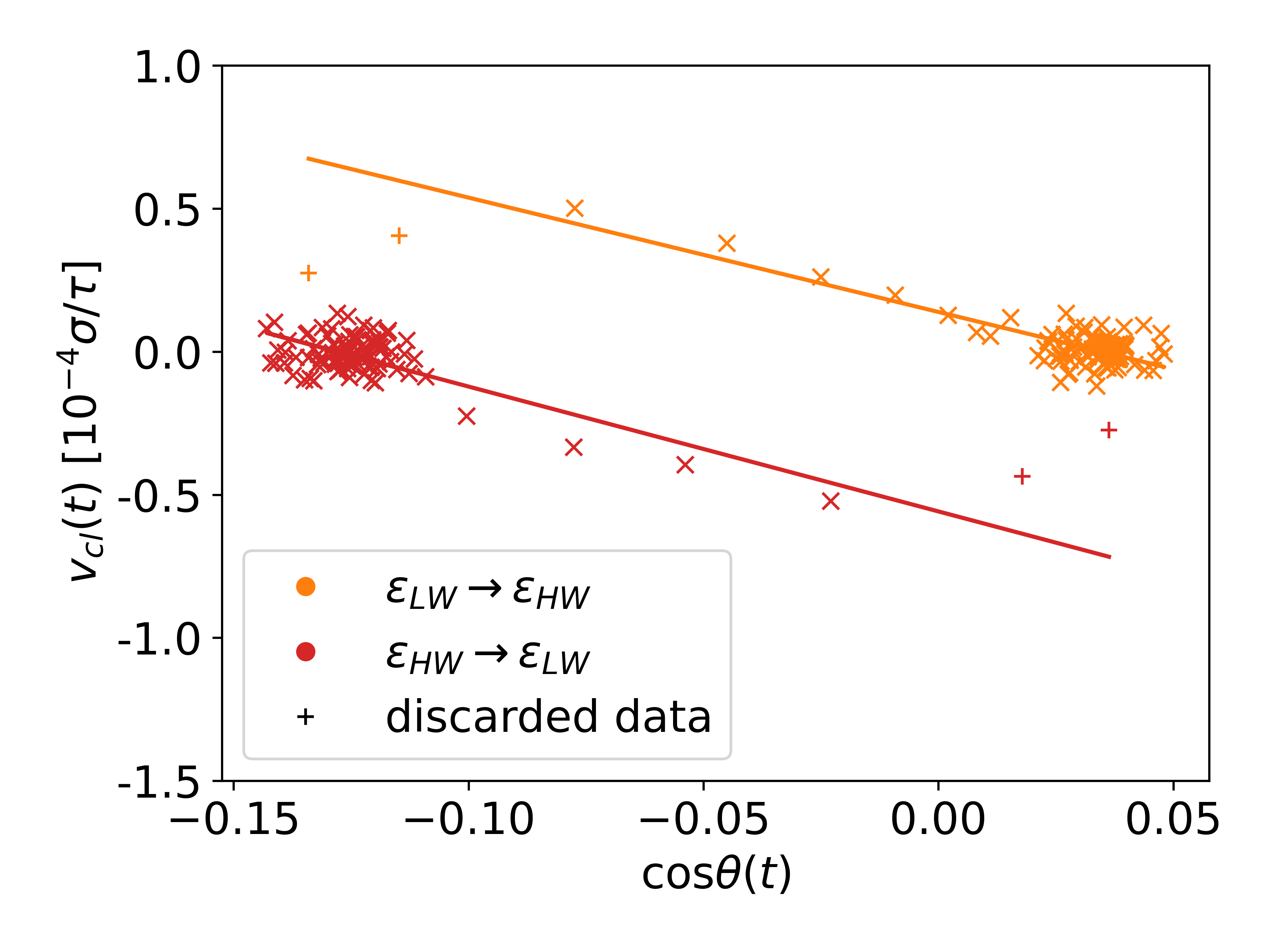}
\caption{Contact line velocity $v_{cl}$ plotted against $\cos(\theta)$ for switching between wettabilities of $\epsilon_{LW} = 0.500$ and $\epsilon_{HW} = 0.548$.}
\label{fig:app_v_vs_theta_025-030}
\end{figure}

\begin{figure}[H]
\centering
\includegraphics[width=0.7\columnwidth]{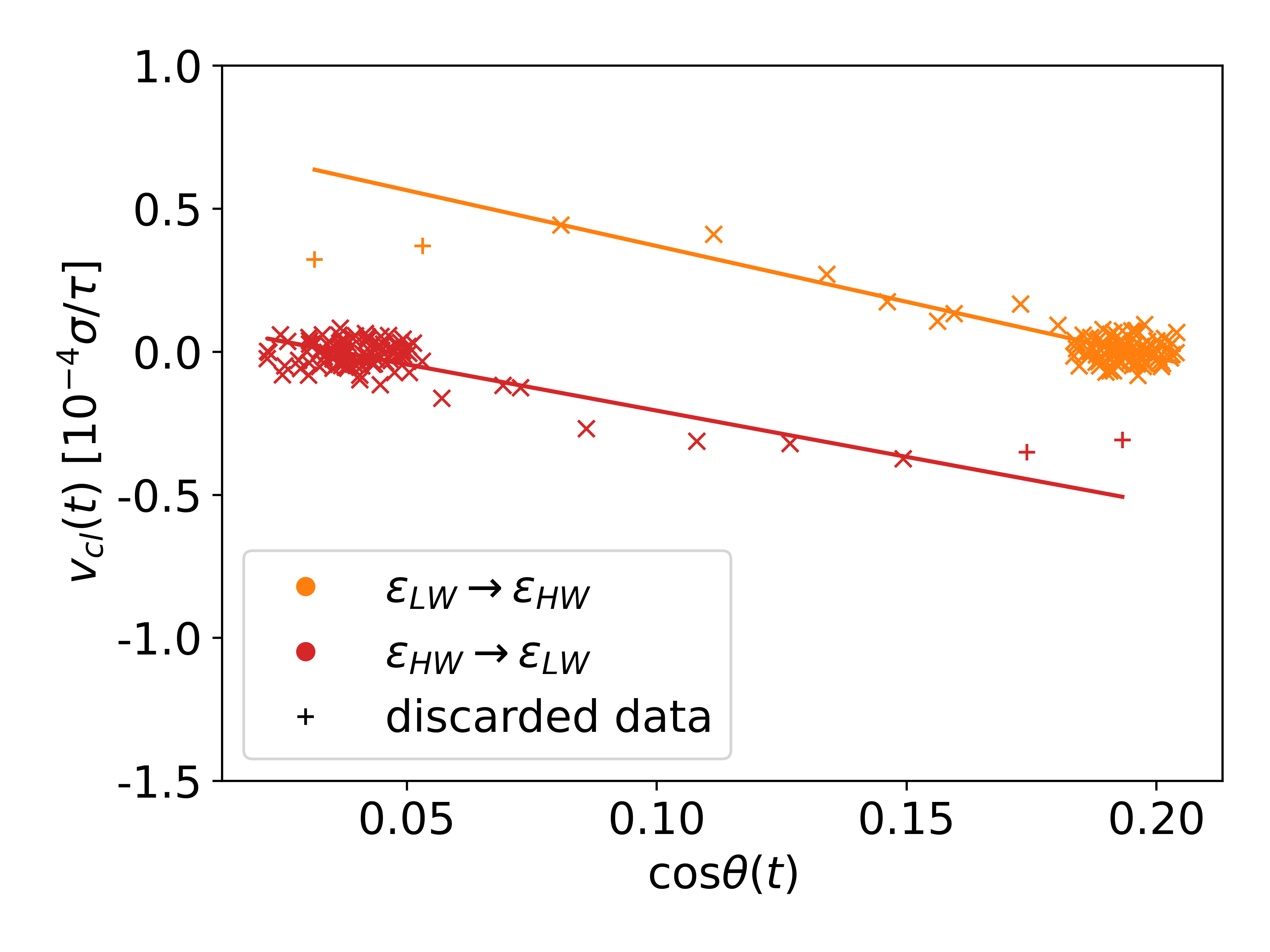}
\caption{Contact line velocity $v_{cl}$ plotted against $\cos(\theta)$ for switching between wettabilities of $\epsilon_{LW} = 0.548$ and $\epsilon_{HW} = 0.592$.}
\label{fig:app_v_vs_theta_030-035}
\end{figure}

\begin{figure}[H]
\centering
\includegraphics[width=0.7\columnwidth]{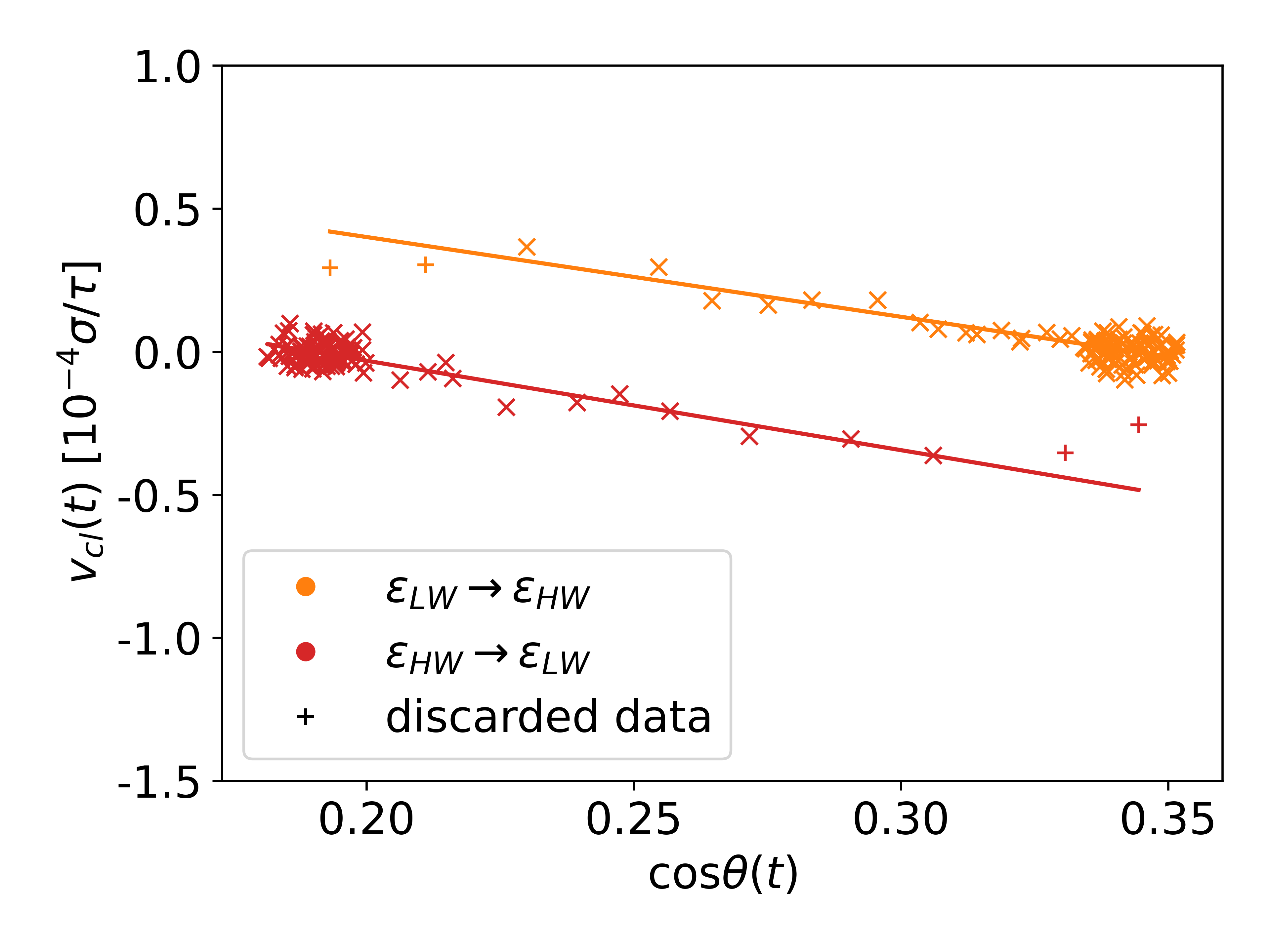}
\caption{Contact line velocity $v_{cl}$ plotted against $\cos(\theta)$ for switching between wettabilities of $\epsilon_{LW} = 0.592$ and $\epsilon_{HW} = 0.632$.}
\label{fig:app_v_vs_theta_035-040}
\end{figure}

\begin{figure}[H]
\centering
\includegraphics[width=0.7\columnwidth]{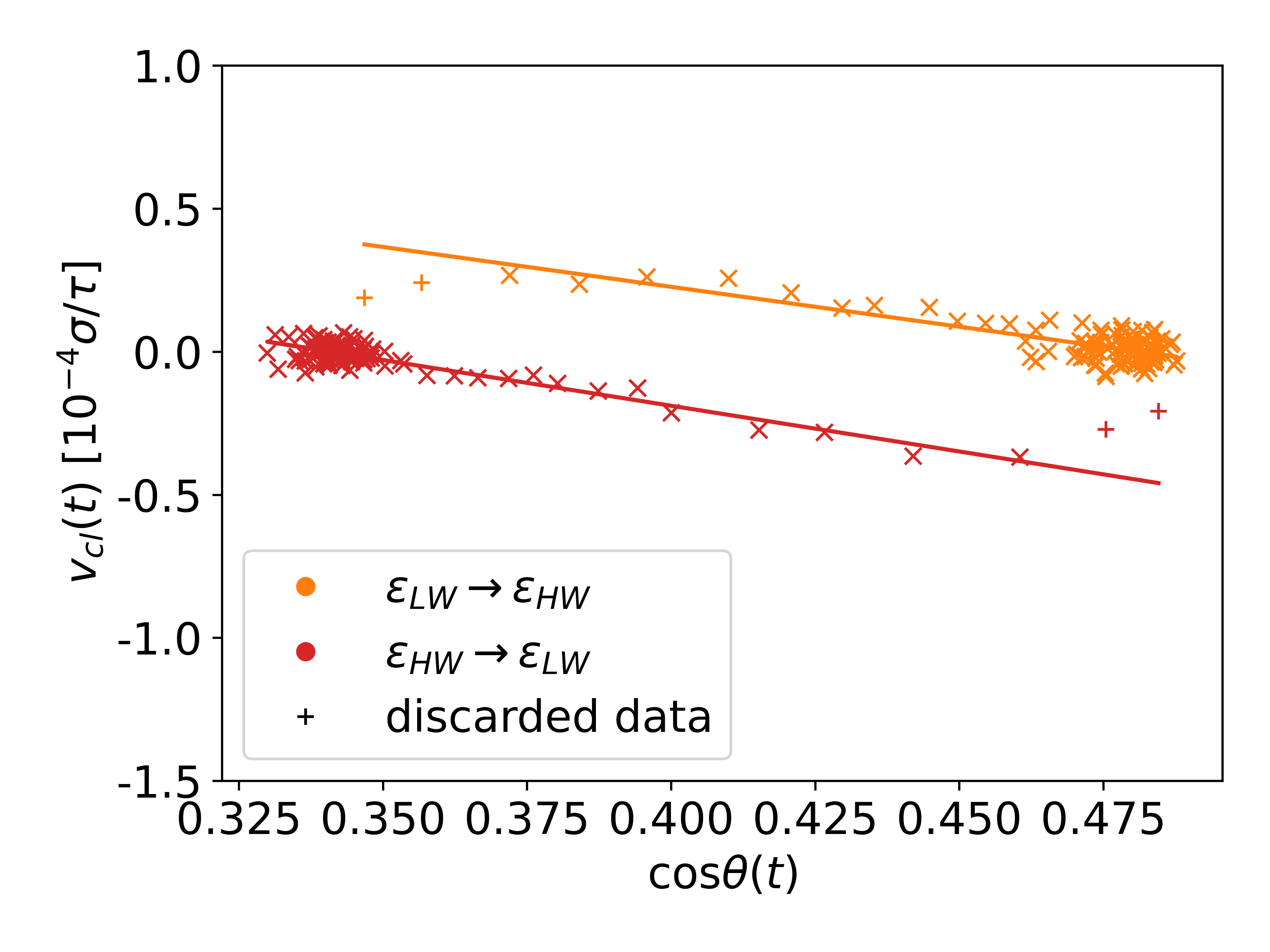}
\caption{Contact line velocity $v_{cl}$ plotted against $\cos(\theta)$ for switching between wettabilities of $\epsilon_{LW} = 0.632$ and $\epsilon_{HW} = 0.671$.}
\label{fig:app_v_vs_theta_040-045}
\end{figure}

Figures \ref{fig:app_v_vs_theta_025-030}, \ref{fig:app_v_vs_theta_030-035}, \ref{fig:app_v_vs_theta_035-040} and \ref{fig:app_v_vs_theta_040-045} show the contact line velocity dependence on the cosine of the contact angle for additional wettabilities not discussed in main manuscript. 
\label{app:plots_vcl_vs_cos}

\section{Stretching/compression of the relaxation}\label{app:beta}

\begin{figure}[H]
\centering
\begin{tabular}{lll}
   (a) & (b)\\
\includegraphics[width = 0.45\columnwidth]{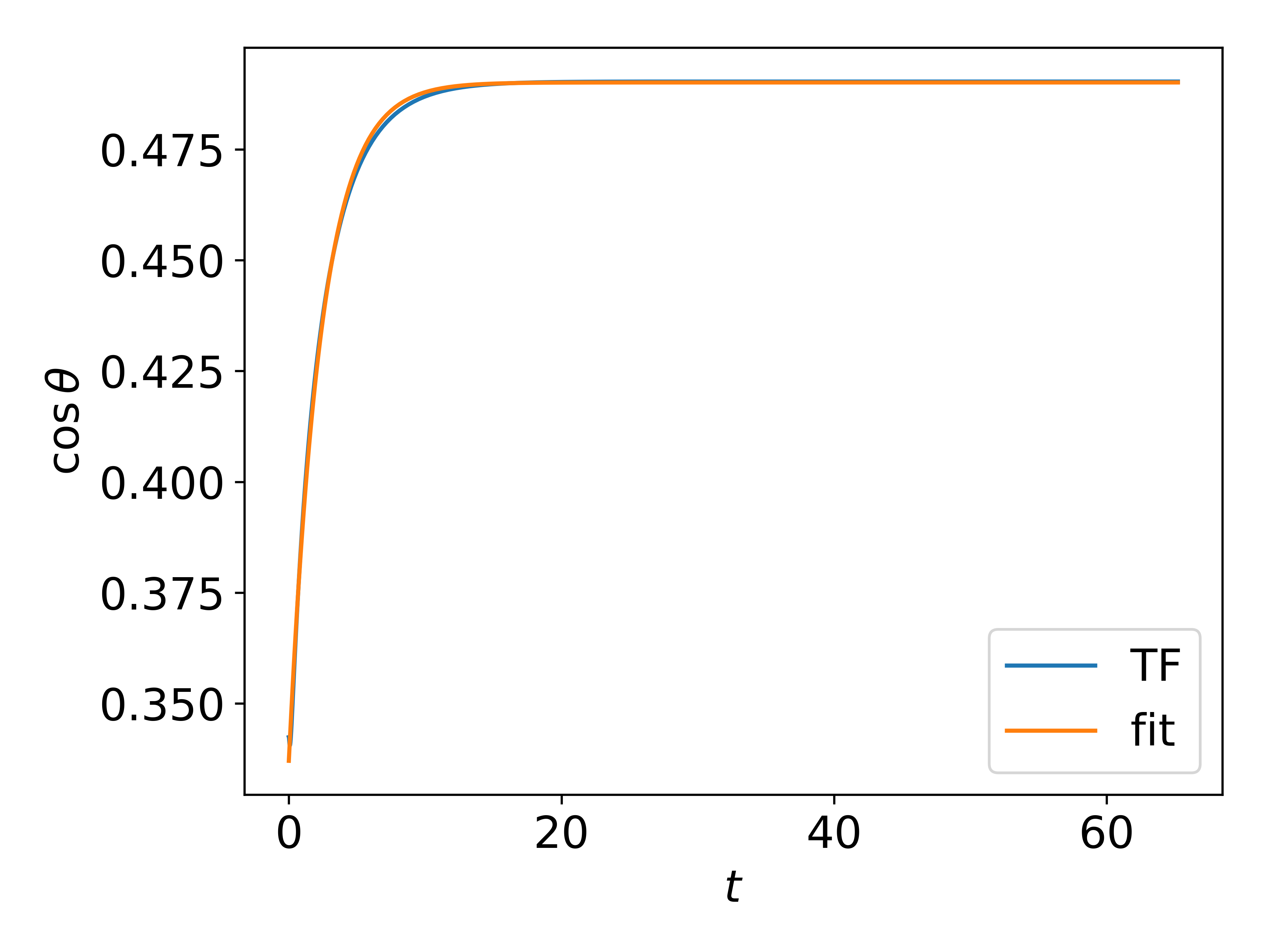} &
\includegraphics[width = 0.45\columnwidth]{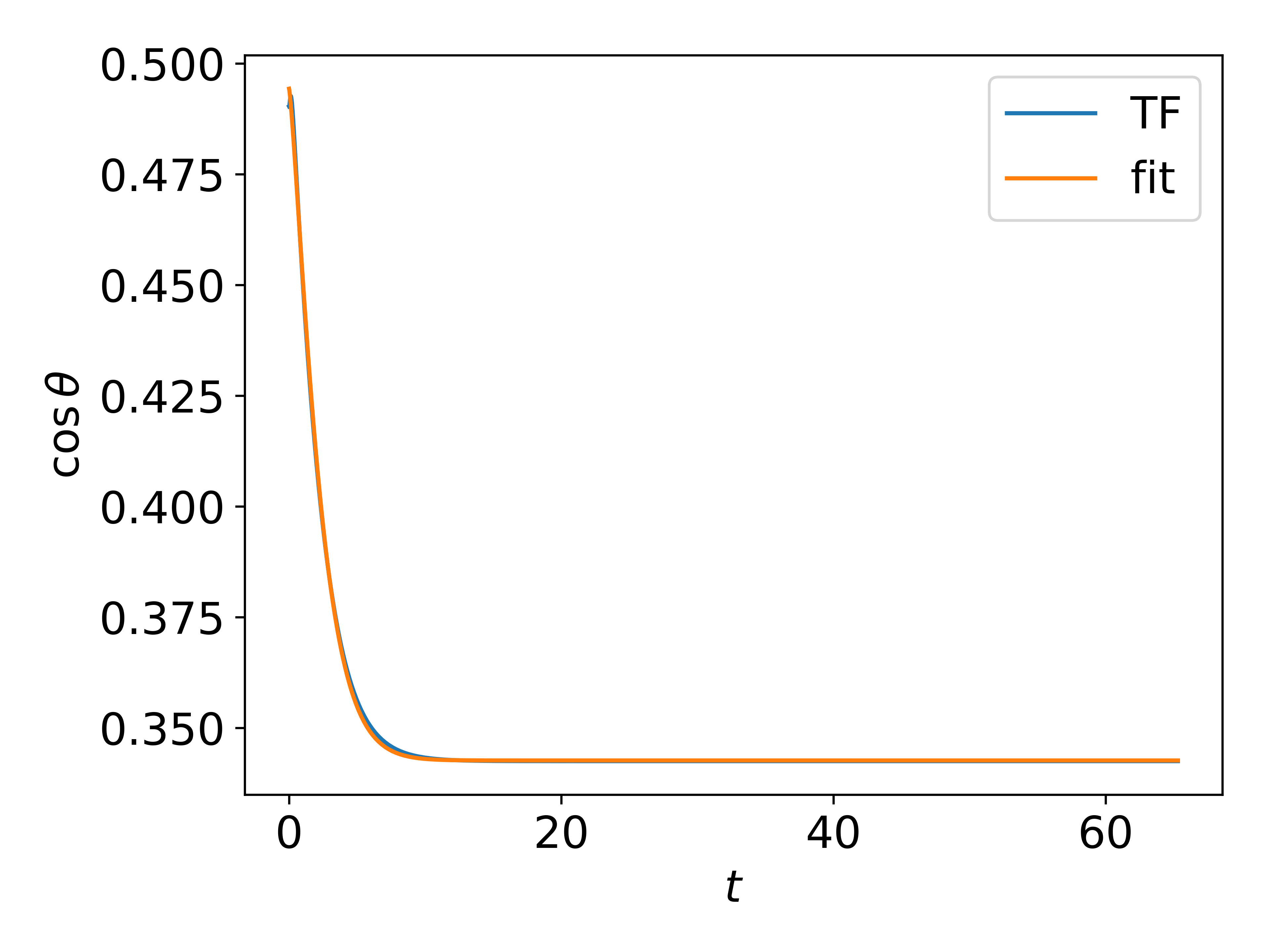}\\
(c) & (d)\\
\includegraphics[width = 0.45\columnwidth]{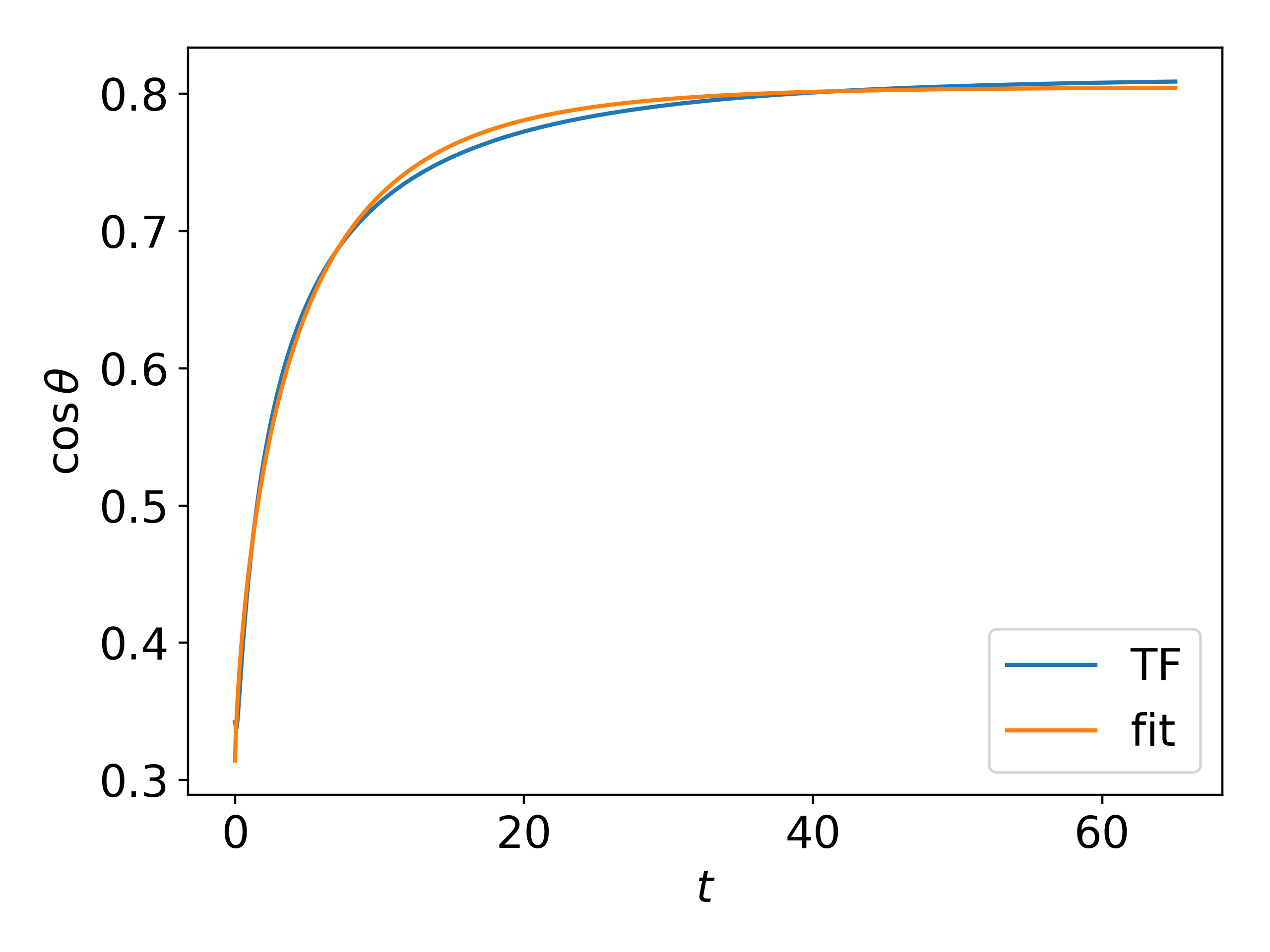} &
\includegraphics[width = 0.45\columnwidth]{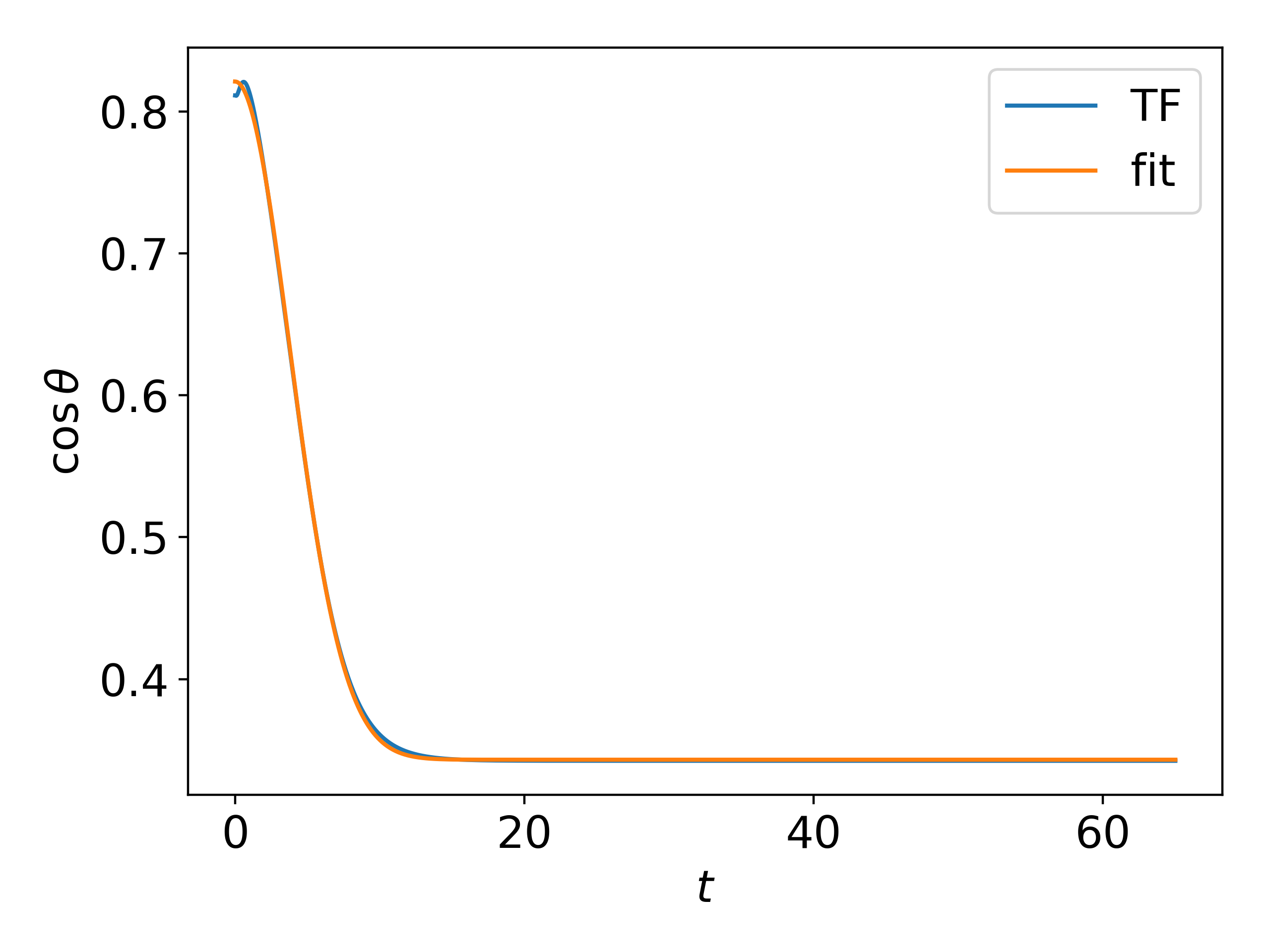}
\end{tabular}
\caption{Relaxation of $\cos \theta$ after an instantaneous change in wettability in the TF model for the wettability values corresponding to different changes in interaction strengths $\epsilon_1\rightarrow\epsilon_2$ in the MD model: a) $0.632 \rightarrow 0.671$, b) $0.671 \rightarrow 0.632$, c) $0.632 \rightarrow 0.0.762$ and d) $0.762 \rightarrow 0.632$. }\label{fig:tf_single_switch_fits}
\end{figure}

\begin{figure}[H]
\centering
\begin{tabular}{lll}
   (a) & (b)\\
\includegraphics[width = 0.45\columnwidth]{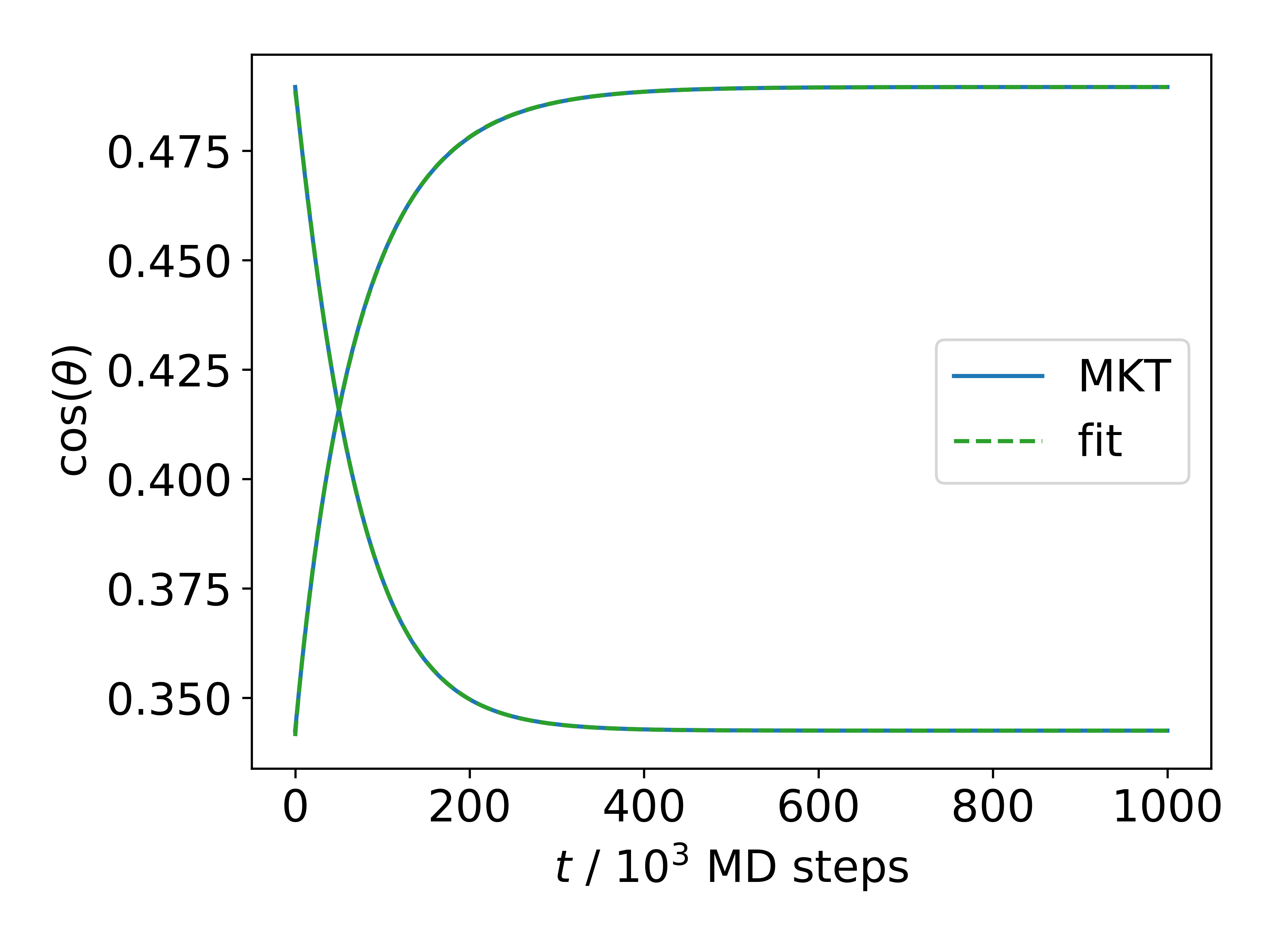} &
\includegraphics[width = 0.45\columnwidth]{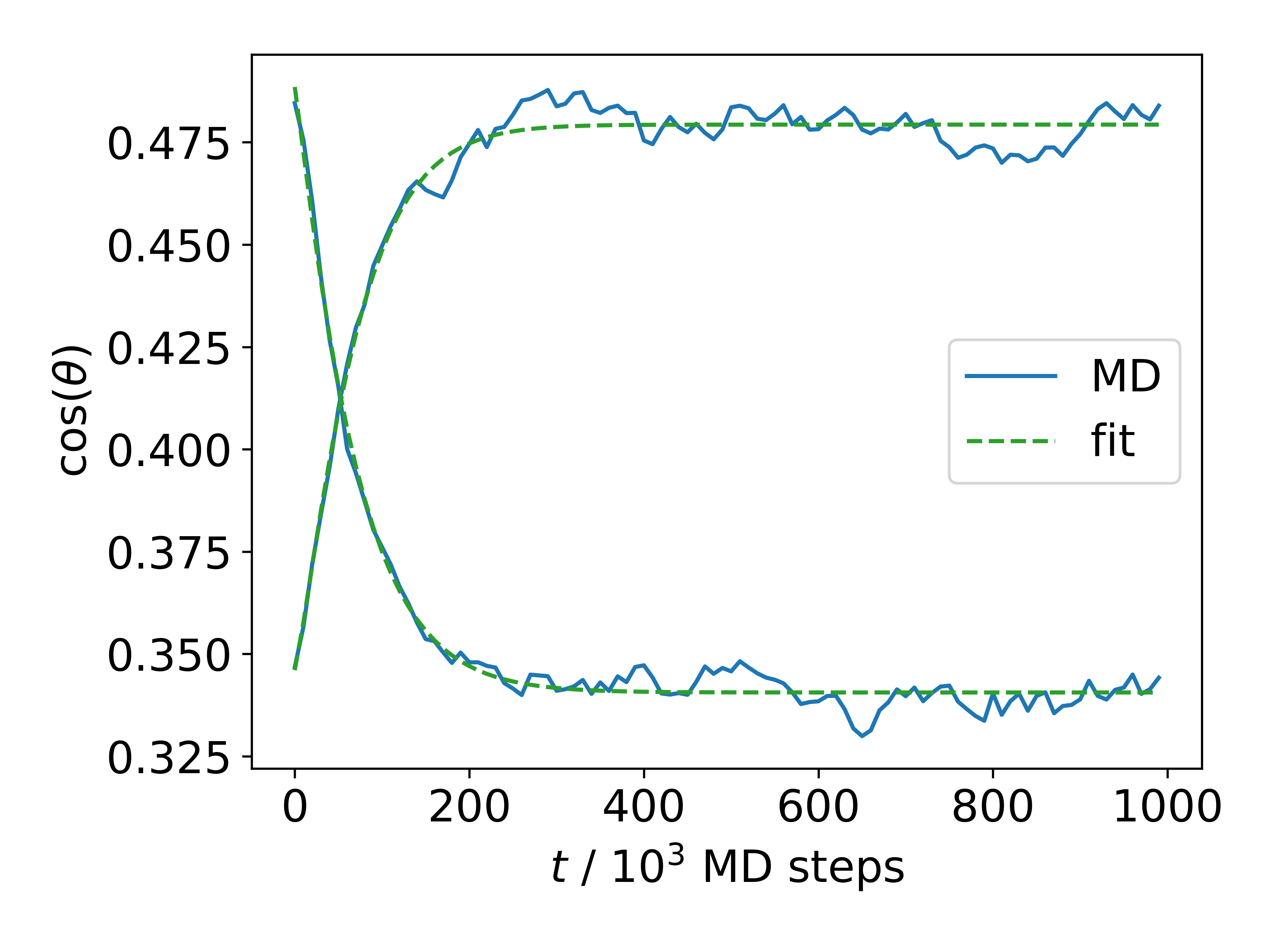}\\
(c) & (d)\\
\includegraphics[width = 0.45\columnwidth]{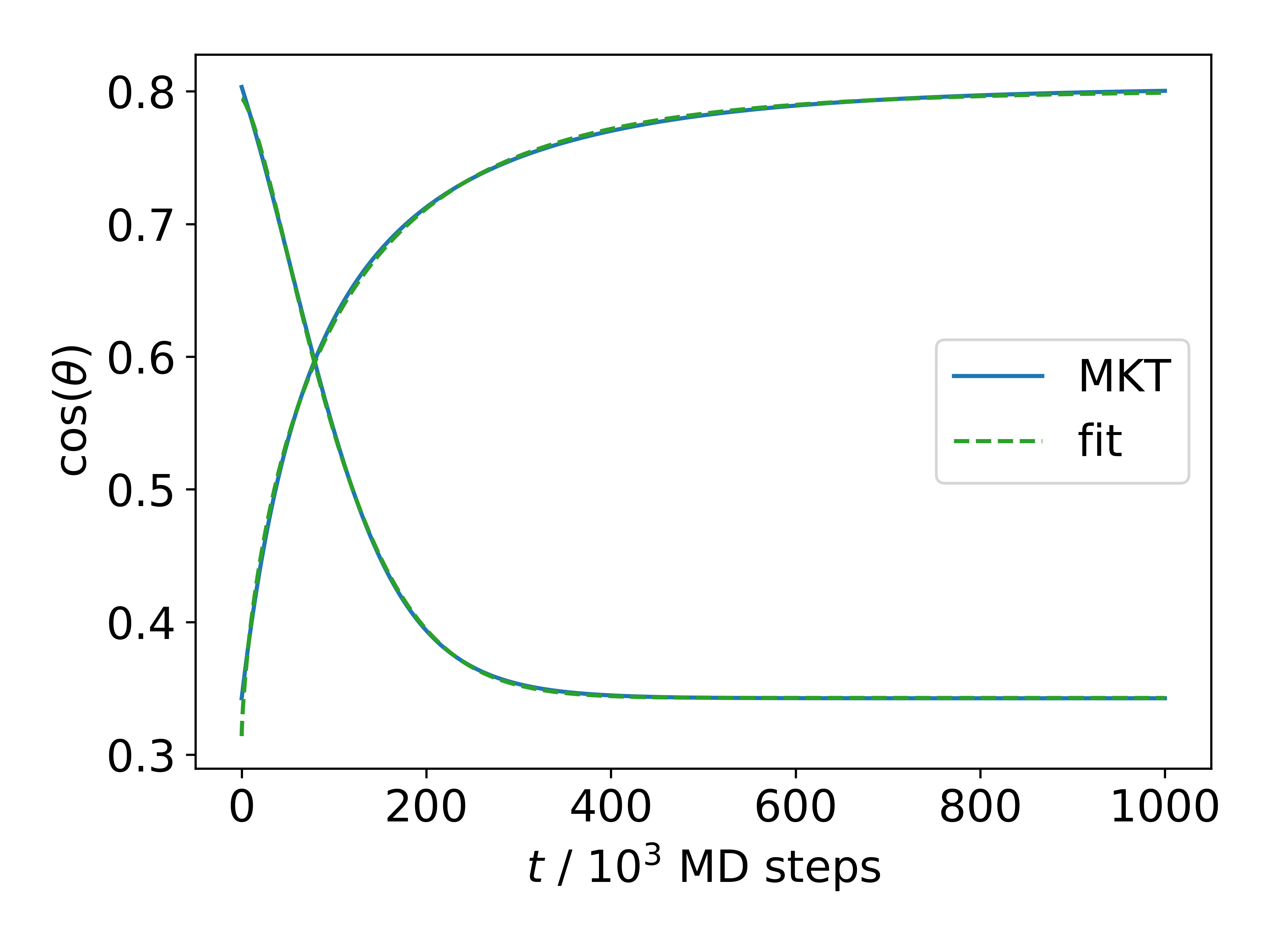} &
\includegraphics[width = 0.45\columnwidth]{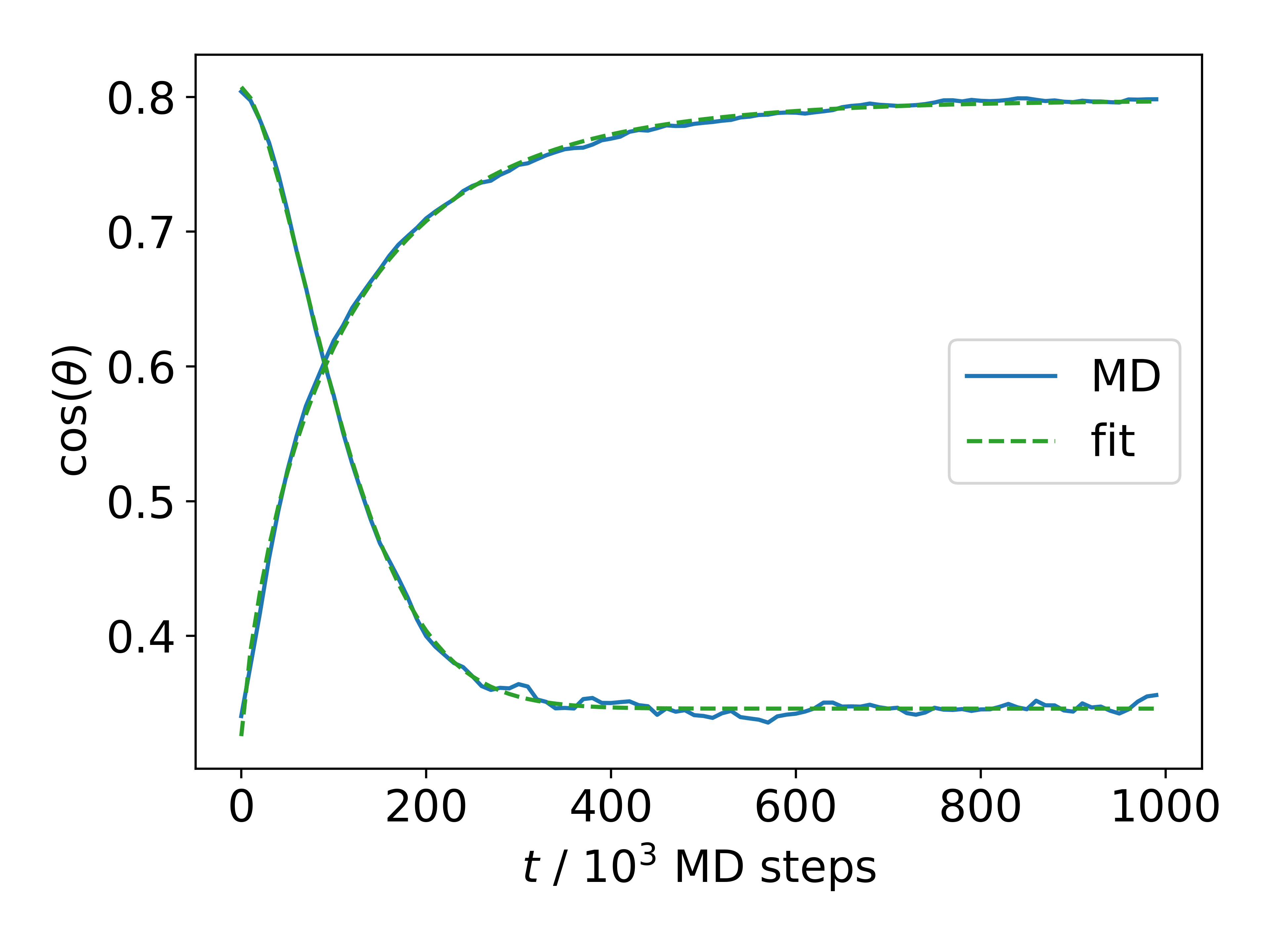}
\end{tabular}
\caption{Relaxation of $\cos \theta$ after an instantaneous change in wettability in the MD and MKT model for the interaction strengths $\epsilon_1\leftrightarrow\epsilon_2$: (a) $0.632 \leftrightarrow 0.671$ (MKT), (b) $0.632 \leftrightarrow 0.671$ (MD), (c) $0.632 \leftrightarrow 0.762$ (MKT) and (d) $0.632 \leftrightarrow 0.762$ (MD). }\label{fig:md_single_switch_fits}
\end{figure}

We have fitted a stretched exponential for the relaxation function after a single switching process which can be seen in Fig.~\ref{fig:tf_single_switch_fits} for TF and in Fig.~\ref{fig:md_single_switch_fits} for MD data. The resulting $\beta$-values for the MD simulations, for the TF analysis, and for the MKT equations (Eq.~\ref {eq:mkt_circle}) are listed in Tab.~\ref{tab:beta_values}.
\begin{table}[h]
\caption{\label{tab:beta_values} Values of $\beta$ obtained from an non-exponential fit to the data for a single switching event from $\epsilon_{1}$ to $\epsilon_{2}$ in MD, MKT and TF simulations.}
\begin{tabular*}{0.98\columnwidth}{@{\extracolsep{\fill}}ccccc}
\mbox{$\epsilon_{1}$} & \mbox{$\epsilon_{2}$} & \mbox{$\beta_{MD}$}& \mbox{$\beta_{MKT}$}& \mbox{$\beta_{TF}$} \\
      \hline
      0.632 & 0.762 & 0.87 & 0.73 & 0.68  \\ 
      0.762 & 0.632 & 1.68 & 1.42 & 2.05  \\ 
   \end{tabular*}
\end{table}

One consistently observes a stretched exponential behavior when increasing the wettability upon switching and a compressed exponential behavior in the opposite case.

For the interpretation of possible non-exponential effects in relaxation functions $y(t)$ we start by defining the
 $n$-th moment of $y(t)$ as
\begin{equation}
    \langle \tau^n\rangle = \frac{\int_0^\infty t^n \cdot y(t) \dd t}{\int_0^\infty y(t)}.
\end{equation}
Now, we can compute the quantity 
\begin{equation}
\frac{\langle \tau^2\rangle}{2\langle \tau \rangle ^2},    \label{eq:moment_relation_for_beta}
\end{equation}
which is equal to one for a purely exponential function $y(t)$. If this quantity is greater than 1, it implies a compressed exponential function, i.\,e. $\beta >1$ whereas in the opposite case it describes a stretched exponential function, i.\,e. $\beta<1$. 

For the $y(t)$, given in Eq.~\eqref{eq:mkt_approx_solved_approximated}, we get 
\begin{equation}
\frac{\langle \tau^2\rangle}{2\langle \tau \rangle ^2} = \frac{\qty(1-\frac{B}{8})\qty(1-\frac{B}{2})}{\qty(1-\frac{B}{4})^2} =
1 - \frac{B}{ 8 (1-\frac{B}{4})^2},\label{eq:moment_ratio_for_y}
\end{equation}
where $B=\frac{x_0-x_{eq}}{1-x_0}$. For switching from higher to lower wettability $x_0> x_{eq}$ holds, which implies $B>0$. Consequently, the expression in Eq.~\eqref{eq:moment_ratio_for_y} has to be smaller than 1 and finally $\beta$ has to be greater 1. Analogously $\beta<1$ follows for the inverse switching direction. This is in accordance with our results from the different models, as shown in Tab.~\ref{tab:beta_values}.


\section{Minimum $y_{plateau}$ value} \label{sec:app_mkt_st}
\begin{figure}[H]
\centering
\includegraphics[width= 0.7\columnwidth]{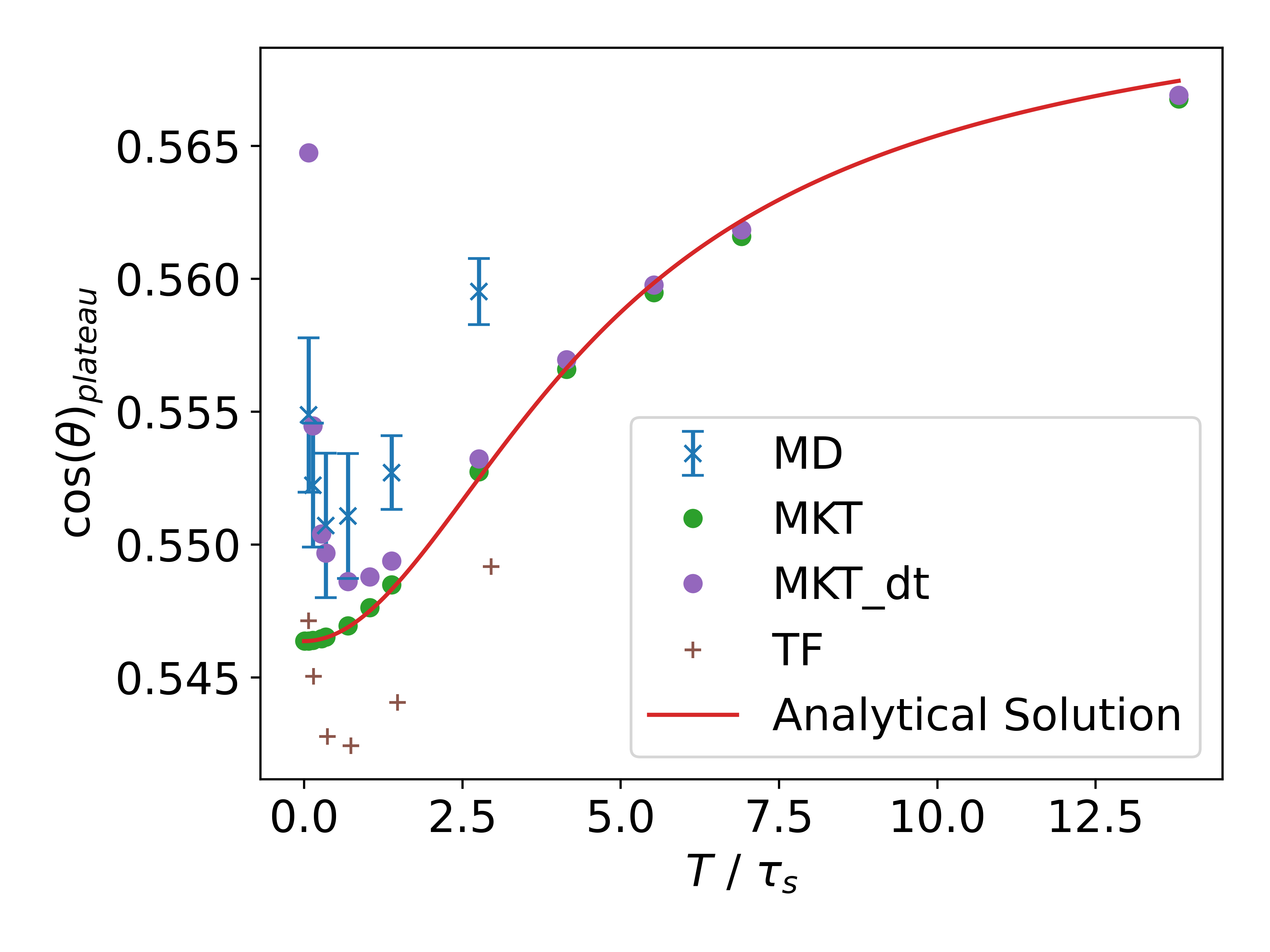}
\caption{(a) $\cos(\theta)_{\text{plateau}}$ obtained from MD, MKT, MKT with a dead time (MKT\_dt), TF and from Eq.~\eqref{eq:y_plateau} plotted against $T$ for switching between $\epsilon_{LW} = 0.632$ and $\epsilon_{HW} = 0.762$.} 
\label{fig:plateau_vs_T_app}
\end{figure}

For calculating the MKT values with short time effects in Fig.~\ref{fig:plateau_vs_T_app} we included a dead time where we set $\gamma/\zeta = 0$ after each switching event for $1500$ time steps when switching to a higher wettability and $\frac{3}{5} \cdot 1500$ time steps when switching to a lower wettability. The factor of $\frac{3}{5}$ results from the relation of the time steps that show the nonlinear behavior in Fig.~\ref{fig:mkt_fit_switch}. 


\begin{thebibliography}{45}%
\makeatletter
\providecommand \@ifxundefined [1]{%
 \@ifx{#1\undefined}
}%
\providecommand \@ifnum [1]{%
 \ifnum #1\expandafter \@firstoftwo
 \else \expandafter \@secondoftwo
 \fi
}%
\providecommand \@ifx [1]{%
 \ifx #1\expandafter \@firstoftwo
 \else \expandafter \@secondoftwo
 \fi
}%
\providecommand \natexlab [1]{#1}%
\providecommand \enquote  [1]{``#1''}%
\providecommand \bibnamefont  [1]{#1}%
\providecommand \bibfnamefont [1]{#1}%
\providecommand \citenamefont [1]{#1}%
\providecommand \href@noop [0]{\@secondoftwo}%
\providecommand \href [0]{\begingroup \@sanitize@url \@href}%
\providecommand \@href[1]{\@@startlink{#1}\@@href}%
\providecommand \@@href[1]{\endgroup#1\@@endlink}%
\providecommand \@sanitize@url [0]{\catcode `\\12\catcode `\$12\catcode
  `\&12\catcode `\#12\catcode `\^12\catcode `\_12\catcode `\%12\relax}%
\providecommand \@@startlink[1]{}%
\providecommand \@@endlink[0]{}%
\providecommand \url  [0]{\begingroup\@sanitize@url \@url }%
\providecommand \@url [1]{\endgroup\@href {#1}{\urlprefix }}%
\providecommand \urlprefix  [0]{URL }%
\providecommand \Eprint [0]{\href }%
\providecommand \doibase [0]{http://dx.doi.org/}%
\providecommand \selectlanguage [0]{\@gobble}%
\providecommand \bibinfo  [0]{\@secondoftwo}%
\providecommand \bibfield  [0]{\@secondoftwo}%
\providecommand \translation [1]{[#1]}%
\providecommand \BibitemOpen [0]{}%
\providecommand \bibitemStop [0]{}%
\providecommand \bibitemNoStop [0]{.\EOS\space}%
\providecommand \EOS [0]{\spacefactor3000\relax}%
\providecommand \BibitemShut  [1]{\csname bibitem#1\endcsname}%
\let\auto@bib@innerbib\@empty
\bibitem [{\citenamefont {Squires}\ and\ \citenamefont
  {Quake}(2005)}]{SqQu2005rev.mod.phys.}%
  \BibitemOpen
  \bibfield  {author} {\bibinfo {author} {\bibfnamefont {T.~M.}\ \bibnamefont
  {Squires}}\ and\ \bibinfo {author} {\bibfnamefont {S.~R.}\ \bibnamefont
  {Quake}},\ }\href {\doibase 10.1103/RevModPhys.77.977} {\bibfield  {journal}
  {\bibinfo  {journal} {Rev. Mod. Phys.}\ }\textbf {\bibinfo {volume} {77}},\
  \bibinfo {pages} {977} (\bibinfo {year} {2005})}\BibitemShut {NoStop}%
\bibitem [{\citenamefont {Ichimura}, \citenamefont {Oh},\ and\ \citenamefont
  {Nakagawa}(2000)}]{Ichimura1624}%
  \BibitemOpen
  \bibfield  {author} {\bibinfo {author} {\bibfnamefont {K.}~\bibnamefont
  {Ichimura}}, \bibinfo {author} {\bibfnamefont {S.-K.}\ \bibnamefont {Oh}}, \
  and\ \bibinfo {author} {\bibfnamefont {M.}~\bibnamefont {Nakagawa}},\ }\href
  {\doibase 10.1126/science.288.5471.1624} {\bibfield  {journal} {\bibinfo
  {journal} {Science}\ }\textbf {\bibinfo {volume} {288}},\ \bibinfo {pages}
  {1624} (\bibinfo {year} {2000})},\ \Eprint
  {http://arxiv.org/abs/https://science.sciencemag.org/content/288/5471/1624.full.pdf}
  {https://science.sciencemag.org/content/288/5471/1624.full.pdf} \BibitemShut
  {NoStop}%
\bibitem [{\citenamefont {Chaudhury}\ and\ \citenamefont
  {Whitesides}(1992)}]{ChWh1992s}%
  \BibitemOpen
  \bibfield  {author} {\bibinfo {author} {\bibfnamefont {M.~K.}\ \bibnamefont
  {Chaudhury}}\ and\ \bibinfo {author} {\bibfnamefont {G.~M.}\ \bibnamefont
  {Whitesides}},\ }\href {\doibase 10.1126/science.256.5063.1539} {\bibfield
  {journal} {\bibinfo  {journal} {Science}\ }\textbf {\bibinfo {volume}
  {256}},\ \bibinfo {pages} {1539} (\bibinfo {year} {1992})}\BibitemShut
  {NoStop}%
\bibitem [{\citenamefont {Karpitschka}\ \emph {et~al.}(2015)\citenamefont
  {Karpitschka}, \citenamefont {Das}, \citenamefont {van Gorcum}, \citenamefont
  {Perrin}, \citenamefont {Andreotti},\ and\ \citenamefont
  {Snoeijer}}]{karpitschka2015droplets}%
  \BibitemOpen
  \bibfield  {author} {\bibinfo {author} {\bibfnamefont {S.}~\bibnamefont
  {Karpitschka}}, \bibinfo {author} {\bibfnamefont {S.}~\bibnamefont {Das}},
  \bibinfo {author} {\bibfnamefont {M.}~\bibnamefont {van Gorcum}}, \bibinfo
  {author} {\bibfnamefont {H.}~\bibnamefont {Perrin}}, \bibinfo {author}
  {\bibfnamefont {B.}~\bibnamefont {Andreotti}}, \ and\ \bibinfo {author}
  {\bibfnamefont {J.~H.}\ \bibnamefont {Snoeijer}},\ }\href@noop {} {\bibfield
  {journal} {\bibinfo  {journal} {Nature communications}\ }\textbf {\bibinfo
  {volume} {6}},\ \bibinfo {pages} {1} (\bibinfo {year} {2015})}\BibitemShut
  {NoStop}%
\bibitem [{\citenamefont {Karpitschka}\ \emph {et~al.}(2016)\citenamefont
  {Karpitschka}, \citenamefont {Pandey}, \citenamefont {Lubbers}, \citenamefont
  {Weijs}, \citenamefont {Botto}, \citenamefont {Das}, \citenamefont
  {Andreotti},\ and\ \citenamefont {Snoeijer}}]{karpitschka2016liquid}%
  \BibitemOpen
  \bibfield  {author} {\bibinfo {author} {\bibfnamefont {S.}~\bibnamefont
  {Karpitschka}}, \bibinfo {author} {\bibfnamefont {A.}~\bibnamefont {Pandey}},
  \bibinfo {author} {\bibfnamefont {L.~A.}\ \bibnamefont {Lubbers}}, \bibinfo
  {author} {\bibfnamefont {J.~H.}\ \bibnamefont {Weijs}}, \bibinfo {author}
  {\bibfnamefont {L.}~\bibnamefont {Botto}}, \bibinfo {author} {\bibfnamefont
  {S.}~\bibnamefont {Das}}, \bibinfo {author} {\bibfnamefont {B.}~\bibnamefont
  {Andreotti}}, \ and\ \bibinfo {author} {\bibfnamefont {J.~H.}\ \bibnamefont
  {Snoeijer}},\ }\href@noop {} {\bibfield  {journal} {\bibinfo  {journal}
  {Proceedings of the National Academy of Sciences}\ }\textbf {\bibinfo
  {volume} {113}},\ \bibinfo {pages} {7403} (\bibinfo {year}
  {2016})}\BibitemShut {NoStop}%
\bibitem [{\citenamefont {Ishihara}\ \emph {et~al.}(1982)\citenamefont
  {Ishihara}, \citenamefont {Okazaki}, \citenamefont {Negishi}, \citenamefont
  {Shinohara}, \citenamefont {Okano}, \citenamefont {Kataoka},\ and\
  \citenamefont {Sakurai}}]{Ishihara1982}%
  \BibitemOpen
  \bibfield  {author} {\bibinfo {author} {\bibfnamefont {K.}~\bibnamefont
  {Ishihara}}, \bibinfo {author} {\bibfnamefont {A.}~\bibnamefont {Okazaki}},
  \bibinfo {author} {\bibfnamefont {N.}~\bibnamefont {Negishi}}, \bibinfo
  {author} {\bibfnamefont {I.}~\bibnamefont {Shinohara}}, \bibinfo {author}
  {\bibfnamefont {T.}~\bibnamefont {Okano}}, \bibinfo {author} {\bibfnamefont
  {K.}~\bibnamefont {Kataoka}}, \ and\ \bibinfo {author} {\bibfnamefont
  {Y.}~\bibnamefont {Sakurai}},\ }\href {\doibase
  https://doi.org/10.1002/app.1982.070270125} {\bibfield  {journal} {\bibinfo
  {journal} {Journal of Applied Polymer Science}\ }\textbf {\bibinfo {volume}
  {27}},\ \bibinfo {pages} {239} (\bibinfo {year} {1982})},\ \Eprint
  {http://arxiv.org/abs/https://onlinelibrary.wiley.com/doi/pdf/10.1002/app.1982.070270125}
  {https://onlinelibrary.wiley.com/doi/pdf/10.1002/app.1982.070270125}
  \BibitemShut {NoStop}%
\bibitem [{\citenamefont {Rosario}\ \emph {et~al.}(2002)\citenamefont
  {Rosario}, \citenamefont {Gust}, \citenamefont {Hayes}, \citenamefont
  {Jahnke}, \citenamefont {Springer},\ and\ \citenamefont
  {Garcia}}]{Rosario2002}%
  \BibitemOpen
  \bibfield  {author} {\bibinfo {author} {\bibfnamefont {R.}~\bibnamefont
  {Rosario}}, \bibinfo {author} {\bibfnamefont {D.}~\bibnamefont {Gust}},
  \bibinfo {author} {\bibfnamefont {M.}~\bibnamefont {Hayes}}, \bibinfo
  {author} {\bibfnamefont {F.}~\bibnamefont {Jahnke}}, \bibinfo {author}
  {\bibfnamefont {J.}~\bibnamefont {Springer}}, \ and\ \bibinfo {author}
  {\bibfnamefont {A.~A.}\ \bibnamefont {Garcia}},\ }\href {\doibase
  10.1021/la025963l} {\bibfield  {journal} {\bibinfo  {journal} {Langmuir}\
  }\textbf {\bibinfo {volume} {18}},\ \bibinfo {pages} {8062} (\bibinfo {year}
  {2002})},\ \Eprint {http://arxiv.org/abs/https://doi.org/10.1021/la025963l}
  {https://doi.org/10.1021/la025963l} \BibitemShut {NoStop}%
\bibitem [{\citenamefont {Brunet}, \citenamefont {Eggers},\ and\ \citenamefont
  {Deegan}(2007)}]{PhysRevLett.99.144501}%
  \BibitemOpen
  \bibfield  {author} {\bibinfo {author} {\bibfnamefont {P.}~\bibnamefont
  {Brunet}}, \bibinfo {author} {\bibfnamefont {J.}~\bibnamefont {Eggers}}, \
  and\ \bibinfo {author} {\bibfnamefont {R.~D.}\ \bibnamefont {Deegan}},\
  }\href {\doibase 10.1103/PhysRevLett.99.144501} {\bibfield  {journal}
  {\bibinfo  {journal} {Phys. Rev. Lett.}\ }\textbf {\bibinfo {volume} {99}},\
  \bibinfo {pages} {144501} (\bibinfo {year} {2007})}\BibitemShut {NoStop}%
\bibitem [{\citenamefont {Sartori}\ \emph {et~al.}(2015)\citenamefont
  {Sartori}, \citenamefont {Quagliati}, \citenamefont {Varagnolo},
  \citenamefont {Pierno}, \citenamefont {Mistura}, \citenamefont {Magaletti},\
  and\ \citenamefont {Casciola}}]{Sartori_2015}%
  \BibitemOpen
  \bibfield  {author} {\bibinfo {author} {\bibfnamefont {P.}~\bibnamefont
  {Sartori}}, \bibinfo {author} {\bibfnamefont {D.}~\bibnamefont {Quagliati}},
  \bibinfo {author} {\bibfnamefont {S.}~\bibnamefont {Varagnolo}}, \bibinfo
  {author} {\bibfnamefont {M.}~\bibnamefont {Pierno}}, \bibinfo {author}
  {\bibfnamefont {G.}~\bibnamefont {Mistura}}, \bibinfo {author} {\bibfnamefont
  {F.}~\bibnamefont {Magaletti}}, \ and\ \bibinfo {author} {\bibfnamefont
  {C.~M.}\ \bibnamefont {Casciola}},\ }\href {\doibase
  10.1088/1367-2630/17/11/113017} {\bibfield  {journal} {\bibinfo  {journal}
  {New Journal of Physics}\ }\textbf {\bibinfo {volume} {17}},\ \bibinfo
  {pages} {113017} (\bibinfo {year} {2015})}\BibitemShut {NoStop}%
\bibitem [{\citenamefont {Mugele}, \citenamefont {Baret},\ and\ \citenamefont
  {Steinhauser}(2006)}]{doi:10.1063/1.2204831}%
  \BibitemOpen
  \bibfield  {author} {\bibinfo {author} {\bibfnamefont {F.}~\bibnamefont
  {Mugele}}, \bibinfo {author} {\bibfnamefont {J.-C.}\ \bibnamefont {Baret}}, \
  and\ \bibinfo {author} {\bibfnamefont {D.}~\bibnamefont {Steinhauser}},\
  }\href {\doibase 10.1063/1.2204831} {\bibfield  {journal} {\bibinfo
  {journal} {Applied Physics Letters}\ }\textbf {\bibinfo {volume} {88}},\
  \bibinfo {pages} {204106} (\bibinfo {year} {2006})},\ \Eprint
  {http://arxiv.org/abs/https://doi.org/10.1063/1.2204831}
  {https://doi.org/10.1063/1.2204831} \BibitemShut {NoStop}%
\bibitem [{\citenamefont {Meier}\ \emph {et~al.}(2000)\citenamefont {Meier},
  \citenamefont {Greune}, \citenamefont {Meyboom},\ and\ \citenamefont
  {Hofmann}}]{Meier2000}%
  \BibitemOpen
  \bibfield  {author} {\bibinfo {author} {\bibfnamefont {W.}~\bibnamefont
  {Meier}}, \bibinfo {author} {\bibfnamefont {G.}~\bibnamefont {Greune}},
  \bibinfo {author} {\bibfnamefont {A.}~\bibnamefont {Meyboom}}, \ and\
  \bibinfo {author} {\bibfnamefont {K.~P.}\ \bibnamefont {Hofmann}},\ }\href
  {\doibase 10.1007/s002490050256} {\bibfield  {journal} {\bibinfo  {journal}
  {European Biophysics Journal}\ }\textbf {\bibinfo {volume} {29}},\ \bibinfo
  {pages} {113} (\bibinfo {year} {2000})}\BibitemShut {NoStop}%
\bibitem [{\citenamefont {Zografov}, \citenamefont {Tankovsky},\ and\
  \citenamefont {Andreeva}(2014)}]{ZOGRAFOV2014351}%
  \BibitemOpen
  \bibfield  {author} {\bibinfo {author} {\bibfnamefont {N.}~\bibnamefont
  {Zografov}}, \bibinfo {author} {\bibfnamefont {N.}~\bibnamefont {Tankovsky}},
  \ and\ \bibinfo {author} {\bibfnamefont {A.}~\bibnamefont {Andreeva}},\
  }\href {\doibase https://doi.org/10.1016/j.colsurfa.2013.12.013} {\bibfield
  {journal} {\bibinfo  {journal} {Colloids and Surfaces A: Physicochemical and
  Engineering Aspects}\ }\textbf {\bibinfo {volume} {460}},\ \bibinfo {pages}
  {351} (\bibinfo {year} {2014})},\ \bibinfo {note} {27th European Colloid and
  Interface Society conference (27th ECIS 2013)}\BibitemShut {NoStop}%
\bibitem [{\citenamefont {Feng}, \citenamefont {Zhai},\ and\ \citenamefont
  {Jiang}(2005)}]{tio22005}%
  \BibitemOpen
  \bibfield  {author} {\bibinfo {author} {\bibfnamefont {X.}~\bibnamefont
  {Feng}}, \bibinfo {author} {\bibfnamefont {J.}~\bibnamefont {Zhai}}, \ and\
  \bibinfo {author} {\bibfnamefont {L.}~\bibnamefont {Jiang}},\ }\href
  {\doibase https://doi.org/10.1002/anie.200501337} {\bibfield  {journal}
  {\bibinfo  {journal} {Angewandte Chemie International Edition}\ }\textbf
  {\bibinfo {volume} {44}},\ \bibinfo {pages} {5115} (\bibinfo {year}
  {2005})},\ \Eprint
  {http://arxiv.org/abs/https://onlinelibrary.wiley.com/doi/pdf/10.1002/anie.200501337}
  {https://onlinelibrary.wiley.com/doi/pdf/10.1002/anie.200501337} \BibitemShut
  {NoStop}%
\bibitem [{\citenamefont {Groten}, \citenamefont {Bunte},\ and\ \citenamefont
  {Rühe}(2012)}]{ruehe2012}%
  \BibitemOpen
  \bibfield  {author} {\bibinfo {author} {\bibfnamefont {J.}~\bibnamefont
  {Groten}}, \bibinfo {author} {\bibfnamefont {C.}~\bibnamefont {Bunte}}, \
  and\ \bibinfo {author} {\bibfnamefont {J.}~\bibnamefont {Rühe}},\ }\href
  {\doibase 10.1021/la302764k} {\bibfield  {journal} {\bibinfo  {journal}
  {Langmuir}\ }\textbf {\bibinfo {volume} {28}},\ \bibinfo {pages} {15038}
  (\bibinfo {year} {2012})},\ \bibinfo {note} {pMID: 22967018},\ \Eprint
  {http://arxiv.org/abs/https://doi.org/10.1021/la302764k}
  {https://doi.org/10.1021/la302764k} \BibitemShut {NoStop}%
\bibitem [{\citenamefont {Wang}\ \emph {et~al.}(1997)\citenamefont {Wang},
  \citenamefont {Hashimoto}, \citenamefont {Fujishima}, \citenamefont
  {Chikuni}, \citenamefont {Kojima}, \citenamefont {Kitamura}, \citenamefont
  {Shimohigoshi},\ and\ \citenamefont {Watanabe}}]{Wang1997}%
  \BibitemOpen
  \bibfield  {author} {\bibinfo {author} {\bibfnamefont {R.}~\bibnamefont
  {Wang}}, \bibinfo {author} {\bibfnamefont {K.}~\bibnamefont {Hashimoto}},
  \bibinfo {author} {\bibfnamefont {A.}~\bibnamefont {Fujishima}}, \bibinfo
  {author} {\bibfnamefont {M.}~\bibnamefont {Chikuni}}, \bibinfo {author}
  {\bibfnamefont {E.}~\bibnamefont {Kojima}}, \bibinfo {author} {\bibfnamefont
  {A.}~\bibnamefont {Kitamura}}, \bibinfo {author} {\bibfnamefont
  {M.}~\bibnamefont {Shimohigoshi}}, \ and\ \bibinfo {author} {\bibfnamefont
  {T.}~\bibnamefont {Watanabe}},\ }\href {\doibase 10.1038/41233} {\bibfield
  {journal} {\bibinfo  {journal} {Nature}\ }\textbf {\bibinfo {volume} {388}},\
  \bibinfo {pages} {431} (\bibinfo {year} {1997})}\BibitemShut {NoStop}%
\bibitem [{\citenamefont {Sun}\ \emph {et~al.}(2001)\citenamefont {Sun},
  \citenamefont {Nakajima}, \citenamefont {Fujishima}, \citenamefont
  {Watanabe},\ and\ \citenamefont {Hashimoto}}]{Sun2001}%
  \BibitemOpen
  \bibfield  {author} {\bibinfo {author} {\bibfnamefont {R.-D.}\ \bibnamefont
  {Sun}}, \bibinfo {author} {\bibfnamefont {A.}~\bibnamefont {Nakajima}},
  \bibinfo {author} {\bibfnamefont {A.}~\bibnamefont {Fujishima}}, \bibinfo
  {author} {\bibfnamefont {T.}~\bibnamefont {Watanabe}}, \ and\ \bibinfo
  {author} {\bibfnamefont {K.}~\bibnamefont {Hashimoto}},\ }\href {\doibase
  10.1021/jp002525j} {\bibfield  {journal} {\bibinfo  {journal} {The Journal of
  Physical Chemistry B}\ }\textbf {\bibinfo {volume} {105}},\ \bibinfo {pages}
  {1984} (\bibinfo {year} {2001})}\BibitemShut {NoStop}%
\bibitem [{\citenamefont {Chan}\ \emph {et~al.}(2017)\citenamefont {Chan},
  \citenamefont {McGraw}, \citenamefont {Salez}, \citenamefont {Seemann},\ and\
  \citenamefont {Brinkmann}}]{chan_mcgraw_salez_seemann_brinkmann_2017}%
  \BibitemOpen
  \bibfield  {author} {\bibinfo {author} {\bibfnamefont {T.~S.}\ \bibnamefont
  {Chan}}, \bibinfo {author} {\bibfnamefont {J.~D.}\ \bibnamefont {McGraw}},
  \bibinfo {author} {\bibfnamefont {T.}~\bibnamefont {Salez}}, \bibinfo
  {author} {\bibfnamefont {R.}~\bibnamefont {Seemann}}, \ and\ \bibinfo
  {author} {\bibfnamefont {M.}~\bibnamefont {Brinkmann}},\ }\href {\doibase
  10.1017/jfm.2017.515} {\bibfield  {journal} {\bibinfo  {journal} {Journal of
  Fluid Mechanics}\ }\textbf {\bibinfo {volume} {828}},\ \bibinfo {pages}
  {271–288} (\bibinfo {year} {2017})}\BibitemShut {NoStop}%
\bibitem [{\citenamefont {Grawitter}\ and\ \citenamefont
  {Stark}(2021{\natexlab{a}})}]{GrSt2021softmatter}%
  \BibitemOpen
  \bibfield  {author} {\bibinfo {author} {\bibfnamefont {J.}~\bibnamefont
  {Grawitter}}\ and\ \bibinfo {author} {\bibfnamefont {H.}~\bibnamefont
  {Stark}},\ }\href {\doibase 10.1039/D0SM02082F} {\bibfield  {journal}
  {\bibinfo  {journal} {Soft Matter}\ }\textbf {\bibinfo {volume} {17}},\
  \bibinfo {pages} {2454} (\bibinfo {year} {2021}{\natexlab{a}})}\BibitemShut
  {NoStop}%
\bibitem [{\citenamefont {Grawitter}\ and\ \citenamefont
  {Stark}(2021{\natexlab{b}})}]{D1SM01113H}%
  \BibitemOpen
  \bibfield  {author} {\bibinfo {author} {\bibfnamefont {J.}~\bibnamefont
  {Grawitter}}\ and\ \bibinfo {author} {\bibfnamefont {H.}~\bibnamefont
  {Stark}},\ }\href {\doibase 10.1039/D1SM01113H} {\bibfield  {journal}
  {\bibinfo  {journal} {Soft Matter}\ }\textbf {\bibinfo {volume} {17}},\
  \bibinfo {pages} {9469} (\bibinfo {year} {2021}{\natexlab{b}})}\BibitemShut
  {NoStop}%
\bibitem [{\citenamefont {Honisch}\ \emph {et~al.}(2015)\citenamefont
  {Honisch}, \citenamefont {Lin}, \citenamefont {Heuer}, \citenamefont
  {Thiele},\ and\ \citenamefont {Gurevich}}]{HLHT2015w}%
  \BibitemOpen
  \bibfield  {author} {\bibinfo {author} {\bibfnamefont {C.}~\bibnamefont
  {Honisch}}, \bibinfo {author} {\bibfnamefont {T.-S.}\ \bibnamefont {Lin}},
  \bibinfo {author} {\bibfnamefont {A.}~\bibnamefont {Heuer}}, \bibinfo
  {author} {\bibfnamefont {U.}~\bibnamefont {Thiele}}, \ and\ \bibinfo {author}
  {\bibfnamefont {S.~V.}\ \bibnamefont {Gurevich}},\ }\href {\doibase
  10.1021/acs.langmuir.5b02407} {\bibfield  {journal} {\bibinfo  {journal}
  {Langmuir}\ }\textbf {\bibinfo {volume} {31}},\ \bibinfo {pages} {10618}
  (\bibinfo {year} {2015})}\BibitemShut {NoStop}%
\bibitem [{\citenamefont {Gauglitz}\ and\ \citenamefont
  {Radke}(1988)}]{GAUGLITZ19881457}%
  \BibitemOpen
  \bibfield  {author} {\bibinfo {author} {\bibfnamefont {P.}~\bibnamefont
  {Gauglitz}}\ and\ \bibinfo {author} {\bibfnamefont {C.}~\bibnamefont
  {Radke}},\ }\href {\doibase https://doi.org/10.1016/0009-2509(88)85137-6}
  {\bibfield  {journal} {\bibinfo  {journal} {Chemical Engineering Science}\
  }\textbf {\bibinfo {volume} {43}},\ \bibinfo {pages} {1457} (\bibinfo {year}
  {1988})}\BibitemShut {NoStop}%
\bibitem [{\citenamefont {Sega}\ \emph {et~al.}(2013)\citenamefont {Sega},
  \citenamefont {Sbragaglia}, \citenamefont {Biferale},\ and\ \citenamefont
  {Succi}}]{C3SM51508G}%
  \BibitemOpen
  \bibfield  {author} {\bibinfo {author} {\bibfnamefont {M.}~\bibnamefont
  {Sega}}, \bibinfo {author} {\bibfnamefont {M.}~\bibnamefont {Sbragaglia}},
  \bibinfo {author} {\bibfnamefont {L.}~\bibnamefont {Biferale}}, \ and\
  \bibinfo {author} {\bibfnamefont {S.}~\bibnamefont {Succi}},\ }\href
  {\doibase 10.1039/C3SM51508G} {\bibfield  {journal} {\bibinfo  {journal}
  {Soft Matter}\ }\textbf {\bibinfo {volume} {9}},\ \bibinfo {pages} {8526}
  (\bibinfo {year} {2013})}\BibitemShut {NoStop}%
\bibitem [{\citenamefont {Barrat}\ and\ \citenamefont
  {Bocquet}(1999)}]{PhysRevLett.82.4671}%
  \BibitemOpen
  \bibfield  {author} {\bibinfo {author} {\bibfnamefont {J.-L.}\ \bibnamefont
  {Barrat}}\ and\ \bibinfo {author} {\bibfnamefont {L.}~\bibnamefont
  {Bocquet}},\ }\href {\doibase 10.1103/PhysRevLett.82.4671} {\bibfield
  {journal} {\bibinfo  {journal} {Phys. Rev. Lett.}\ }\textbf {\bibinfo
  {volume} {82}},\ \bibinfo {pages} {4671} (\bibinfo {year}
  {1999})}\BibitemShut {NoStop}%
\bibitem [{\citenamefont {Bertrand}\ \emph {et~al.}(2007)\citenamefont
  {Bertrand}, \citenamefont {Blake}, \citenamefont {Ledauphin}, \citenamefont
  {Ogonowski}, \citenamefont {De~Coninck}, \citenamefont {Fornasiero},\ and\
  \citenamefont {Ralston}}]{doi:10.1021/la062920m}%
  \BibitemOpen
  \bibfield  {author} {\bibinfo {author} {\bibfnamefont {E.}~\bibnamefont
  {Bertrand}}, \bibinfo {author} {\bibfnamefont {T.~D.}\ \bibnamefont {Blake}},
  \bibinfo {author} {\bibfnamefont {V.}~\bibnamefont {Ledauphin}}, \bibinfo
  {author} {\bibfnamefont {G.}~\bibnamefont {Ogonowski}}, \bibinfo {author}
  {\bibfnamefont {J.}~\bibnamefont {De~Coninck}}, \bibinfo {author}
  {\bibfnamefont {D.}~\bibnamefont {Fornasiero}}, \ and\ \bibinfo {author}
  {\bibfnamefont {J.}~\bibnamefont {Ralston}},\ }\href {\doibase
  10.1021/la062920m} {\bibfield  {journal} {\bibinfo  {journal} {Langmuir}\
  }\textbf {\bibinfo {volume} {23}},\ \bibinfo {pages} {3774} (\bibinfo {year}
  {2007})},\ \bibinfo {note} {pMID: 17328565},\ \Eprint
  {http://arxiv.org/abs/https://doi.org/10.1021/la062920m}
  {https://doi.org/10.1021/la062920m} \BibitemShut {NoStop}%
\bibitem [{\citenamefont {Blake}\ and\ \citenamefont
  {Haynes}(1969)}]{BLAKE_mkt}%
  \BibitemOpen
  \bibfield  {author} {\bibinfo {author} {\bibfnamefont {T.}~\bibnamefont
  {Blake}}\ and\ \bibinfo {author} {\bibfnamefont {J.}~\bibnamefont {Haynes}},\
  }\href {\doibase https://doi.org/10.1016/0021-9797(69)90411-1} {\bibfield
  {journal} {\bibinfo  {journal} {Journal of Colloid and Interface Science}\
  }\textbf {\bibinfo {volume} {30}},\ \bibinfo {pages} {421} (\bibinfo {year}
  {1969})}\BibitemShut {NoStop}%
\bibitem [{\citenamefont {de~Ruijter}, \citenamefont {Blake},\ and\
  \citenamefont {De~Coninck}(1999{\natexlab{a}})}]{ruijter1999}%
  \BibitemOpen
  \bibfield  {author} {\bibinfo {author} {\bibfnamefont {M.~J.}\ \bibnamefont
  {de~Ruijter}}, \bibinfo {author} {\bibfnamefont {T.~D.}\ \bibnamefont
  {Blake}}, \ and\ \bibinfo {author} {\bibfnamefont {J.}~\bibnamefont
  {De~Coninck}},\ }\href {\doibase 10.1021/la990171l} {\bibfield  {journal}
  {\bibinfo  {journal} {Langmuir}\ }\textbf {\bibinfo {volume} {15}},\ \bibinfo
  {pages} {7836} (\bibinfo {year} {1999}{\natexlab{a}})},\ \Eprint
  {http://arxiv.org/abs/https://doi.org/10.1021/la990171l}
  {https://doi.org/10.1021/la990171l} \BibitemShut {NoStop}%
\bibitem [{\citenamefont {Seveno}\ \emph {et~al.}(2009)\citenamefont {Seveno},
  \citenamefont {Vaillant}, \citenamefont {Rioboo}, \citenamefont {Adão},
  \citenamefont {Conti},\ and\ \citenamefont {De~Coninck}}]{seveno2009}%
  \BibitemOpen
  \bibfield  {author} {\bibinfo {author} {\bibfnamefont {D.}~\bibnamefont
  {Seveno}}, \bibinfo {author} {\bibfnamefont {A.}~\bibnamefont {Vaillant}},
  \bibinfo {author} {\bibfnamefont {R.}~\bibnamefont {Rioboo}}, \bibinfo
  {author} {\bibfnamefont {H.}~\bibnamefont {Adão}}, \bibinfo {author}
  {\bibfnamefont {J.}~\bibnamefont {Conti}}, \ and\ \bibinfo {author}
  {\bibfnamefont {J.}~\bibnamefont {De~Coninck}},\ }\href {\doibase
  10.1021/la901125a} {\bibfield  {journal} {\bibinfo  {journal} {Langmuir}\
  }\textbf {\bibinfo {volume} {25}},\ \bibinfo {pages} {13034} (\bibinfo {year}
  {2009})},\ \bibinfo {note} {pMID: 19845346},\ \Eprint
  {http://arxiv.org/abs/https://doi.org/10.1021/la901125a}
  {https://doi.org/10.1021/la901125a} \BibitemShut {NoStop}%
\bibitem [{\citenamefont {Duvivier}, \citenamefont {Blake},\ and\ \citenamefont
  {De~Coninck}(2013)}]{duvivier2013}%
  \BibitemOpen
  \bibfield  {author} {\bibinfo {author} {\bibfnamefont {D.}~\bibnamefont
  {Duvivier}}, \bibinfo {author} {\bibfnamefont {T.~D.}\ \bibnamefont {Blake}},
  \ and\ \bibinfo {author} {\bibfnamefont {J.}~\bibnamefont {De~Coninck}},\
  }\href {\doibase 10.1021/la4017917} {\bibfield  {journal} {\bibinfo
  {journal} {Langmuir}\ }\textbf {\bibinfo {volume} {29}},\ \bibinfo {pages}
  {10132} (\bibinfo {year} {2013})},\ \bibinfo {note} {pMID: 23844877},\
  \Eprint {http://arxiv.org/abs/https://doi.org/10.1021/la4017917}
  {https://doi.org/10.1021/la4017917} \BibitemShut {NoStop}%
\bibitem [{\citenamefont {Grawitter}\ and\ \citenamefont
  {Stark}(2018)}]{grawitter2018feedback}%
  \BibitemOpen
  \bibfield  {author} {\bibinfo {author} {\bibfnamefont {J.}~\bibnamefont
  {Grawitter}}\ and\ \bibinfo {author} {\bibfnamefont {H.}~\bibnamefont
  {Stark}},\ }\href@noop {} {\bibfield  {journal} {\bibinfo  {journal} {Soft
  Matter}\ }\textbf {\bibinfo {volume} {14}},\ \bibinfo {pages} {1856}
  (\bibinfo {year} {2018})}\BibitemShut {NoStop}%
\bibitem [{\citenamefont {Stieneker}\ \emph {et~al.}(2021)\citenamefont
  {Stieneker}, \citenamefont {Topp}, \citenamefont {Gurevich},\ and\
  \citenamefont {Heuer}}]{StienekerToppEtAl}%
  \BibitemOpen
  \bibfield  {author} {\bibinfo {author} {\bibfnamefont {M.}~\bibnamefont
  {Stieneker}}, \bibinfo {author} {\bibfnamefont {L.}~\bibnamefont {Topp}},
  \bibinfo {author} {\bibfnamefont {S.}~\bibnamefont {Gurevich}}, \ and\
  \bibinfo {author} {\bibfnamefont {A.}~\bibnamefont {Heuer}},\ }\href@noop {}
  {\enquote {\bibinfo {title} {Multiscale perspective on wetting on switchable
  substrates: mapping between microscopic and mesoscopic models},}\ } (\bibinfo
  {year} {2021}),\ \Eprint {http://arxiv.org/abs/2108.00641} {arXiv:2108.00641
  [physics.flu-dyn]} \BibitemShut {NoStop}%
\bibitem [{\citenamefont {Anderson}, \citenamefont {Glaser},\ and\
  \citenamefont {Glotzer}(2020)}]{anderson_hoomd}%
  \BibitemOpen
  \bibfield  {author} {\bibinfo {author} {\bibfnamefont {J.~A.}\ \bibnamefont
  {Anderson}}, \bibinfo {author} {\bibfnamefont {J.}~\bibnamefont {Glaser}}, \
  and\ \bibinfo {author} {\bibfnamefont {S.~C.}\ \bibnamefont {Glotzer}},\
  }\href {\doibase https://doi.org/10.1016/j.commatsci.2019.109363} {\bibfield
  {journal} {\bibinfo  {journal} {Computational Materials Science}\ }\textbf
  {\bibinfo {volume} {173}},\ \bibinfo {pages} {109363} (\bibinfo {year}
  {2020})}\BibitemShut {NoStop}%
\bibitem [{\citenamefont {Phillips}, \citenamefont {Anderson},\ and\
  \citenamefont {Glotzer}(2011)}]{phillips_hoomd_dpd}%
  \BibitemOpen
  \bibfield  {author} {\bibinfo {author} {\bibfnamefont {C.~L.}\ \bibnamefont
  {Phillips}}, \bibinfo {author} {\bibfnamefont {J.~A.}\ \bibnamefont
  {Anderson}}, \ and\ \bibinfo {author} {\bibfnamefont {S.~C.}\ \bibnamefont
  {Glotzer}},\ }\href {\doibase https://doi.org/10.1016/j.jcp.2011.05.021}
  {\bibfield  {journal} {\bibinfo  {journal} {Journal of Computational
  Physics}\ }\textbf {\bibinfo {volume} {230}},\ \bibinfo {pages} {7191 }
  (\bibinfo {year} {2011})}\BibitemShut {NoStop}%
\bibitem [{\citenamefont {Hoogerbrugge}\ and\ \citenamefont
  {Koelman}(1992)}]{Hoogerbrugge_1992}%
  \BibitemOpen
  \bibfield  {author} {\bibinfo {author} {\bibfnamefont {P.~J.}\ \bibnamefont
  {Hoogerbrugge}}\ and\ \bibinfo {author} {\bibfnamefont {J.~M. V.~A.}\
  \bibnamefont {Koelman}},\ }\href {\doibase 10.1209/0295-5075/19/3/001}
  {\bibfield  {journal} {\bibinfo  {journal} {Europhysics Letters ({EPL})}\
  }\textbf {\bibinfo {volume} {19}},\ \bibinfo {pages} {155} (\bibinfo {year}
  {1992})}\BibitemShut {NoStop}%
\bibitem [{\citenamefont {Oron}, \citenamefont {Davis},\ and\ \citenamefont
  {Bankoff}(1997)}]{OrDB1997rmp}%
  \BibitemOpen
  \bibfield  {author} {\bibinfo {author} {\bibfnamefont {A.}~\bibnamefont
  {Oron}}, \bibinfo {author} {\bibfnamefont {S.~H.}\ \bibnamefont {Davis}}, \
  and\ \bibinfo {author} {\bibfnamefont {S.~G.}\ \bibnamefont {Bankoff}},\
  }\href {\doibase 10.1103/RevModPhys.69.931} {\bibfield  {journal} {\bibinfo
  {journal} {Reviews of Modern Physics}\ }\textbf {\bibinfo {volume} {69}},\
  \bibinfo {pages} {931} (\bibinfo {year} {1997})}\BibitemShut {NoStop}%
\bibitem [{\citenamefont {Heil}\ and\ \citenamefont {Hazel}(2006)}]{HeHa2006}%
  \BibitemOpen
  \bibfield  {author} {\bibinfo {author} {\bibfnamefont {M.}~\bibnamefont
  {Heil}}\ and\ \bibinfo {author} {\bibfnamefont {A.~L.}\ \bibnamefont
  {Hazel}},\ }in\ \href@noop {} {\emph {\bibinfo {booktitle} {Fluid-Structure
  Interaction}}},\ \bibinfo {editor} {edited by\ \bibinfo {editor}
  {\bibfnamefont {H.-J.}\ \bibnamefont {Bungartz}}\ and\ \bibinfo {editor}
  {\bibfnamefont {M.}~\bibnamefont {Sch{\"a}fer}}}\ (\bibinfo  {publisher}
  {Springer Berlin Heidelberg},\ \bibinfo {address} {Berlin, Heidelberg},\
  \bibinfo {year} {2006})\ pp.\ \bibinfo {pages} {19--49}\BibitemShut {NoStop}%
\bibitem [{\citenamefont {Thiele}(2010)}]{Thie2010}%
  \BibitemOpen
  \bibfield  {author} {\bibinfo {author} {\bibfnamefont {U.}~\bibnamefont
  {Thiele}},\ }\href {\doibase 10.1088/0953-8984/22/8/084019} {\enquote
  {\bibinfo {title} {Thin film evolution equations from (evaporating) dewetting
  liquid layers to epitaxial growth},}\ } (\bibinfo {year} {2010})\BibitemShut
  {NoStop}%
\bibitem [{\citenamefont {Engelnkemper}(2017)}]{Engelnkemper2017}%
  \BibitemOpen
  \bibfield  {author} {\bibinfo {author} {\bibfnamefont {S.}~\bibnamefont
  {Engelnkemper}},\ }\emph {\bibinfo {title} {{N}ichtlineare {A}nalyse
  physikochemisch getriebener {E}ntnetzung - {S}tatik und {D}ynamik}},\
  \href@noop {} {Ph.D. thesis},\ \bibinfo  {school} {Westfälische
  Wilhelms-Universität Münster} (\bibinfo {year} {2017})\BibitemShut
  {NoStop}%
\bibitem [{\citenamefont {Bonn}\ \emph {et~al.}(2009)\citenamefont {Bonn},
  \citenamefont {Eggers}, \citenamefont {Indekeu}, \citenamefont {Meunier},\
  and\ \citenamefont {Rolley}}]{BEIM2009romp}%
  \BibitemOpen
  \bibfield  {author} {\bibinfo {author} {\bibfnamefont {D.}~\bibnamefont
  {Bonn}}, \bibinfo {author} {\bibfnamefont {J.}~\bibnamefont {Eggers}},
  \bibinfo {author} {\bibfnamefont {J.}~\bibnamefont {Indekeu}}, \bibinfo
  {author} {\bibfnamefont {J.}~\bibnamefont {Meunier}}, \ and\ \bibinfo
  {author} {\bibfnamefont {E.}~\bibnamefont {Rolley}},\ }\href {\doibase
  10.1103/RevModPhys.81.739} {\bibfield  {journal} {\bibinfo  {journal}
  {Reviews of Modern Physics}\ }\textbf {\bibinfo {volume} {81}},\ \bibinfo
  {pages} {739} (\bibinfo {year} {2009})}\BibitemShut {NoStop}%
\bibitem [{\citenamefont {Kubochkin}\ and\ \citenamefont
  {Gambaryan-Roisman}(2021)}]{Kubochkin2021}%
  \BibitemOpen
  \bibfield  {author} {\bibinfo {author} {\bibfnamefont {N.}~\bibnamefont
  {Kubochkin}}\ and\ \bibinfo {author} {\bibfnamefont {T.}~\bibnamefont
  {Gambaryan-Roisman}},\ }\href {\doibase 10.1103/PhysRevFluids.6.093603}
  {\bibfield  {journal} {\bibinfo  {journal} {Phys. Rev. Fluids}\ }\textbf
  {\bibinfo {volume} {6}},\ \bibinfo {pages} {093603} (\bibinfo {year}
  {2021})}\BibitemShut {NoStop}%
\bibitem [{\citenamefont {de~Ruijter}, \citenamefont {Blake},\ and\
  \citenamefont {De~Coninck}(1999{\natexlab{b}})}]{mkt_zeta_calc}%
  \BibitemOpen
  \bibfield  {author} {\bibinfo {author} {\bibfnamefont {M.~J.}\ \bibnamefont
  {de~Ruijter}}, \bibinfo {author} {\bibfnamefont {T.~D.}\ \bibnamefont
  {Blake}}, \ and\ \bibinfo {author} {\bibfnamefont {J.}~\bibnamefont
  {De~Coninck}},\ }\href {\doibase 10.1021/la990171l} {\bibfield  {journal}
  {\bibinfo  {journal} {Langmuir}\ }\textbf {\bibinfo {volume} {15}},\ \bibinfo
  {pages} {7836} (\bibinfo {year} {1999}{\natexlab{b}})},\ \Eprint
  {http://arxiv.org/abs/https://doi.org/10.1021/la990171l}
  {https://doi.org/10.1021/la990171l} \BibitemShut {NoStop}%
\bibitem [{\citenamefont {Ingebrigtsen}\ and\ \citenamefont
  {Toxvaerd}(2007)}]{Toxvaerd2007}%
  \BibitemOpen
  \bibfield  {author} {\bibinfo {author} {\bibfnamefont {T.}~\bibnamefont
  {Ingebrigtsen}}\ and\ \bibinfo {author} {\bibfnamefont {S.}~\bibnamefont
  {Toxvaerd}},\ }\href {\doibase 10.1021/jp0676235} {\bibfield  {journal}
  {\bibinfo  {journal} {The Journal of Physical Chemistry C}\ }\textbf
  {\bibinfo {volume} {111}},\ \bibinfo {pages} {8518} (\bibinfo {year}
  {2007})},\ \Eprint {http://arxiv.org/abs/https://doi.org/10.1021/jp0676235}
  {https://doi.org/10.1021/jp0676235} \BibitemShut {NoStop}%
\bibitem [{\citenamefont {Walton}\ \emph {et~al.}(1983)\citenamefont {Walton},
  \citenamefont {Tildesley}, \citenamefont {Rowlinson},\ and\ \citenamefont
  {Henderson}}]{calc_surf_tension}%
  \BibitemOpen
  \bibfield  {author} {\bibinfo {author} {\bibfnamefont {J.}~\bibnamefont
  {Walton}}, \bibinfo {author} {\bibfnamefont {D.}~\bibnamefont {Tildesley}},
  \bibinfo {author} {\bibfnamefont {J.}~\bibnamefont {Rowlinson}}, \ and\
  \bibinfo {author} {\bibfnamefont {J.}~\bibnamefont {Henderson}},\ }\href
  {\doibase 10.1080/00268978300100971} {\bibfield  {journal} {\bibinfo
  {journal} {Molecular Physics}\ }\textbf {\bibinfo {volume} {48}},\ \bibinfo
  {pages} {1357} (\bibinfo {year} {1983})},\ \Eprint
  {http://arxiv.org/abs/https://doi.org/10.1080/00268978300100971}
  {https://doi.org/10.1080/00268978300100971} \BibitemShut {NoStop}%
\bibitem [{\citenamefont {Rotenberg}, \citenamefont {Boruvka},\ and\
  \citenamefont {Neumann}(1983)}]{ROTENBERG1983169}%
  \BibitemOpen
  \bibfield  {author} {\bibinfo {author} {\bibfnamefont {Y.}~\bibnamefont
  {Rotenberg}}, \bibinfo {author} {\bibfnamefont {L.}~\bibnamefont {Boruvka}},
  \ and\ \bibinfo {author} {\bibfnamefont {A.}~\bibnamefont {Neumann}},\ }\href
  {\doibase https://doi.org/10.1016/0021-9797(83)90396-X} {\bibfield  {journal}
  {\bibinfo  {journal} {Journal of Colloid and Interface Science}\ }\textbf
  {\bibinfo {volume} {93}},\ \bibinfo {pages} {169} (\bibinfo {year}
  {1983})}\BibitemShut {NoStop}%
\bibitem [{\citenamefont {Kwok}\ and\ \citenamefont
  {Neumann}(1999)}]{KWOK1999167}%
  \BibitemOpen
  \bibfield  {author} {\bibinfo {author} {\bibfnamefont {D.}~\bibnamefont
  {Kwok}}\ and\ \bibinfo {author} {\bibfnamefont {A.}~\bibnamefont {Neumann}},\
  }\href {\doibase https://doi.org/10.1016/S0001-8686(98)00087-6} {\bibfield
  {journal} {\bibinfo  {journal} {Advances in Colloid and Interface Science}\
  }\textbf {\bibinfo {volume} {81}},\ \bibinfo {pages} {167} (\bibinfo {year}
  {1999})}\BibitemShut {NoStop}%
\bibitem [{\citenamefont {Tewes}\ \emph {et~al.}(2017)\citenamefont {Tewes},
  \citenamefont {Buller}, \citenamefont {Heuer}, \citenamefont {Thiele},\ and\
  \citenamefont {Gurevich}}]{TBHT2017tjocp}%
  \BibitemOpen
  \bibfield  {author} {\bibinfo {author} {\bibfnamefont {W.}~\bibnamefont
  {Tewes}}, \bibinfo {author} {\bibfnamefont {O.}~\bibnamefont {Buller}},
  \bibinfo {author} {\bibfnamefont {A.}~\bibnamefont {Heuer}}, \bibinfo
  {author} {\bibfnamefont {U.}~\bibnamefont {Thiele}}, \ and\ \bibinfo {author}
  {\bibfnamefont {S.~V.}\ \bibnamefont {Gurevich}},\ }\href {\doibase
  10.1063/1.4977739} {\bibfield  {journal} {\bibinfo  {journal} {The Journal of
  Chemical Physics}\ }\textbf {\bibinfo {volume} {146}},\ \bibinfo {pages}
  {094704} (\bibinfo {year} {2017})},\ \Eprint
  {http://arxiv.org/abs/https://doi.org/10.1063/1.4977739}
  {https://doi.org/10.1063/1.4977739} \BibitemShut {NoStop}%
\end{thebibliography}
\end{document}